\documentclass[a4paper,11pt,table]{article}

\pdfoutput=1 % if your are submitting a pdflatex (i.e. if you have
\usepackage{jheppub} % for details on the use of the package, please
\usepackage[T1]{fontenc} % if needed
\usepackage[utf8]{inputenc}

\usepackage{youngtab}
\usepackage{young}

\usepackage{xcolor}
\usepackage{draft} 
\usepackage{hyperref}
\usepackage{graphicx,color}
\usepackage{tcolorbox}
\usepackage{tikz-cd} 
\usepackage[all]{xy}
\usepackage{array,multirow}\usepackage{float}

\usetikzlibrary{arrows,chains,matrix,positioning,scopes}
\makeatletter
\tikzset{join/.code=\tikzset{after node path={%
			\ifx\tikzchainprevious\pgfutil@empty\else(\tikzchainprevious)%
			edge[every join]#1(\tikzchaincurrent)\fi}}}
\makeatother
\tikzset{>=stealth',every on chain/.append style={join},
	every join/.style={->}}
\tikzstyle{labeled}=[execute at begin node=$\scriptstyle,
execute at end node=$]

\title{\boldmath $4d$ $\cN=2$ SCFT from Complete Intersection Singularity}

\author[a,c]{Yifan Wang}
\author[b,c]{Dan Xie}
\author[d]{Stephen S.-T. Yau}
\author[b,c]{Shing-Tung Yau}

\affiliation[a]{Center for Theoretical Physics, Massachusetts Institute of Technology,Cambridge, 02139, USA}
\affiliation[b]{Center of Mathematical Sciences and Applications, Harvard University, Cambridge, 02138, USA}
\affiliation[c]{Jefferson Physical Laboratory, Harvard University, Cambridge, MA 02138, USA} 
\affiliation[d]{Department of Mathematical Sciences, Tsinghua  University, Beijing, 100084, P.R. China}

\emailAdd{yifanw@mit.edu}
\emailAdd{dxie@cmsa.fas.harvard.edu}
\emailAdd{yau@uic.edu}
\emailAdd{yau@math.harvard.edu}

\abstract{Detailed studies of four dimensional $\mathcal{N}=2$ superconformal field theories (SCFT) defined by isolated complete intersection singularities are performed: we compute 
the Coulomb branch spectrum, Seiberg-Witten solutions and central charges. Most of our theories have exactly marginal deformations and we identify
the weakly coupled gauge theory descriptions for many of them,  which involve (affine) $D$ and $E$ shaped quiver gauge theories and theories formed from Argyres-Douglas matters. 
These investigations provide strong evidence for 
the singularity approach in classifying 4d $\mathcal{N}=2$ SCFTs. }

\begin{document} 
	\rightline{MIT-CTP/4813 }
	\maketitle
	
	\flushbottom
	
	%\tableofcontents
	
	%

\section{Introduction}\label{sect:I}
Six dimensional $(2,0)$ theory has a remarkable ADE classification, which can actually be derived from the classification of two dimensional {\it isolated rational Gorenstein} singularities \cite{Witten:1995zh}. By compactifiying the
$6d$ theory on torus \cite{Witten:2009at}, one can derive the classification of four dimensional $\mathcal{N}=4$ superconformal field theories (SCFT)\footnote{There are some subtleties involving non-local objects \cite{Kapustin:2005py,Aharony:2013hda,Tachikawa:2013hya,Xie:2013vfa}.}. 
The natural next step is to classify four dimensional $\mathcal{N}=2$ SCFTs \cite{Seiberg:1994aj,Seiberg:1994rs} and it transpires that the space of such theories is surprisingly large due to less supersymmetry and the possibility of non-Lagrangian theories \cite{Argyres:1995jj,Argyres:1995xn,Eguchi:1996vu,Minahan:1996fg,Minahan:1996cj,Cecotti:2010fi}.
Those intrinsically strongly coupled theories make the classification much more difficult, and the traditional field theory tools become inadequate.
 
 It turns out that the geometric tools are more suitable to implement classification: one can use wrapped M5 branes to engineer a large class of new $\mathcal{N}=2$ SCFTs \cite{Witten:1997sc,Gaiotto:2009we,Gaiotto:2009hg,Xie:2012hs,Wang:2015mra}, and the classification of these theories is reduced to that of 
 the punctures \cite{Witten:2007td,Chacaltana:2012zy,Wang:2015mra}; another seemingly much larger space of theories can be constructed using singularity theory \cite{Shapere:1999xr, Xie:2015rpa}\footnote{See  \cite{arnold2012singularitytheory,Arnold1993,arnold1985singularitiesv1,arnold2012singularitiesv2} for an introduction to singularity theory.}, and the classification of SCFTs boils down to the classification of three dimensional {\it isolated rational Gorenstein} singularities \cite{Xie:2015rpa}.\footnote{See recent work of \cite{Argyres:2015ffa,Argyres:2015ffa,Argyres:2016xua} for classifications of rank 1 theories based on the Kodaira classification.}  
 The isolated hypersurface singularities corresponding to $\mathcal{N}=2$ SCFTs have been classified in \cite{Yau:2003uk,Xie:2015rpa}, and recently this program has been extended to the isolated complete intersection singularities (ICIS) \cite{Chen:2016bzh}. 
 The major result of \cite{Chen:2016bzh} is that to obtain an $\mathcal{N}=2$ SCFT, the ICIS must be defined by at most two polynomials $(f_1, f_2)$, and 
 there are in total 303 classes of singularities (many classes involve infinite sequences of singularities).
	
The purpose of this paper is to study properties of these new $\mathcal{N}=2$ SCFTs engineered using ICISs in \cite{Chen:2016bzh}. We will explore various physical properties of these theories such as 
the Coulomb branch spectrum, Seiberg-Witten solutions \cite{Seiberg:1994aj,Seiberg:1994rs}, central charges \cite{Shapere:2008zf}, etc. using the data of the singularities. We also identify the weakly coupled gauge theory 
descriptions for some of these theories and compute various quantities from field theory techniques. The results are in complete agreements with those derived from singularity theory. Many of these 
theories take the form of (affine) $D$ or $E$ shaped quivers.  We view these checks 
as compelling evidence for the power of our approach of employing singularity theory to classify four dimensional $\mathcal{N}=2$ SCFTs. 	

The theory defined by an ICIS has some interesting and new features compared with the theory defined by a hypersurface singularity, for example, the multiplicity of the Coulomb branch operators (scalar chiral primaries)
with the maximal scaling dimension can be larger than one; the number of $A_1$ singularities (co-dimension one singularities on the Coulomb branch or the base of the Milnor fibration) is different from the Milnor number, etc. These models 
imply that some of the features of the SCFTs from hypersurface singularities may not be generic. 
	
This paper is organized as follows: section \ref{ICIS-SCFT} explains how to derive physical properties of the SCFTs from the data of singularities; section \ref{quiver} describes the weakly coupled gauge theory descriptions for 
theories engineered using the singularities; we conclude with a short summary and discussion for future directions in section \ref{dis}.

\section{$4d$ $\mathcal{N}=2$ SCFT and ICIS}\label{ICIS-SCFT}
In this section, we will start by reviewing general properties of $4d$ $\cN=2$ SCFTs, and then explain the geometric constructions of a large class of such theories from ICISs. In particular, we will describe the necessary conditions on the ICISs to give rise to $4d$ $\cN=2$ SCFTs, and
 how to read off the Coulomb branch spectrum and conformal central charges of the $4d$ theory from the singularity.
	
\subsection{Generality of $4d$ $\mathcal{N}=2$ SCFT}	
The $\mathcal{N}=2$ superconformal group $SU(2,2|2)$ contains the bosonic conformal group $SO(4,2)$ and $SU(2)_R\times U(1)_R$ R-symmetry. The unitary irreducible representations of $\mathcal{N}=2$ superconformal algebra have been classified in \cite{Dolan:2002zh}, and a highest weight state
is labeled as $|\Delta, R, r, j_1, j_2\rangle$, where $\Delta$ is the scaling dimension, $R$ labels the representation of $SU(2)_R$ symmetry,  $r$ is the charge for $U(1)_R$ symmetry, and $j_1, j_2$ 
are the left and right spins. The important half-BPS operators (short multiplets) are ${\cal E}_{r, (0,0)}$ ($\Delta=r$) and $\hat{B}_R$ ($\Delta=2R$), whose expectation values parametrize the Coulomb branch and the Higgs branch of the vacuum moduli space of the $\cN=2$ SCFT respectively. Moreover if $1<\Delta[{\cal E}_{r, (0,0)}]\leq 2$, one can turn on the following relevant or exactly marginal deformations\footnote{We denote the $4+4$ supercharges by $Q,\widetilde Q$, suppressing the spacetime and R-symmetry indices.}
\begin{equation}
\delta S= \lambda \int d^4x\, \widetilde Q^4 {\cal E}_{r, (0,0)} +c.c 
\end{equation}
One can assign the scaling dimension  to $\lambda$ which satisfies the condition $\Delta[\lambda]+\Delta[{\cal E}_{r, (0,0)}]=2$. If there is an operator $\hat{B}_1$ in the spectrum, one can also have the following relevant deformation
\begin{equation}
\delta S=m\int d^4x \,\widetilde Q^2  \hat B_1+c.c
\end{equation}
and $m$ has scaling dimension one, which in Lagrangian theories gives the usual $\cN=2$ mass deformation. The above deformations are all of the $\mathcal{N}=2$ preserving relevant or marginal deformations \cite{Argyres:2015ffa,Cordova:2016xhm}. The IR physics on the Coulomb branch depends on the parameters 
$(m, \lambda, u)$ where $u\equiv \langle {\cal E}_{r, (0,0)} \rangle$ is the expectation value of the Coulomb branch operator. To solve the Coulomb branch, we need to achieve the following goals
\begin{itemize}
\item Determine the Coulomb branch spectrum, namely the scaling dimensions of the parameters $(m, \lambda, u)$.
\item Determine the Seiberg-Witten solution, which is often described by a family of geometric objects 
\begin{equation}
F(z,m,\lambda, u)=0.
\end{equation}
$F$ can be a one-fold or a three-fold. 
\end{itemize}

As for the Higgs branch, parametrized by the operators $\hat{B}_1$, the question is to determine the flavor symmetry group $G$ and the corresponding affine ring, namely to 
identify the generators and relations for the Higgs branch. 

\subsection{Geometric engineering and 2d/4d correspondence}
One can engineer a four dimensional $\mathcal{N}=2$ SCFT by starting with a three dimensional  {\it graded rational Gorenstein}  singularity \cite{Xie:2015rpa}. Graded implies that the three-fold singularity should 
have a $\mC^{*}$ action which is required by the $U(1)_R$ symmetry of $\mathcal{N}=2$ SCFTs. Gorenstein means that there is a distinguished top form $\Omega$ which will be identified with the Seiberg-Witten differential. The rational 
condition ensure that the Coulomb branch operators have the positive $U(1)_R$ charge, or the top form $\Omega$ has positive charge. 

In the case of ICISs defined by $f=(f_1,f_2,\dots,f_k) \in \mC[x_1,x_2,\dots, x_{n}]$ in $\mC^n$ with $n=k+3\geq 5$, the above conditions become 
\begin{itemize}
\item Gorenstein: this is automatic for ICISs.
\item $\mC^*$ action: a set of positive weights $w_a$ and $d_i$ such that $f_i(\eta^{w_a} x_a)=\eta^{d_i}f_i(x_a)$
\item Rationality: $\sum_{a=1}^n w_a > \sum_{i=1}^k d_i$
\end{itemize}
It was recently proved in \cite{Chen:2016bzh} that the above conditions demand $k=2$.

Consider now graded  ICISs which are defined by two polynomials $f=(f_1, f_2)$, which defines the map $f: (\mC^5,0)\rightarrow (\mC^2,0)$. We can denote the charge of the coordinates $x_a, a=1,\ldots 5$ and 
the degree of two polynomials by $(w_1, w_2, w_3, w_4, w_5; d_1, d_2)$ (up to an overall normalization). The rational condition is simply 
\begin{equation}
\sum_{i=1}^5 w_i >\sum_{i=1}^2 d_i.
\end{equation}
We can normalize the charges such that $d_1=1$ and $d_2\leq 1$ without losing any generality. 

The above constraint can be interpreted from string theory. Considering type II string probing an ICIS. To decouple gravity, we send the string coupling $g_s\rightarrow 0$ while keeping the string scale $\ell_s$ fixed, this way we could end up with a non-gravitational 4$d$ little string theory (LST) whose holographic dual is described by type II string theory in the background
	\ie
	\mR^{3,1}\times \mR_{\phi} \times (S^1\times LG(W))/\Gamma
	\fe
	with a suitable GSO projection $\Gamma$ that acts as an orbifold to ensure  4$d$ $\cN=2$ spacetime supersymmetry \cite{Ooguri:1995wj,Giveon:1999zm,Kutasov:2001uf}. Here $\mR_{\phi}$ denotes the $\cN=1$ linear dilaton SCFT with dilaton profile $\varphi=-{Q\over 2}\phi$. 
	For  an ICIS, $LG(W)$ is the $\cN=2$ Landau-Ginzburg (LG) theory with chiral superfields $z_a$ and $\Lambda$.\footnote{This has been considered for Calabi-Yau case in \cite{Greene:1988ut,greene1990} which extends naturally to the Fano case we consider here.} Assuming $d_1\geq d_2$ and normalizing the weights on $z_a$ such that $d_1=1$, the superpotential is given by:
	\ie
	W=f_1(z_a)+\Lambda  f_2 (z_a),
	\fe
	where  the $\mC^*$ action on $z_a$ is identified with the $U(1)_R$ charge of $z_a$ in the LG model. 
	The $U(1)_R$ charges of $\Lambda$ is fixed to be $1-d_2$ so that $W$ is quasi-homogeneous with charge $1$.  In the low energy limit ($\ell_s\rightarrow 0$), we expect to recover the 4$d$ SCFT from the LST.
	
	Now the $\cN=1$ linear dilaton theory has central charge $3(1/2+Q^2)$, whereas the $\cN=2$ LG model has central charge $3\hat c$. Consistency of the type II string theory on this background requires the worldsheet theory to have a total central charge of 15 which implies $Q^2=2-\hat c$ thus $\hat c<2$.
	On the other hand, the central charge of the LG theory is determined by the $U(1)_R$ charges of the chiral superfields,
	\ie
	\hat c=\sum_{a=1}^5 \left(1-{2w_a}\right)+(1-2(1-d_2))=4-2\sum_{a=1}^5w_a+2d_2.
	\fe
	thus the condition $\hat c<2$ is equivalent to 
	\begin{equation}
	 \sum_{a=1}^5w_a>1+d_2,
	\end{equation}
	which is exactly the rationality condition for the singularity. Such $2d$ LG models are much less studied in the context of 2d $(2,2)$ SCFTs, in particular, the 
	chiral ring structure is little explored. It is definitely useful to have a deeper understanding of these $2d$ theories so that the corresponding $4d$ theories can be understood better. 
	In this paper, we will rely on the intrinsic properties of singularities to study 4d $\cN=2$ theories. 
	
	\subsection{Mini-versal deformations and Seiberg-Witten solution}
	The  Seiberg-Witten (SW) solution of the $\mathcal{N}=2$ SCFT defined by an ICIS is identified with the mini-versal deformations of the singularity. 
	Given a complete intersection singularity specified by polynomials $f=(f_2, f_2)$, the mini-versal deformations are captured by the Jacobi module
	\ie
	{\cJ}={\mC^2[x_1,x_2,\dots, x_5]\over \left({\pa f_i\over \pa x_a}\right)}.
	\label{jacobi}
	\fe 
	We denote a monomial basis of the Jacobi module by  $\phi_{\alpha}$  which are $2\times 1$ column vectors with only one non-zero entry. 
	The mini-versal deformation of the ICIS is defined as 
	\ie
	F(\lambda, z)=f(z)+\sum_{\alpha=1}^\mu \lambda_{\alpha} \phi_{\alpha},
	\label{mf}
	\fe
	with the holomophic 3-form 
	\ie
	\Omega={d x_1\wedge dx_2\wedge\dots\wedge dx_5\over dF_1 \wedge d F_2},
	\label{h3}
	\fe
	which describes the Milnor fibration of deformed 3-folds over the space of parameters $\lambda_\A$.
	Here $\mu$ is the dimension of the Jacobi module and is also called Milnor number which counts the middle homology cycles (which are $S^3$ topologically) of the deformed 3-fold. The basis $\phi_\A$ of the Jacobi module for ICIS can be computed 
	using the software Singular \cite{DGPS}, and the result is also listed in \cite{Chen:2016bzh}.

As explained in the previous subsection, type IIB string theory on this singular background gives rise to a $4d$ $\cN=2$ SCFT. 
The coefficients $\lambda_\alpha$ in \eqref{mf} are identified with the Coulomb branch parameters of  $4d$ theory. 
The (assumed) $\mC^*$ action on the singularity descends to the Jacobi module and is interpreted as (proportional to) the $U(1)_R$ symmetry of the $4d$ SCFT. The rank of the BPS charge lattice is given by $\m$. The BPS particles in the $4d$ theory comes from D3 branes wrapping special Lagrangian 3-cycles in the deformed 3-fold. Their masses are determined by the integral of $\Omega$ over the corresponding homology cycles. 

Demanding the mass to have scaling dimension 1, we fix the relative normalization between the $\mC^*$ and $U(1)_R$ charges .
	The scaling dimension ($U(1)_R$ charge) of a Coulomb branch parameter is then given by
	\ie
	\Delta[\lambda_{\alpha}]={d_j-Q[\phi_\alpha]\over \sum_{a=1}^{5} w_a-d_1-d_2}.
	\label{scale}
	\fe 
	where $w_a>0$ are the $\mC^*$ charges of $x_a$. 	Here only the $j$-th entry of $\phi_\alpha$ is nonzero. The Coulomb branch scaling dimensions $\Delta[\lambda_{\alpha}]$ are symmetric around $1$ as shown in \cite{wahl1987jacobian}, which is 
	in agreement with field theory result.
	
	The Jacobi module of a graded ICIS will be captured by the following Poincare polynomial 
	\begin{equation}
	P(z)=\sum z^{\alpha} \text{dim} H_{\alpha}.
	\end{equation}
	Here $\dim H_{\alpha}$ counts the number of basis elements in Jacobi module whose $\mC^*$ charge is $\alpha$. The Poincare polynomial has the following simple form \cite{Greuel1978} for the 3-fold ICISs here:
	\ie
	P(z)=
	{\rm res}_{t=0} {1\over t^{4}(1+t)} \left[{\prod_{a=1}^5 (1+tz^{w_a})\prod_{i=1}^2 (1-z^{d_i})\over \prod_{a=1}^5 (1-z^{w_a})\prod_{i=1}^2 (1+tz^{d_i})} +t
	\right]
	\fe
 %	\ie
%	P(z)={\prod_{i=1}^2 (1-z^{d_i})\over\prod_{a=1}^5 (1-z^{w_a})}
%	+z^{-c}\left[
%	1+{\rm res}_{t=0} {1\over t^{2}(1+t)} {\prod_{a=1}^5 (1+tz^{w_a})\prod_{i=1}^2 (1-z^{d_i})\over \prod_{a=1}^5 (1-z^{w_a})\prod_{i=1}^2 (1-tz^{d_i})}
%	\right]
%	\fe
	The Milnor number can be computed from the Poincare polynomial by setting 
	$z=1$:
\ie
\m=P(1)=\begin{cases}
\prod_{a=1}^5 \left({d\over w_a}-1\right)\left(4+\sum_{a=1}^5  { w_i\over d-w_a }\right)
%\sum_{1\leq  \n_1<\dots <\n_4 \leq 5}\prod_{\lambda=1}^4 \left({d\over w_{\n_\lambda}}-1\right)
%+4\sum_{1\leq  \n_1<\dots <\n_5 \leq 5}\prod_{\lambda=1}^5 \left({d\over w_{\n_\lambda}}-1\right)
%\sum_{4\leq \ell\leq 5, 1\leq  \n_1<\dots <\n_l \leq 5} \begin{pmatrix} \ell-1 \\3\end{pmatrix}\prod_{\lambda=1}^\ell \left({d\over w_{\n_\lambda}}-1\right)
 & d_1=d_2=d
\\
  \prod_{a=1}^5 \left({d_1\over w_a}-1\right) {d_2 \over d_1-d_2}+\prod_{a=1}^5 \left({d_2\over w_a}-1\right) {d_1 \over d_2-d_1}
  & d_1\neq d_2
  \end{cases}
  \label{milnor}
\fe
 
%\ie
%	&\mu=P(1)
%	\\
%	&={1\over {w_1 w_2 w_3 w_4 w_5} } \times
%	\\
%	&
%\Bigg(d_2 d_1^3 (d_2-w_1-w_2-w_3-w_4-w_5)+d_2 d_1^2 \Bigg(-d_2 (w_1+w_2+w_3+w_4+w_5)
%\\
%&+d_2^2+w_2 w_3+w_2 w_4+w_3 w_4+w_2 w_5+w_3 w_5+w_4 w_5+w_1 (w_2+w_3+w_4+w_5)\Bigg)
%\\
%&+d_2 d_1 \left(-d_2^2 (w_1+w_2+w_3+w_4+w_5)+d_2 (w_3 w_4+w_5 w_4+w_3 w_5+w_2 (w_3+w_4+w_5)\right.
%\\
%&+w_1 (w_2+w_3+w_4+w_5))+d_2^3-w_2 w_3 w_4-w_2 w_3 w_5-w_2 w_4 w_5-w_3 w_4 w_5
%\\
%&\left.-w_1 (w_4 w_5+w_3 (w_4+w_5)+w_2 (w_3+w_4+w_5))\right)+d_2 d_1^4+w_1 w_2 w_3 w_4 w_5 \Bigg)
%\fe

\subsection{Central charges $a$ and $c$}	
Knowing the full spectrum of Coulomb branch parameters, we can compute the conformal central charges of the 4d $\cN=2$ SCFT using the following formula \cite{Shapere:2008zf}:
\begin{equation}
\label{eq:acformula}
a={R(A)\over 4}+{R(B)\over6}+{5 r\over 24},\quad c={R(B)\over 3}+{r\over 6},
\end{equation}
where $r$ is the rank of the Coulomb branch and we have used the fact that the generic fibre of the Milnor fibration \eqref{mf} has only non-vanishing middle cohomology, thus there is no free massless hypermultiplets at a generic 
point on the Coulomb branch. $R(A)$ can be computed straightforwardly from the Coulomb branch spectrum which can be solved using the SW solution given above. 
The key point is to compute $R(B)$. In the hypersurface case, $R(B)$ takes 
an elegant form \cite{Xie:2015rpa}:
\begin{equation}
R(B)={\mu u_{\max}\over 4};
\end{equation}
Here $\mu$ is the Milnor number which is equal to the dimension of the charge lattice of our field theory, and $u_{max}$ is the maximal scaling dimension of our theory. 

For SCFTs defined by ICIS $(f_1, f_2)$ with weights $(w_1, w_2, w_3, w_4, w_5; 1, d_2)$ and $d_2\leq 1$,  we propose that $R(A)$ and $R(B)$ are given by the following formula:
\begin{equation}
R(A)=\sum_{\Delta[u_i]>1}([u_i]-1),\quad R(B)={1\over 4}\mu' \alpha.
\label{c2}
\end{equation}
Here $\mu'$ counts the ``effective'' number of $A_1$ singularities after generic deformations, which differs from the Milnor number $\m$, unlike in the case of hypersurface singularities, and $\alpha$ is the scaling dimension 
of one of the Coulomb branch operator which can be expressed as follows (with $d_2\leq 1$):
\begin{equation}
\alpha={1\over \sum_{i=1}^5 w_i-1-d_2}.
\label{alpha}
\end{equation}
Notice that this is the scaling dimension of the operator associated with the deformation $(f_1+t,f_2)$. 
The effective number of $A_1$ singularities is given by the following formula
\ie
	\mu'=\mu+\mu_{1},
	\label{RB}
	\fe
where $\mu$ is the Milnor number of the ICIS, and $\mu_1$ is the Milnor number for $f_1$ as an isolated hypersurface singularity in $\mC^5$,
\begin{equation}
\mu_1=\prod_{i=1}^5({1\over w_i}-1). 
\label{c3}
\end{equation}

Let's explain how the effective number of $A_1$ singularities are computed. We have an ICIS $(f_1, f_2)$ with weights  $(w_1, w_2, w_3, w_4, w_5; 1, d_2)$ and $d_2\leq 1$. We first deal with the situation where $f_1$ defines an isolated hypersurface singularity, and consider $f_2$ over the coordinates restricted on the variety $X_{f_1}=\{f_1(x_a)=0\}\subset \mC^5$. By deforming $f_2$ to $f_2'= f_2+t$,  
the critical points of $f_2'$ over $X_{f_1}$ are Morse (the critical points are $A_1$ singularities). The number of such $A_1$ singularities is $\mu'=\mu+\mu_{1}$ \cite{Arnold1993}, and $\mu_1$ is the 
Milnor number of $f_1$. The scaling dimension of the coordinates ({i.e Coulomb branch parameters, not to be confused with $x_a$}) near the Morse critical points are determined by the scaling dimension of the polynomial $f_1$, which is given by   
\begin{equation}
\alpha={1\over \sum_{i=1}^5 w_i-1-d_2};
\end{equation}
This gives the $U(1)_R$ charge of the coordinates near the $A_1$ singularities, and explains the formula for $R(B)$ using the result in \cite{Shapere:2008zf}.

What is quite amazing is  that even in the case that neither $f_1$ nor $f_2$ defines an isolated singularity, one can still use the above formula to compute 
the central charge although now $\mu^{'}$ is fractional, and the results match with the field theory expectations as we will see in the subsequent section.

\section{Gauge Theory Descriptions}\label{quiver}
In the previous section, we have explained how to extract Coulomb branch data from the geometric information associated with the singularities (ICISs). In this section, we will use these information to identify the gauge theory descriptions for a subclass of such constructions, thus providing nontrivial evidences for the general geometric constructions of $4d$ $\cN=2$ SCFTs from ICISs.

 \subsection{Strategy of finding gauge theory descriptions}
 For some of the theories engineered using ICISs, the Coulomb branch spectrum contains operators with scaling dimension 2, which give rise to exactly marginal deformations.
 It appears that for $4d$ $\mathcal{N}=2$ SCFTs with such deformations, one can always 
 find weakly coupled gauge theory descriptions.\footnote{It is interesting to prove or disprove this statement.} We would like 
 to find at least one weakly coupled gauge theory descriptions for our SCFTs with exactly marginal deformations. Unfortunately we 
 do not have a systematic procedure at the moment and will take a brutal force approach: we compute the Coulomb branch spectrum from the ICIS
 and then try to guess a consistent quiver gauge theory such that the Coulomb branch spectrum matches (with additional consistency conditions such as central charges\footnote{We would like to emphasize here that there are examples of 4d $\cN=2$ SCFTs with the same Coulomb branch spectrum and flavor symmetries yet different conformal central charges.}). Even with this naive approach, we 
 do find many interesting quiver gauge theories, and we compute the central charge from the quiver gauge theory which agrees with the result from that 
 using singularity theory.

\begin{table}
	\begin{center}
		\begin{tabular}{ |c|c| c|c|c|c| }
			\hline
			$G$ & 
			$\dim G$ & $h^\vee$  & $\{d_i\}_{i=1,\dots,\,\rank(G)}$\\ \hline
			$A_{n-1} $  & $n^2-1$ & $n$ & $2,3,\dots,n$ \\     \hline
			$B_{n} $  & $n(2n+1)$ & $2n-1$ & $2,4,\dots,2n$ \\     \hline
			$C_{n} $  & $n(2n+1)$ & $n+1$ & $2,4,\dots,2n$ \\     \hline
			$D_n$   & $n(2n-1)$ & $2n-2$ & $2,4,\dots,2n-2;n$ \\     \hline
			$E_6$  & $78$ & $12$ & $2,5,6,8,9,12$\\     \hline
			$E_7$  & $133$ & $18$ &$2,6,8,10,12,14,18$ \\     \hline
			$E_8$   & $248$ & $30$ & $2,8,12,14,18,20,24,30$ \\     \hline
			$F_4$  & $52$ & $9$ & $2,6,8,12$\\     \hline
			$G_2$  & $14$ & $4$ & $2,6$\\     \hline
		\end{tabular}
	\end{center}
	\caption{Relevant Lie algebra data: $h^\vee$ denotes the Coxeter number and $\{d_i\}$ are the degrees of the fundamental invariants.}
	\label{liedata}
\end{table}

 The essential strategy to identify gauge theory descriptions from the Coulomb branch spectrum consists of the following:
 \begin{enumerate}
 \item Identify candidate gauge groups $G$:
 among the integral scaling dimensions (greater than 1) in the CB spectrum, find the sequences that coincide with the set of fundamental degrees for various Lie groups listed in Table~\ref{liedata}.
 \item Identify candidate $\cN=2$ matter: 
 group the remaining scaling dimensions (those that are larger than 1) into isolated $\cN=2$ SCFTs (many Argyres-Douglas type theories) which serve as non-Lagrangian matters with the necessary flavor symmetries to couple to the gauge groups.
 \item Gluing the pieces: put together the gauge groups and matter by conformal gauging, check beta function
 \ie
 \B_G=2 h^\vee(G)-\sum_\A \kappa_\A(G)
 \fe
 vanishes for all gauge groups $G$. Here we use $\kappa_\A(G)$ to denote the flavor central charges of the matter theory for the symmetry $G$. We choose the normalization that for $G=A,B,C,D$, the flavor central charge $\kappa(G)=1$ for one fundamental hypermultiplet in the $G=A,C$ case and one half-hypermultiplet in the $G=B, D$ case. 
  In the case when $G$ is a subgroup of a larger non-abelian flavor 
 symmetry $J$ in the matter theory,
 the flavor central charges for $G$ 
 is related that of $J$ by $\kappa(G)= I_{G\hookrightarrow J}\kappa(J)$ where $I_{G\hookrightarrow J}$ is known as the Dynkin index of embedding  \cite{Argyres:2007cn}.\footnote{Recall from \cite{Argyres:2007cn} that the Dynkin index of embedding for $G\subset J$ is computed by 
 \ie
 I_{G\hookrightarrow J}={\sum_i T({\bf r}_i)\over T({\bf r})}
 \fe	
where ${\bf r}$ denotes a representation of $J$  which decomposes into $\oplus_i {\bf r}_i$ under $G$, and $T(\cdot)$ computes the quadratic index of the representation (which can be found for example in \cite{mckay1981tables}).
}
 
\item Consistency check: make sure that the rank of remaining flavor symmetry matches with the the CB spectrum from singularity; in addition, if the quiver has a Lagrangian description, we compute the conformal central charges $a$ and $c$ from the gauge theory description using\footnote{For strongly coupled matter theories (e.g. Argyres-Douglas matters), we can think of $n_h$ and $n_v$ as counting the {\it effective} number of vector multiplets and hypermultiplets. To get $a$ and $c$ in those cases, the formula \eqref{eq:acformula} is more useful.}
\ie
a={5n_v+ n_h\over 24},\quad c={2n_v+ n_h \over 12}
\fe    
and compare with that computed from \eqref{eq:acformula}. If the matter system includes the strongly coupled constituents, we also need to include the contributions from these subsectors. 
 \end{enumerate}

We believe that these quiver gauge theory descriptions provide compelling evidence that 
our program of using singularity theory to classify $\mathcal{N}=2$ SCFTs is correct. In the following subsection, we will list some of the interesting 
examples. 

\subsection{Matter system}
The obvious matter system are  free hypermultiplets. We also need to use strongly coupled matter systems such as theories defined by three punctured spheres \cite{Gaiotto:2009hg,Chacaltana:2012zy} and 
Argyres-Douglas matters \cite{Xie:2012hs,Wang:2015mra}. For later convenience, we will summarize some properties about these strongly coupled matter systems. 

\subsubsection{Gaiotto theory}
We can have the strongly coupled matter defined by six dimensional $(2,0)$ type $J=ADE$ SCFTs on a sphere with regular punctures \cite{Gaiotto:2009hg}. We are going to mainly use $J=A_{N-1}$ in which case the regular punctures 
are classified by Young Tableaux $Y= [n_s^{h_s},\ldots, n_2^{h_2}, n_1^{h_1} ]$ with the ordering $n_s>n_{s-1}>\ldots>n_1$.  The flavor symmetry of a regular puncture is given by
\begin{equation}
G_{Y}=[\prod_{i=1}^s U(h_i)]/U(1).
\end{equation}
The flavor central charge of a non-abelian factor $SU(h_i)$ is given by the following formula \cite{Chacaltana:2012zy}: 
\begin{equation}
k_{SU(h_i)}=\sum_{j\leq i} m_js_j,
\end{equation}
here $s_i$ is defined by the transposed Young Tableaux $Y^T=[s_1^{m_1},s_2^{m_2},\ldots, s_{n_s}^{m_{n_s}}]$.

The flavor symmetry of a theory defined by a three punctured sphere is $G=G_{Y_1}\times G_{Y_2} \times G_{Y_3}$, and it can be enhanced to a larger symmetry group using 
the $3d$ mirror as described in \cite{Benini:2010uu}. 

The Coulomb branch spectrum of these theories can be computed as follows: label the boxes of $Y_j$ from $1,2,\ldots n$ row by row, and define the number (pole order associated with $i$-th fundamental invariant) 
\begin{equation}
p_i^{(j)}= i-s_i
\end{equation}
where $s_i$ is the height of the $i$th box in $Y_j$. The number of degree (scaling dimension) $i$ Coulomb branch operators are given by 
\begin{equation}
d_i=\sum_{j=1}^3 p_i^{(j)}-2i+1.
\end{equation}
The conformal central charges $a$ and $c$ can be found (or rather $n_v$ and $n_h$) using the puncture data \cite{Chacaltana:2012zy} or equivelently the method described in \cite{Xie:2013jc}.

\subsubsection{Argyres-Douglas matters} \label{sec:ADmatters}

The class of  Argyres-Douglas (AD) matters (general $\cN=2$ SCFTs with large flavor symmetry) is huge but as we will see in the next subsection, many of the AD matters that show up in the gauge theory descriptions of our theories from ICISs, fall in the simple subclass of $D_p(G)$ theories\footnote{This is denoted by $(G^{(b)}[k],F)$ in the more general classification of \cite{Wang:2015mra} with the restriction $b=h^\vee(G)$ and $p=k+h^\vee (G)$.} \cite{Xie:2012hs,Cecotti:2013lda} and its twisted analog \cite{Wang:2015mra}.  

One type of $D_p(G)$ matter theory that shows up often in our analysis is $D_2 SU(N)$, and its basic property is summarized as follows (more details can be found in \cite{Xie:2012hs,Cecotti:2013lda}):
\begin{itemize}
\item When $N=2n$, $D_2SU(N)$ represents the $SU(n)$ SQCD with $2n$ hypermultiplets in the fundamental representation, and we have the quiver description
\begin{center}
\begin{tabular}{ccc}
	\xymatrix{
		SU(n) \ar@{-}[r] & \boxed{2n}
	}
\end{tabular}
\end{center}
with flavor symmetry $SU(N)_{N\over 2}\times U(1)$.\footnote{For simplicity of notation, we will use subscript to denote the flavor central charge for the global symmetries.}
%
%
%\begin{equation}
%SU(n)-\boxed{2n}
%\end{equation}
\item When $N=2n+1$, $D_2SU(N)$ is an isolated SCFT with flavor symmetry $SU(N)_{N\over 2}$. The Coulomb branch spectrum is 
\ie
{N\over 2}, {N-2\over 2}, \ldots, {3\over2},\underbrace{1,\ldots 1}_{N-1}.
\fe
The conformal central charges are given by
 \begin{equation}
a[D_2(SU(2n+1))]=\frac{7}{24} n (n+1), \quad c[D_2(SU(2n+1)]=\frac{1}{3} n (n+1).
\label{D2SUNac}
 \end{equation}
\end{itemize} 
  We also use $D_n SU(n)$ to denote the following quiver tail
 \begin{center}
 	\begin{tabular}{ccc}
 		\xymatrix{
 		\boxed{1}  \ar@{-}[r] & SU(2) \ar@{-}[r] & SU(3)  \ar@{-}[r] &	\dots  \ar@{-}[r] &	SU(n-1)  \ar@{-}[r] & \boxed{n}
 		}
 	\end{tabular}
 \end{center} 
 with flavor symmetry $SU(n)_{n-1}\times U(1)^{n-1}$.

More general Argyres-Douglas matter can be engineered using 6d $J=ADE$ type $(2,0)$ theory on a sphere with one irregular puncture and one regular puncture \cite{Wang:2015mra}. These theories are denoted by $(J^{(b)}[k], Y)$ where $J^{(b)}[k]$ specifies the irregular puncture and $Y$ labels the regular puncture (which are certain types of Young-Tableaux for classical Lie groups and Bala-Carter labels for exceptional groups \cite{collingwood1993nilpotent,Chacaltana:2012zy}). The simplest cases correspond to $Y=F$ by which we mean the regular puncture is of the full (maximal, principal) type thus enjoys a maximal $J$ flavor symmetry but as we shall see, AD matters from degenerations of the full punctures also show up as building blocks of the $4d$ theories from ICISs.

In addition, we will make use of two matter theories from twisted $D$-type punctures in \cite{Wang:2015mra}. These theories are constructed by introducing a $\mZ_2$ outer-automorphism twist in the compactification of $D$-type $(2,0)$ SCFT on a sphere with one irregular twisted puncture and one full regular twisted puncture. We denote such theories by $(D_n^{(b)}[k],\widetilde F)$. 
\begin{itemize}
	\item The parameters $b,k$ specifies the irregular singularity and satisfy \textbf{i.} $k\in \mZ$ and $b=2n-2$ or \textbf{ii.} $k\in \mZ+1/2$ and $b=n$ \cite{Wang:2015mra}.
	\item  $\widetilde F$ indicates that the regular twisted puncture is full (maximal) which gives rise to $USp(2n-2)$ flavor symmetry with central charge $\kappa=n-{b\over 2(b+k)}$.
	\item  Moreover for $b=2n-2$ and $k$ odd, we have an additional $U(1)$ flavor symmetry. 
\end{itemize}

 \subsection{Examples}
We adopt the labeling of \cite{Chen:2016bzh} to keep track of the ICISs that we are going to study below. For each case, we list the quiver gauge theory description and compute the conformal central charges. The meaning of 
the symbols appearing in each example is 
\begin{itemize}
\item \textbf{ICIS \boldmath$(i)$} denotes the $(i)$-th ICIS appearing in \cite{Chen:2016bzh}, and we record the defining equations here.
\item {\boldmath$(w_1,w_2,w_3,w_4,w_5;1,d)$} are the weights of the coordinates and degrees of the polynomials. We do not normalize the weights here such that $d\leq 1$. 
\item {\boldmath$\mu$} denotes the Milnor number of the ICIS, which is also the dimension of the charge lattice of the SCFT. 
\item {\boldmath$\mathbf{\mu_1}+\mu$} gives the effective number of $A_1$ singularities.
\item {\boldmath$\alpha$} is the scaling dimension near the $A_1$ singularities. 
\item {\boldmath$r$} is the dimension of Coulomb branch, namely counting the part of the spectrum with scaling dimension larger than one.
\item {\boldmath$f$} denotes the number of mass parameters.
\item {\boldmath$a,c$} denote the conformal central charges. 
\end{itemize}

\subsection*{ICIS (1)}
 \begin{flalign*}
\begin{split}
\left\{ \begin{array}{l} x_1^{2}+x_2^{2}+x_3^{2}+x_4^{2}+x_5^{n}=0\\x_1^{2}+2x_2^{2}+3x_3^{2}+4x_4^{2}+5x_5^{n}=0\end{array}\right.
\end{split}&
\end{flalign*}
\\$n\ge2$
\\$(w_1,w_2,w_3,w_4,w_5;1,d)=(\frac{1}{2},\frac{1}{2},\frac{1}{2},\frac{1}{2},\frac{1}{n};1,1)$
\\$\mu=-7 + 8 n,
\quad \m_1=n-1,
\quad \A=n,
\quad r=\begin{cases}
4n-6 & n \in 2\mZ
\\
4n-4 & n \in 2\mZ+1
\end{cases},
\quad
f=\begin{cases}
5 & n \in 2\mZ
\\
1 & n \in 2\mZ+1
\end{cases},
$\\
$
   a=\begin{cases}
{3\over 4}n^2-{5\over 4} & n \in 2\mZ
\\
{3\over 4}n^2-{17\over 24} & n \in 2\mZ+1
\end{cases},\quad
c=\begin{cases}
{3\over 4}n^2-1 & n \in 2\mZ
\\
{3\over 4}n^2-{2\over 3} & n \in 2\mZ+1
\end{cases}	
$
 \begin{figure}[H]
 \centering
\includegraphics[scale=0.9,keepaspectratio]{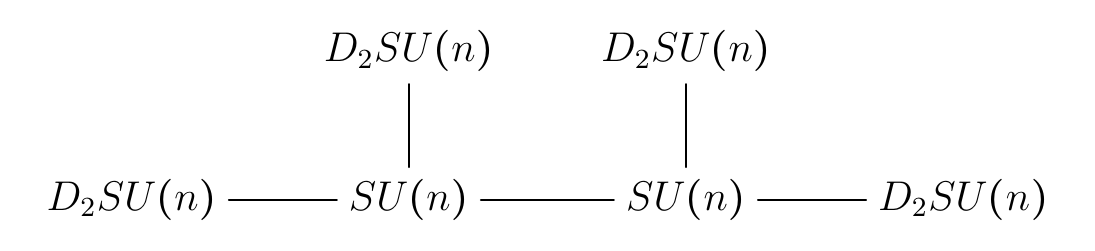}
\end{figure}

\begin{figure}[H]
	\centering
	\includegraphics[scale=.9,keepaspectratio]{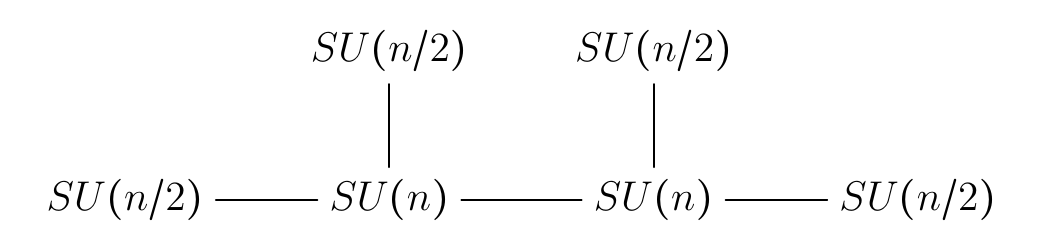}\caption{Affine $D_5$ quiver from ICIS (1) with $n\in 2\mZ $ }
	\label{fig:D5}
\end{figure}

For illustration, let's discuss how we identify the quiver. To start, we see from the singularity data (using the program Singular \cite{DGPS}) the Coulomb branch spectrum is given by (including positive scaling dimensions only)
\ie
&n\in 2\mZ+1:\quad n,n,n-1,n-1,\dots,{n+1\over 2},{n+1\over 2},{n\over 2},{n\over 2},{n\over 2},{n\over 2},{n-1\over 2},{n-1\over 2},\dots, {3\over 2},{3\over 2},{3\over 2},{3\over 2},1
\\
&n\in 2\mZ:\quad n,n,n-1,n-1,\dots,{n+2\over 2},{n+2\over 2},{n\over 2},{n\over 2},{n\over 2},{n\over 2},{n-2\over 2},{n-2\over 2},\dots ,2,2,2,2,1,1,1,1,1
\label{spec:ICIS1}
\fe
We see immediately that this is the affine $D_5$ theory when $n$ is even (Figure~\ref{fig:D5})
while more generally we have
\begin{figure}[H]
	\centering
	\includegraphics[scale=.9,keepaspectratio]{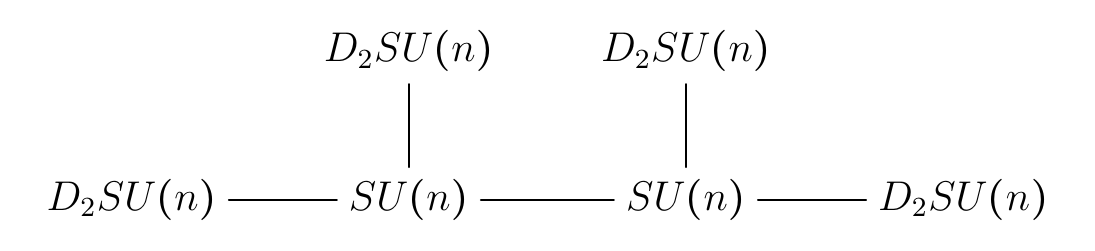}
%	\caption{ICIS (1): $\m=-7+8n$ }
\end{figure}
We can compute the central charges from the spectrum \eqref{spec:ICIS1} using \eqref{eq:acformula}
\ie
%\m=&-7+8n,\quad 
a=&\begin{cases}
{3\over 4}n^2-{5\over 4} & n \in 2\mZ
\\
{3\over 4}n^2-{17\over 24} & n \in 2\mZ+1
\end{cases},
\quad 
c=&\begin{cases}
{3\over 4}n^2-1 & n \in 2\mZ
\\
{3\over 4}n^2-{2\over 3} & n \in 2\mZ+1
\end{cases}
\fe	
It's easy to check that they agree with those computed from the quiver description (using \eqref{D2SUNac}). 

For most of the examples below we will omit unnecessary details. The interested reader is welcome to follow the strategy outlined in the previous subsections to double check these gauge theory descriptions in comparison to the singularity theory point of view.

\subsection*{ICIS (2)}
$\left\{ \begin{array}{l} x_1^{2}+x_2^{2}+x_3^{2}+x_4^{3}+x_5^{3}=0\\x_1^{2}+2x_2^{2}+3x_3^{2}+4x_4^{3}+5x_5^{3}=0\end{array}\right.$
\\$(w_1,w_2,w_3,w_4,w_5;1,d)=(\frac{1}{2},\frac{1}{2},\frac{1}{2},\frac{1}{3},\frac{1}{3};1,1)$
\\$\mu=32,\quad \m_1=4,\quad \A=6,\quad r=12,\quad f=6,\quad a={473\over 24},\quad c={121\over 6}
$
 \begin{figure}[H]
 	\centering
 	\includegraphics[scale=0.9,keepaspectratio]{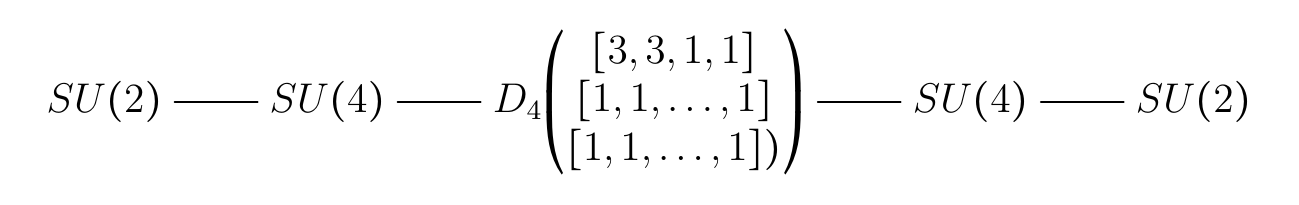}
 \end{figure}

The middle conformal matter theory $D_4([3,3,1,1],[1,1,\dots,1]^2)$ is defined in the ordinary type $D_4$ class S construction, by two full $SO(8)$ punctures and one puncture of the type $[3,3,1,1]$ on $S^2$. This conformal matter supplies three CB operators $6,4,3$ and flavor symmetry $(SO(8)_6)^2\times U(1)^2$. Note that the index of embedding $I_{SU(4)\hookrightarrow SO(8)}=1$ from the decomposition~$\bf 8 \to 4+\overline 4$. Therefore the conformal matter theory supply current central charge $\kappa_{SU(4)}=6$ for both $SU(4)$ gauge groups, making them conformal
\ie
\B_{SU(4)}=8-6-2=0.
\fe
The left over flavor symmetry has rank 6 from the quiver which agrees with that from the singularity. 

Furthermore, we can compute the conformal central charges from the quiver as follows. The conformal matter theory supplies $n_v=41, n_h=72$ (or $a={277\over 24},~c={77\over 6}$). Including the contributions from the $SU(2)$ and $SU(4)$ vector and bifundamental multiplets we have
\ie
n_v=41+2(3+15)=77,\quad n_h=72+2\times 8=88
\fe
or
\ie
a={473\over 24},\quad c={121\over 6}.
\fe

\subsection*{ICIS (7)}
$\left\{ \begin{array}{l} x_1x_3+x_4x_5=0\\x_1^{2}+x_2^{2}+x_3^{n}+x_4^{2}+2x_5^{n}=0\end{array}\right.$
\\$n\ge3$
\\$(w_1,w_2,w_3,w_4,w_5;1,d)=(\frac{n}{2 + n},\frac{n}{2 + n},\frac{2}{2 + n},\frac{n}{2 + n},\frac{2}{2 + n};1,\frac{2 n}{2 + n})$
\\$\mu=(n+1)^2,\quad \m_1=(n-1)^2,\quad \A={n-2\over 2},\quad
 r=\begin{cases}
{n(n+1)-2\over 2} & n \in 2\mZ
\\
{n(n+1)\over 2} & n \in 2\mZ+1
 \end{cases}
 ,\quad f=
 \begin{cases}
n+3& n \in 2\mZ
\\
n+1 & n \in 2\mZ+1
 \end{cases}
 $
\\$
a=\begin{cases}
{(n+2)(2n-1)(2n+3)\over 48} & n \in 2\mZ
\\
{(n+2)(2n-1)(2n+3)+13\over 48} & n \in 2\mZ+1
\end{cases},\quad
c=\begin{cases}
{n(n+1)(n+2)\over 12} & n \in 2\mZ
\\
{n(n+1)(n+2)+2\over 12}  & n \in 2\mZ+1
\end{cases}
$

\begin{figure}[H]
	\centering
	\includegraphics[scale=.9,keepaspectratio]{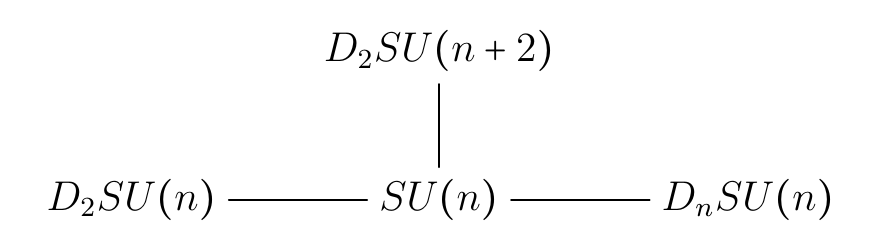}
%	\caption{ICIS (7): 
%		$
%		\m=(n+1)^2
%		%,a=,c=
%		$}
\end{figure}
 
 \subsection*{ICIS (8)}
 $\left\{ \begin{array}{l} x_1x_3+x_4x_5=0\\x_1^{2}+x_2^{2}+x_3x_4^{2}+x_3^{3 n}+2x_5^{2 n}=0\end{array}\right.$
\\$n\ge2$
\\$(w_1,w_2,w_3,w_4,w_5;1,d)=(\frac{3 n}{2 + 3 n},\frac{3 n}{2 + 3 n},\frac{2}{2 + 3 n},\frac{-1 + 3 n}{2 + 3 n},\frac{3}{2 + 3 n};1,\frac{6 n}{2 + 3 n})$
\\$\mu=3 + 7 n + 6 n^2,
\quad \m_1=(3n+1)(2n-1),
\quad \A={3n+2\over 2},\quad
 r=3n(n+1)
 ,\quad f=n+3
 ,\newline a= \frac{1}{16} n (n (24 n+37)+11)+\frac{5}{6},\quad
c= \frac{1}{4} n \left(6 n^2+9 n+5\right)+\frac{1}{6}
 $
 \begin{figure}[H]
	\centering
	\includegraphics[scale=.9,keepaspectratio]{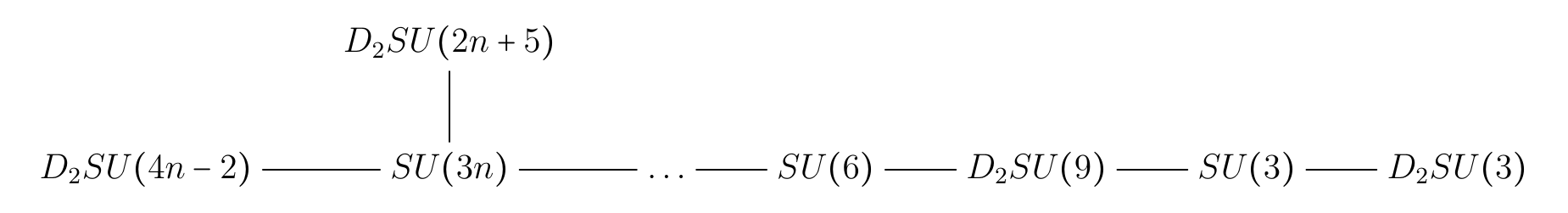}
%	\caption{ICIS (8): 
%		$
%		\m=3+7n+6n^2
%		%,a=,c=
%		$}
\end{figure}

\subsection*{ICIS (9)}
(9)$\left\{ \begin{array}{l} x_1x_3+x_4x_5=0\\x_1^{2}+x_2^{2}+x_3x_4^{2}+x_3^{1 + 3 n}+2x_3x_5^{2 n}=0\end{array}\right.$
\\$n\ge1$
\\$(w_1,w_2,w_3,w_4,w_5;1,d)=(\frac{1 + 3 n}{3 (1 + n)},\frac{1 + 3 n}{3 (1 + n)},\frac{2}{3 (1 + n)},\frac{n}{1 + n},\frac{1}{1 + n};1,\frac{2 (1 + 3 n)}{3 (1 + n)})$
\\$\mu=6 + 11 n + 6 n^2,\quad
\m_1=6 n^2+3 n-\frac{2}{3},\quad
\A={3(n+1)\over 2},\quad
r=n (3 n+5)+1,\quad
f=n+4,\quad
\newline
a=\frac{1}{8} n (6 n (2 n+5)+25)+\frac{2}{3},\quad
c=\frac{1}{4}   n (3 n (2 n+5)+13)+{5\over 6}.
$
 
	\begin{figure}[H]
		\centering
		\includegraphics[scale=.9,keepaspectratio]{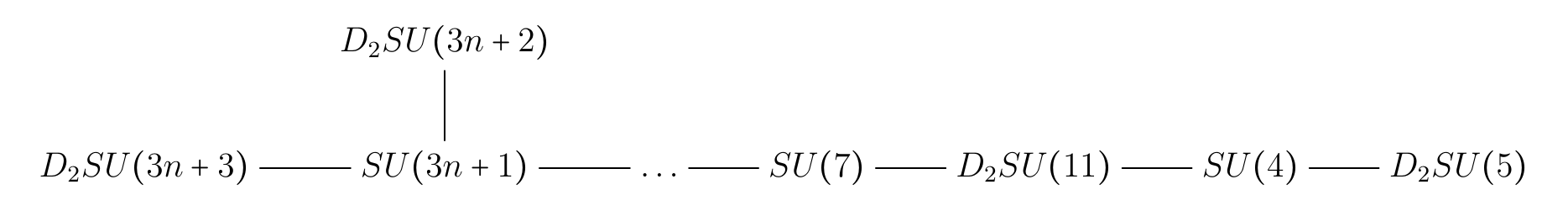}
%		\caption{ICIS (9):  
%			$
%			\m=6+11n+6n^2
%			%,a=,c=
%			$}
	\end{figure}

\subsection*{ICIS (10)}
$\left\{ \begin{array}{l} x_1x_3+x_4x_5=0\\x_1^{2}+x_2^{2}+x_3x_4^{3}+x_3^{16}+x_5^{4}=0\end{array}\right.$
\\$(w_1,w_2,w_3,w_4,w_5;1,d)=(\frac{8}{9},\frac{8}{9},\frac{1}{9},\frac{5}{9},\frac{4}{9};1,\frac{16}{9})$
\\$\mu=127,\quad
\m_1=99,\quad
\A=9,\quad
r=59,\quad
f=9,\quad
a={4285\over 24},\quad
c={538\over 3}.
$
 
	\begin{figure}[H]
		\centering
		\includegraphics[scale=.9,keepaspectratio]{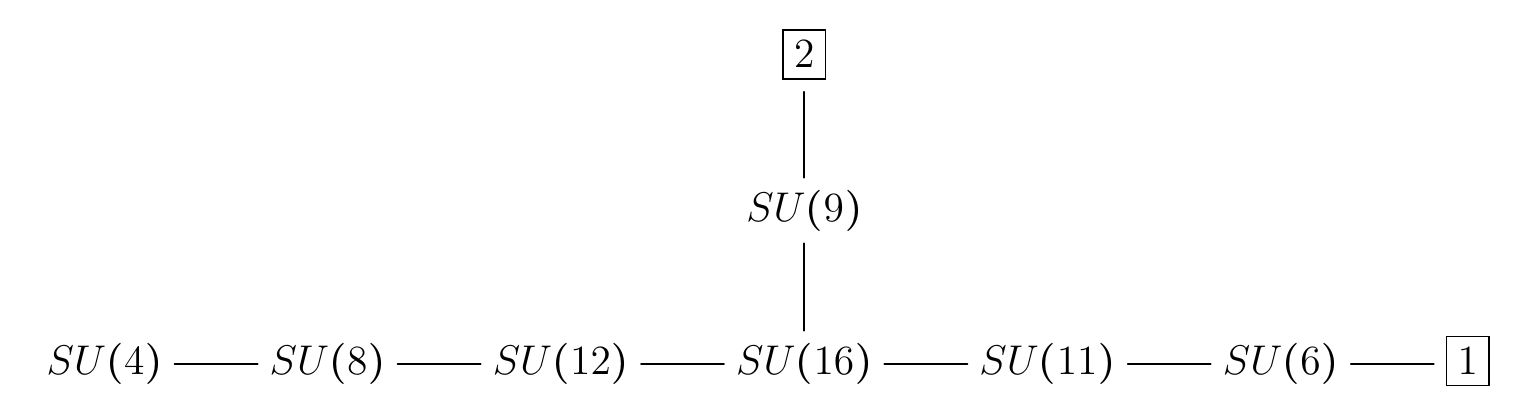}
%		\caption{ICIS (10):  
%			$
%			\m=127
%			%,a=,c=
%			$}
	\end{figure}

\subsection*{ICIS (11)}
$\left\{ \begin{array}{l} x_1x_3+x_4x_5=0\\x_1^{2}+x_2^{2}+x_3x_4^{3}+x_3^{40}+x_5^{5}=0\end{array}\right.$
\\$(w_1,w_2,w_3,w_4,w_5;1,d)=(\frac{20}{21},\frac{20}{21},\frac{1}{21},\frac{13}{21},\frac{8}{21};1,\frac{40}{21})$
\\$\mu=358,\quad
\m_1=324,\quad
\A=21,\quad
r=174,\quad
f=10,\quad
a=1221,\quad
c={2445\over 2}.
$ 
\begin{figure}[H]
		\centering
		\includegraphics[scale=.9,keepaspectratio]{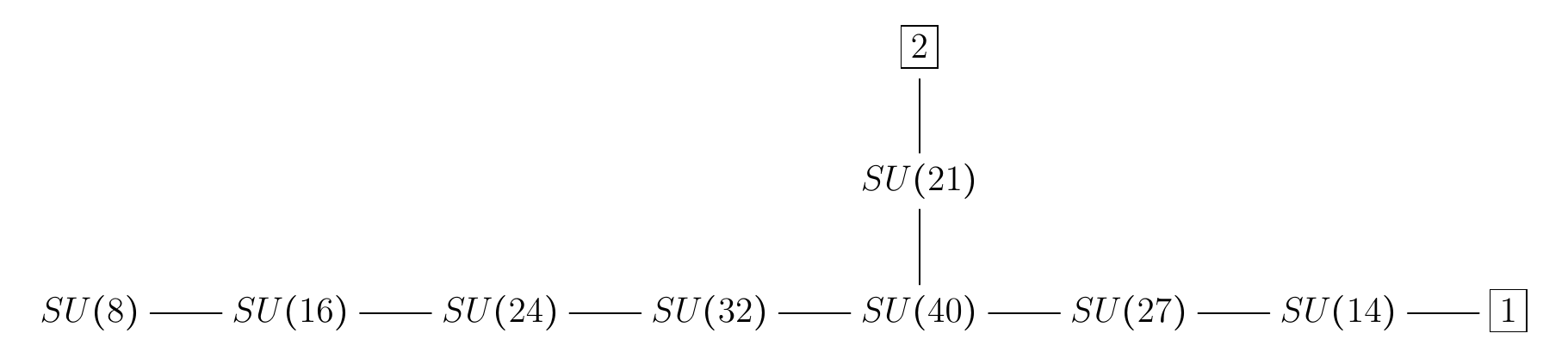}
%		\caption{ICIS (11):  
%			$
%			\m=358
%			%,a=,c=
%			$}
\end{figure}

\subsection*{ICIS (12)}
$\left\{ \begin{array}{l} x_1x_3+x_4x_5=0\\x_1^{2}+x_2^{2}+x_3x_4^{3}+x_3^{10}+x_3x_5^{3}=0\end{array}\right.$
\\$(w_1,w_2,w_3,w_4,w_5;1,d)=(\frac{5}{6},\frac{5}{6},\frac{1}{6},\frac{1}{2},\frac{1}{2};1,\frac{5}{3})$
\\$\mu=73,\quad
\m_1=49,\quad
\A=6,\quad
r=32,\quad
f=9,\quad
a={197\over 3},\quad
c={199\over 3}.
$ 
\begin{figure}[H]
		\centering
		\includegraphics[scale=.9,keepaspectratio]{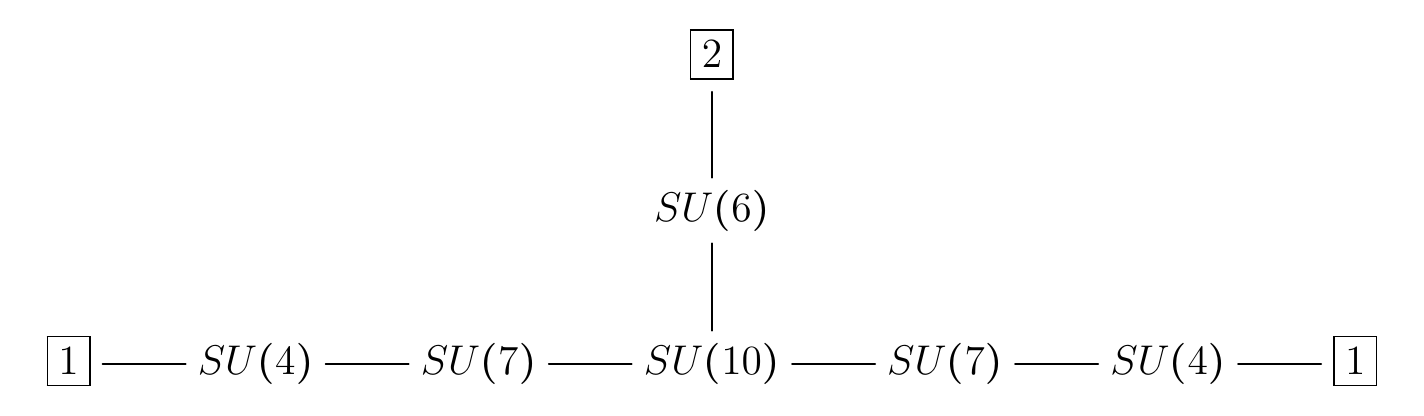}
%		\caption{ICIS (12):  
%			$
%			\m=73
%			%,a=,c=
%			$}
\end{figure}

\subsection*{ICIS (13)}
$\left\{ \begin{array}{l} x_1x_3+x_4x_5=0\\x_1^{2}+x_2^{2}+x_3x_4^{3}+x_3^{19}+x_3x_5^{4}=0\end{array}\right.$
\\$(w_1,w_2,w_3,w_4,w_5;1,d)=(\frac{19}{21},\frac{19}{21},\frac{2}{21},\frac{4}{7},\frac{3}{7};1,\frac{38}{21})$
\\$\mu=155,\quad
\m_1={377\over 3},\quad
\A={21\over 2},\quad
r=74,\quad
f=7,\quad
a={3085\over 12},\quad
c={3095\over 12}.
$ 
\begin{figure}[H]
		\centering
		\includegraphics[scale=.9,keepaspectratio]{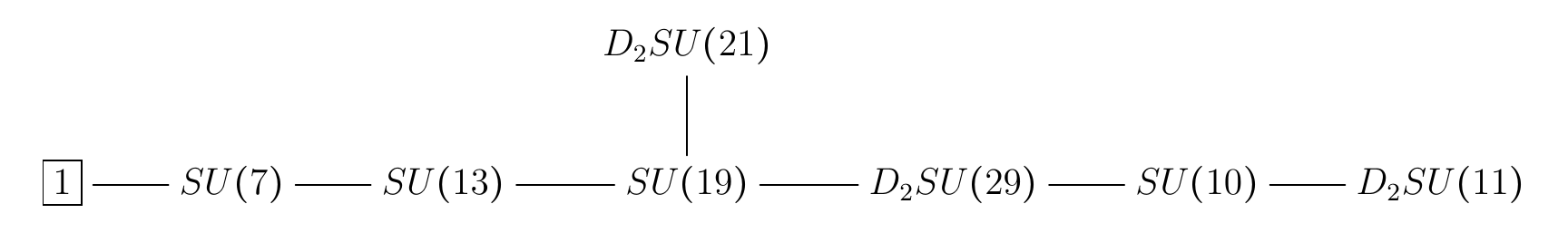}
%		\caption{ICIS (13):  
%			$
%			\m=155
%			%,a=,c=
%			$}
\end{figure}

\subsection*{ICIS (14)}
$\left\{ \begin{array}{l} x_1x_3+x_4x_5=0\\x_1^{2}+x_2^{2}+x_3x_4^{3}+x_3^{46}+x_3x_5^{5}=0\end{array}\right.$
\\$(w_1,w_2,w_3,w_4,w_5;1,d)=(\frac{23}{24},\frac{23}{24},\frac{1}{24},\frac{5}{8},\frac{3}{8};1,\frac{23}{12})$
\\$\mu=417,\quad
\m_1={1147\over 3},\quad
\A=24,\quad
r=203,\quad
f=11,\quad
a={13045\over 8},\quad
c={3265\over 2}.
$ 
\begin{figure}[H]
		\centering
		\includegraphics[scale=.8,keepaspectratio]{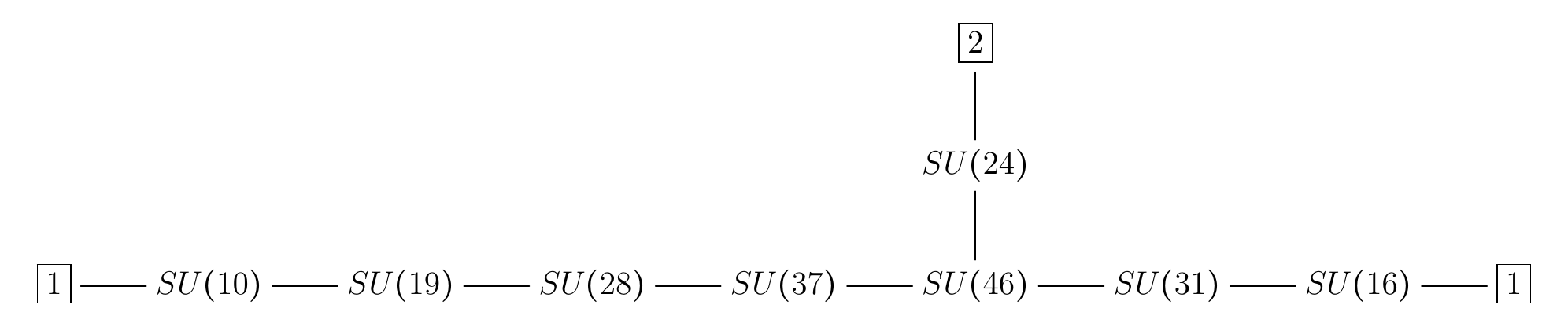}
%		\caption{ICIS (14):  
%			$
%			\m=417
%			%,a=,c=
%			$}
\end{figure}

\subsection*{ICIS (15)}
$\left\{ \begin{array}{l} x_1x_3+x_4x_5=0\\x_1^{2}+x_2^{2}+x_4^{3}+x_3^{6}+x_5^{3}=0\end{array}\right.$
\\$(w_1,w_2,w_3,w_4,w_5;1,d)=(\frac{3}{4},\frac{3}{4},\frac{1}{4},\frac{1}{2},\frac{1}{2};1,\frac{3}{2})$
\\$\mu=39,\quad
\m_1=20,\quad
\A=4,\quad
r=16,\quad
f=7,\quad
a={267\over 12},\quad
c={67\over 3}.
$ 
\begin{figure}[H]
		\centering
		\includegraphics[scale=.9,keepaspectratio]{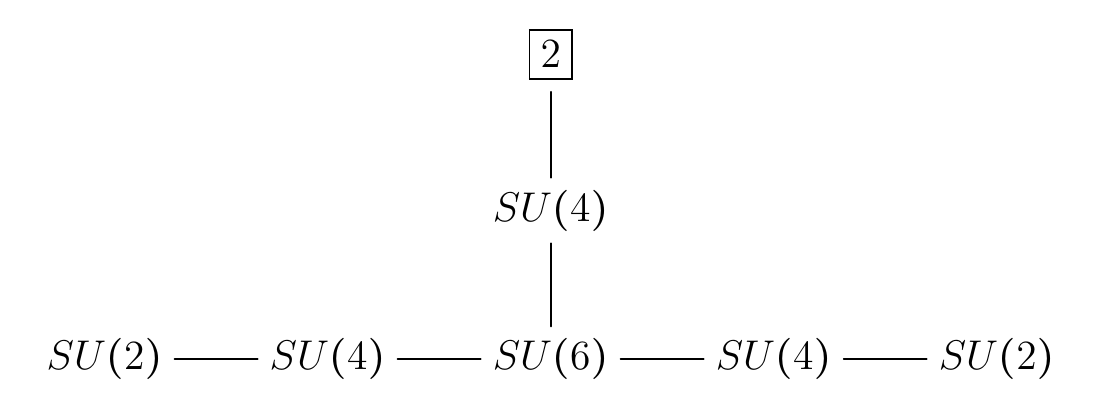}
%		\caption{ICIS (15):  
%			$
%			\m=39
%			%,a=,c=
%			$}
\end{figure}

\subsection*{ICIS (16)}
$\left\{ \begin{array}{l} x_1x_3+x_4x_5=0\\x_1^{2}+x_2^{2}+x_4^{3}+x_3^{12}+x_5^{4}=0\end{array}\right.$
\\$(w_1,w_2,w_3,w_4,w_5;1,d)=(\frac{6}{7},\frac{6}{7},\frac{1}{7},\frac{4}{7},\frac{3}{7};1,\frac{12}{7})$
\\$\mu=92,\quad
\m_1=66,\quad
\A=7,\quad
r=42,\quad
f=8,\quad
a={1183\over 12},\quad
c={595\over 6}.
$ 
\begin{figure}[H]
		\centering
		\includegraphics[scale=.9,keepaspectratio]{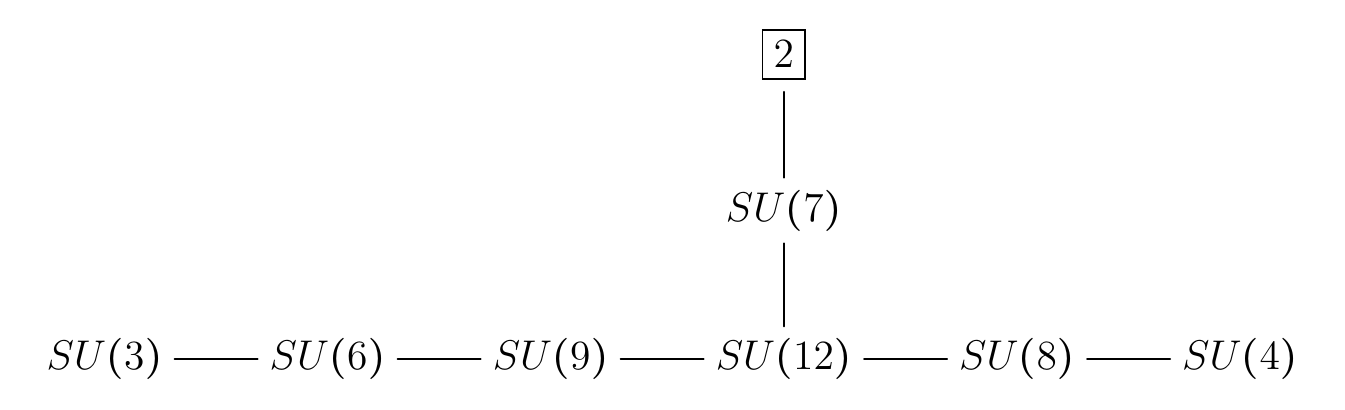}
%		\caption{ICIS (16):  
%			$
%			\m=92
%			%,a=,c=
%			$}
\end{figure}

\subsection*{ICIS (17)}
$\left\{ \begin{array}{l} x_1x_3+x_4x_5=0\\x_1^{2}+x_2^{2}+x_4^{3}+x_3^{30}+x_5^{5}=0\end{array}\right.$
\\$(w_1,w_2,w_3,w_4,w_5;1,d)=(\frac{15}{16},\frac{15}{16},\frac{1}{16},\frac{5}{8},\frac{3}{8};1,\frac{15}{8})$
\\$\mu=265,\quad
\m_1=232,\quad
\A=16,\quad
r=128,\quad
f=9,\quad
a=683,\quad
c=684.
$ 
\begin{figure}[H]
		\centering
		\includegraphics[scale=.9,keepaspectratio]{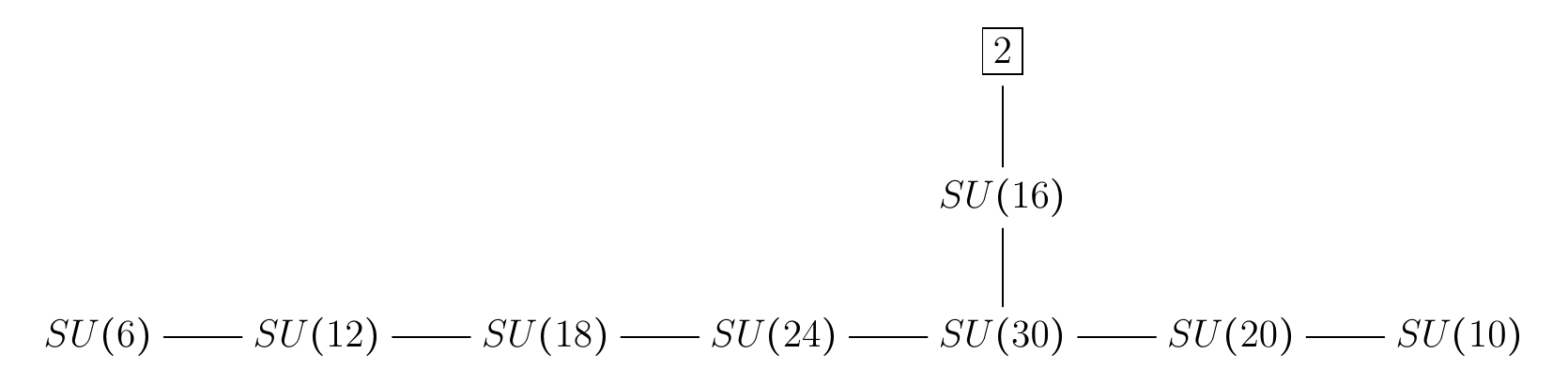}
%		\caption{ICIS (17):  
%			$
%			\m=265
%			%,a=,c=
%			$}
\end{figure}

\subsection*{ICIS (18)}
$\left\{ \begin{array}{l} x_1x_3+x_4x_5=0\\x_1^{2}+x_2^{2}+x_4^{3}+x_3^{8}+x_3x_5^{3}=0\end{array}\right.$
\\$(w_1,w_2,w_3,w_4,w_5;1,d)=(\frac{4}{5},\frac{4}{5},\frac{1}{5},\frac{8}{15},\frac{7}{15};1,\frac{8}{5})$
\\$\mu=56,\quad
\m_1=34,\quad
\A=5,\quad
r=26,\quad
f=4,\quad
a={83\over 2},\quad
c={251\over 6}.
$ 
\begin{figure}[H]
		\centering
		\includegraphics[scale=.9,keepaspectratio]{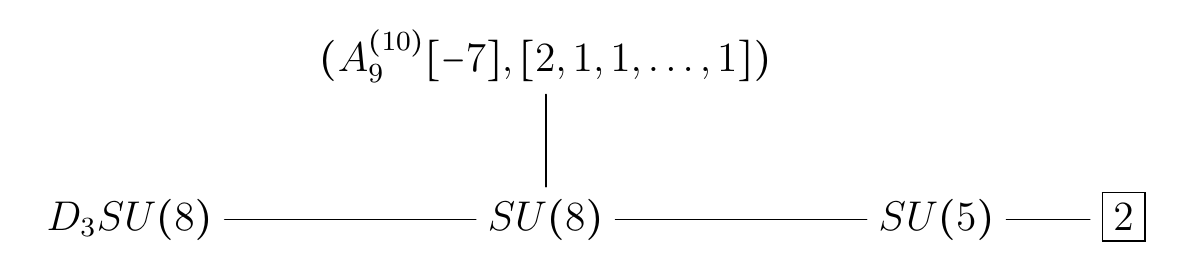}
%		\caption{ICIS (18):  
%			$
%			\m=56
%			%,a=,c=
%			$}
\end{figure}
The theory $(A_{9}^{(10)}[-7],[2,1^6])$ is defined by six dimensional $A_{9}$ $(2,0)$ theory 
on a sphere with irregular puncture $ A_{9}^{(10)}[-7]$  and a regular puncture labeled by Young Tableaux $[2,1,\dots,1]$. This 
theory has flavor symmetry $SU(8)\times U(1)$, and the flavor central charge of $SU(8)$ is ${17\over 3}$. The Coulomb branch 
spectrum is $({17\over3},{14\over3},{11\over3},{10\over3},{8\over3},{7\over3},{5\over3},{4\over3})$, and the central charges are $a=\frac{83}{12}, ~c=\frac{15}{2}$. 
 The theory $D_3 SU(8)=(A_7^{(8)}[-5],[1,1,\dots,1])$ 
has flavor symmetry $SU(8)$, and the flavor central charge is ${16\over 3}$.  The Coulomb branch spectrum of this theory is $({16\over3},{13\over3},{10\over3},{8\over3},{7\over3},{5\over3},{4\over3})$.
Using the above data, one can check that the 
gauging of the $SU(8)$ gauge group is conformal,
\ie
\B_{SU(8)}=16-{16\over 3}-{17\over 3}-5=0
\fe
Futhermore, using \eqref{eq:acformula}, we find the central charges for  $(A_{9}^{(10)}[-7],[2,1^6])$ is $a={35\over 3},~c={38\over 3}$; while for $D_3 SU(8)$ we have $a={77\over 8},~c={21\over 2}$, hence the central charges of the quiver is
\ie
a={35\over 3}+{77\over 8}+{5(24+63)+50\over 24}={83\over 2},\quad
c={38\over 3}+{21\over 2}+{2(24+63)+50\over 12}={251\over 6}
\fe
which agrees with that from the singularity.

\subsection*{ICIS  (19)}
$\left\{ \begin{array}{l} x_1x_3+x_4x_5=0\\x_1^{2}+x_2^{2}+x_4^{3}+x_3^{15}+x_3x_5^{4}=0\end{array}\right.$
\\$(w_1,w_2,w_3,w_4,w_5;1,d)=(\frac{15}{17},\frac{15}{17},\frac{2}{17},\frac{10}{17},\frac{7}{17};1,\frac{30}{17})$
\\$\mu=120,\quad
\m_1=92,\quad
\A={17\over 2},\quad
r=57,\quad
f=6,\quad
a={1909\over 12},\quad
c={479\over 3}.
$ 
\begin{figure}[H]
		\centering
		\includegraphics[scale=.9,keepaspectratio]{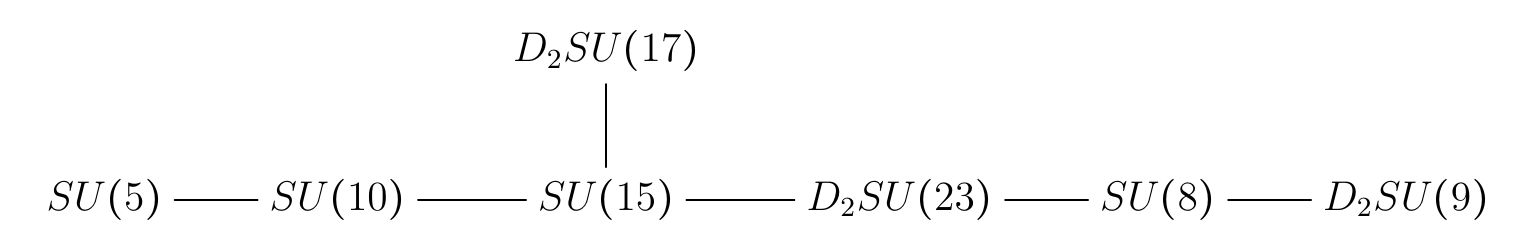}
%		\caption{ICIS (19):  
%			$
%			\m=120
%			%,a=,c=
%			$}
\end{figure}

\subsection*{ICIS (20)}
$\left\{ \begin{array}{l} x_1x_3+x_4x_5=0\\x_1^{2}+x_2^{2}+x_4^{3}+x_3^{36}+x_3x_5^{5}=0\end{array}\right.$
\\$(w_1,w_2,w_3,w_4,w_5;1,d)=(\frac{18}{19},\frac{18}{19},\frac{1}{19},\frac{12}{19},\frac{7}{19};1,\frac{36}{19})$
\\$\mu=324,\quad
\m_1=290,\quad
\A=19,\quad
r=157,\quad
f=10,\quad
a={23929\over 24},\quad
c={2995\over 3}.
$ 
\begin{figure}[H]
		\centering
		\includegraphics[scale=.9,keepaspectratio]{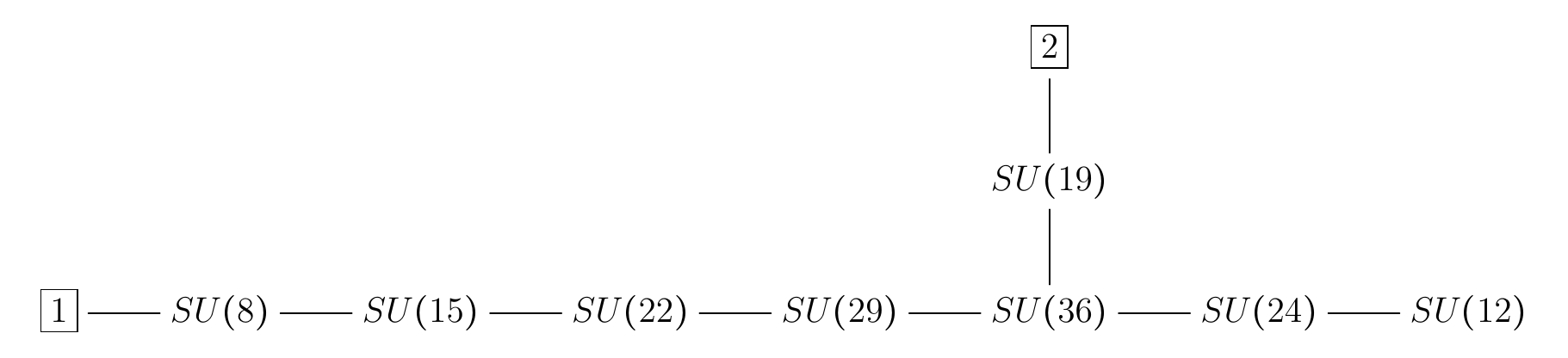}
%		\caption{ICIS (20):  
%			$
%			\m=324
%			%,a=,c=
%			$}
\end{figure}

\subsection*{ICIS (41)}
$\left\{ \begin{array}{l} x_2^{2}+x_3^{2}+x_4^{2}+x_5^{2}=0\\x_1^{2}+x_2^{3}+x_3^{3}+x_4^{3}=0\end{array}\right.$
\\$(w_1,w_2,w_3,w_4,w_5;1,d)=(\frac{3}{4},\frac{1}{2},\frac{1}{2},\frac{1}{2},\frac{1}{2};1,\frac{3}{2})$
\\$\mu=31,\quad
\m_1=16,\quad
\A=4,\quad
r=15,\quad
f=1,\quad
a={437\over 24},\quad
c={109\over 6}.$
\begin{figure}[H]
		\centering
		\includegraphics[scale=.9,keepaspectratio]{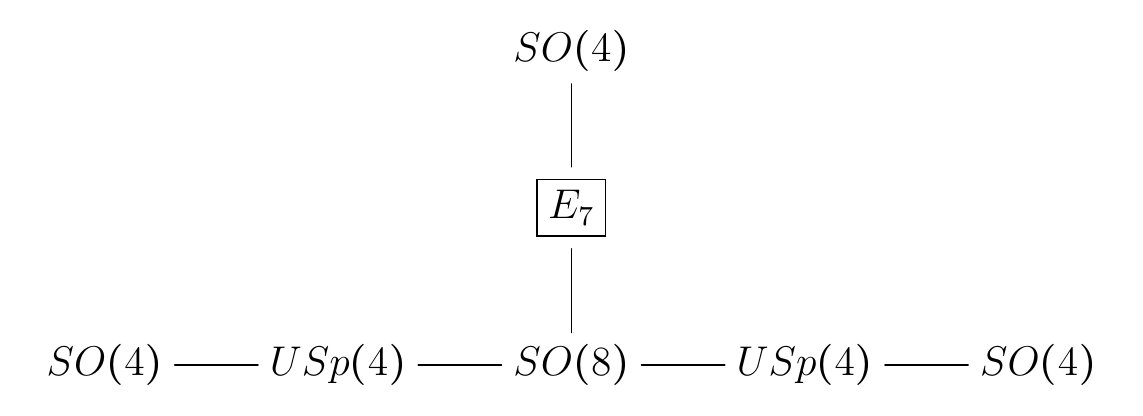}
%		\caption{ICIS (41):  
%			$
%			\m=31
%			%,a=,c=
%			$}
\end{figure}
Here the conformal matter is provided by the $E_7$ Minahan-Nemeschansky theory with a single coulomb branch chiral primary of dimension 4 and $(E_7)_4$ flavor symmetry \cite{Minahan:1996cj}.  From the decomposition of {\bf 56} under $SO(8)\subset E_7$,
\ie
{\bf 56} \to 8\cdot {\bf 1}+2\cdot {\bf 8}_v+2\cdot {\bf 8}_s+2\cdot {\bf 8}_c
\fe
the embedding index
\ie
I_{SO(8)\hookrightarrow E_7}={2T({\bf 8}_v)+2T({\bf 8}_s)+2T({\bf 8}_c)\over T({\bf 56})}
={6\cdot 1  \over 6}=1
\fe
Hence the $E_7$ theory supplies $SO(8)_4$ for the $SO(8)$ gauge coupling, leading to conformal gauging, similarly for the $SO(4)$ gauge node.

The central charges can also be verified. The $E_7$ theory's central charges are characterized by $n_v=7$ and $n_h=24$. Together with the rest of the quiver, we have
\ie
n_v=7+3\cdot 6+2\cdot 10+28=73,\quad n_h=24+{2\cdot 16+ 2\cdot 32 \over 2}=72
\fe
or
\ie
a={437\over 24},\quad c={109\over 6}
\fe
 in agreement with the singularity.

\subsection*{ICIS (42)}
$\left\{ \begin{array}{l} x_2^{2}+x_3^{2}+x_4^{2}+x_5^{3}=0\\x_1^{2}+x_2^{3}+x_3^{3}+x_4^{3}=0\end{array}\right.$
\\$(w_1,w_2,w_3,w_4,w_5;1,d)=(\frac{3}{4},\frac{1}{2},\frac{1}{2},\frac{1}{2},\frac{1}{3};1,\frac{3}{2})$
\\$\mu=54,\quad
\m_1=28,\quad
\A=12,\quad
r=27,\quad
f=0,\quad
a={689\over 8},\quad
c={173\over 2}.$
\begin{figure}[H]
		\centering
		\includegraphics[scale=.9,keepaspectratio]{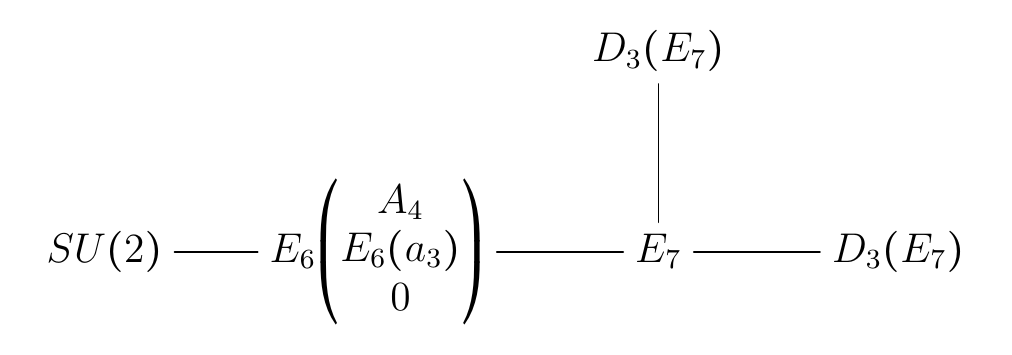}
%		\caption{ICIS (42):  
%			$
%			\m=54
%			%,a=,c=
%			$}
\end{figure}
Here the $\cN=2$ SCFT in the weakly coupled frame is described by $E_7$ and $SU(2)$ $\cN=2$ vector multiplets coupled conformally to matter of the $D_p(G)$ type and Gaiotto type. 

The $E_6(A_4, E_6(a_3),0)$ theory has Coulomb branch spectrum $(12,8,6,4,3)$ and flavor symmetry $(E_7)_{12}\times SU(2)_4$ \cite{Chacaltana:2014jba}. It's conformal central charges are $a={139\over 8}~,c={39\over 2}$ (or $n_v=61,~n_h=112$).

The $D_3(E_7)$ theory has Coulomb branch $(12,8,6,6,4,2,2)$ and flavor symmetry $(E_7)_{12}$.  It's conformal central charges are $a={485\over 24}~,c={133\over 6}$ (or $n_v=73,~n_h=120$). 

We can check immediately that the gauge groups are conformal
\ie
\B_{E_7}=2\cdot 18-2\cdot 12-12=0
\fe
similarly for the $SU(2)$ node.

The conformal central charges can also be verified. The $E_7$ theory's central charges are characterized by $n_v=7$ and $n_h=24$. Together with the rest of the quiver, we have 
\ie
n_v=3+61+133+2\cdot 73=343,\quad 
n_h=112+2\cdot 120=352
\fe
or
\ie
a={689\over 8},\quad
c={173\over 2}.  \fe
 in agreement with the singularity.

%  
%\subsection*{ICIS (50)}
%$\left\{ \begin{array}{l} x_2x_3+x_4x_5=0\\x_1^{2}+x_2^{2}x_5+x_3^{8}x_5+x_4^{3}+x_3x_5^{5}=0\end{array}\right.$
%\\$(w_1,w_2,w_3,w_4,w_5;1,d)=(\frac{39}{40},\frac{4}{5},\frac{1}{5},\frac{13}{20},\frac{7}{20};1,\frac{39}{20})$
%\\$\mu=121$

\subsection*{ICIS 
(55)}
$\left\{ \begin{array}{l} x_1x_2+x_3^{2}+x_4^{2}+x_5^{2}=0\\x_1x_3+2x_2^{5}+x_2x_4^{2}=0\end{array}\right.$
\\$(w_1,w_2,w_3,w_4,w_5;1,d)=(\frac{3}{4},\frac{1}{4},\frac{1}{2},\frac{1}{2},\frac{1}{2};1,\frac{5}{4})$
\\$\mu=31,\quad
\m_1=9,\quad
\A=4,\quad
r=12,\quad
f=7,\quad
a={179\over 12},\quad
c={46\over 3}.$
\begin{figure}[H]
		\centering
		\includegraphics[scale=.9,keepaspectratio]{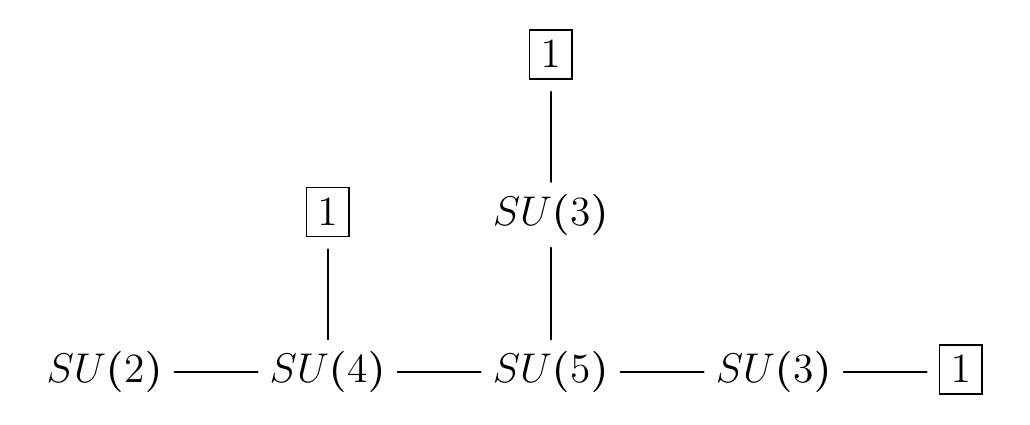}
%		\caption{ICIS (55):  
%			$
%			\m=31
%			%,a=,c=
%			$}
\end{figure}

\subsection*{ICIS 
(56)}
$\left\{ \begin{array}{l} x_1x_2+x_3^{2}+x_4^{2}+x_5^{3}=0\\2x_1x_3+x_2^{5}+x_2x_4^{2}=0\end{array}\right.$
\\$(w_1,w_2,w_3,w_4,w_5;1,d)=(\frac{3}{4},\frac{1}{4},\frac{1}{2},\frac{1}{2},\frac{1}{3};1,\frac{5}{4})$
\\$\mu=56,\quad
\m_1={33\over 2},\quad
\A=12,\quad
r=27,\quad
f=2,\quad
a={613\over 8},\quad
c={77}.$
\begin{figure}[H]
		\centering
		\includegraphics[scale=.9,keepaspectratio]{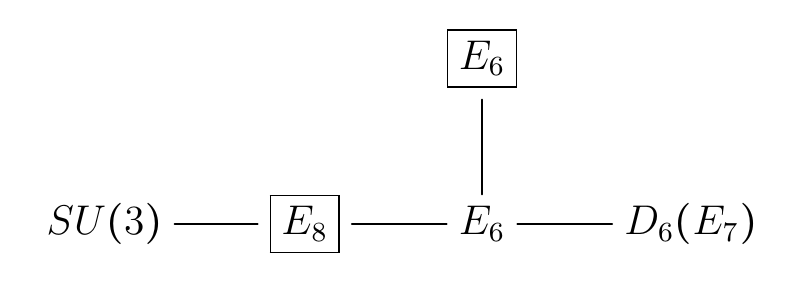}
%		\caption{ICIS (56):  
%			$
%			\m=56
%			%,a=,c=
%			$}
\end{figure}
The conformal matter theories here are provided by $E_6$ and $E_8$ Minahan-Nemeschansky theories which have flavor symmetry $(E_6)_3$ and $(E_8)_6$ respectively \cite{Minahan:1996fg,Minahan:1996cj}.,  as well as an AD matter theory $D_6(E_7)$. 

The $D_6(E_7)$ theory has Coulomb branch spectrum $(15,12,11,9,9,8,7,6,6,5,5,3,3)$ and flavor symmetry $(E_7)_{15}\times U(1)$. Its central charges are $a={1273\over 24},~c={166\over 3}$ (or $n_v=203,~n_h=258$).

Since index of embedding are $I_{E_6 \hookrightarrow E_8}=I_{E_6 \hookrightarrow E_7}=I_{SU(3) \hookrightarrow E_8}=1$, it's easy to verify that the gauge couplings are finite. The central charges can also be matched with the singularity: together with $E_6$ theory ($n_v=5,~n_h=16$) and $E_8$ theory ($n_v=11,~n_h=40$),
\ie
n_v=8+11+5+78+203=305,\quad
n_h=40+16+258=314
\fe
or
\ie
a={613\over 8},\quad
c={77}.
\fe

\subsection*{ICIS 
(57)}
\begin{figure}[H]
		\centering
		\includegraphics[scale=.9,keepaspectratio]{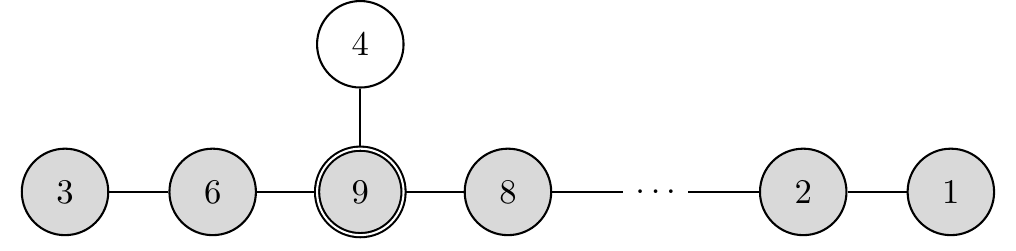}
		\caption{$3d$ $\cN=4$ mirror of $A_8([1,1,\dots,1],[5,4],[3,3,3])$. The circle nodes denotes $U(n)$ gauge groups and the double circle node denotes an $SU(n)$ gauge group. The {\it balanced} nodes are shaded.}
		\label{fig:A8mirror}
%		\caption{ICIS (57):  
%			$
%			\m=43
%			%,a=,c=
%			$}
\end{figure}

$\left\{ \begin{array}{l} x_1x_2+x_3^{2}+x_4^{2}+x_5^{4}=0\\x_1x_3+x_2^{3}+x_2x_5^{3}=0\end{array}\right.$
\\$(w_1,w_2,w_3,w_4,w_5;1,d)=(\frac{5}{8},\frac{3}{8},\frac{1}{2},\frac{1}{2},\frac{1}{4};1,\frac{9}{8})$
\\$\mu=43,\quad
\m_1={35\over 4},\quad
\A=8,\quad
r=19,\quad
f=5,\quad
a={893\over 24},\quad
c={113\over 3}.$
\begin{figure}[H]
		\centering
		\includegraphics[scale=.9,keepaspectratio]{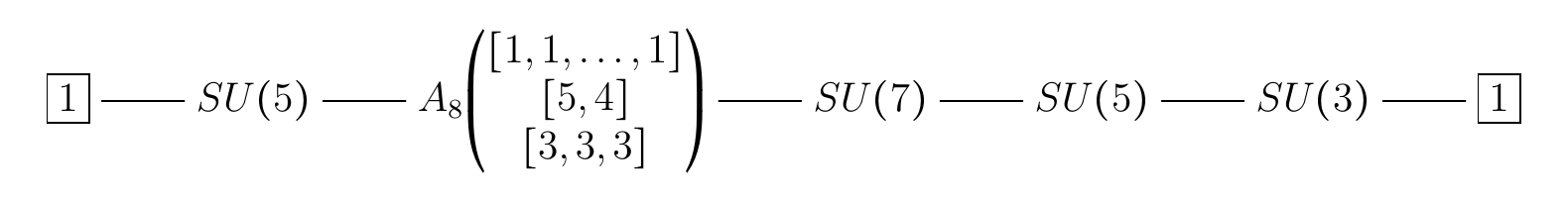}
%		\caption{ICIS (57):  
%			$
%			\m=43
%			%,a=,c=
%			$}
\end{figure}
Here the conformal matter is a Gaiotto type theory $A_8([1,1,\dots,1],[5,4],[3,3,3])$ with Coulomb branch spectrum $(9,8,6)$ and flavor symmetry $SU(12)_9$. Note that from the regular punctures, one can immediately read off the subgroup $SU(9)_9\times SU(3)_9\times U(1)$. The flavor symmetry enhancement can be seen from its $3d$ mirror in Figure~\ref{fig:A8mirror}.
Recall that the enhancement is due to $3d$ monopole operators hitting the unitarity bound. This happens when the corresponding $3d$ quiver node is {\it balanced}, meaning $N_f-2N_c=0$ \cite{Gaiotto:2008ak}. In the above $3d$ mirror quiver, we see that the balanced nodes form the Dynkin diagram of $A_{11}$ thus the enhanced flavor symmetry.

\subsection*{ICIS 
(58)}
\begin{figure}[H]
		\centering
		\includegraphics[scale=.9,keepaspectratio]{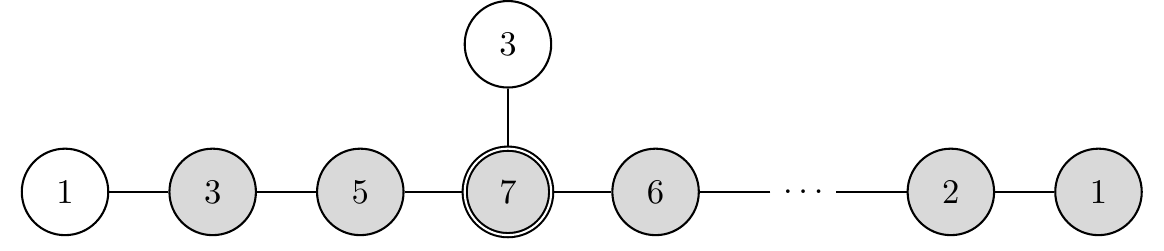}
		\caption{$3d$ $\cN=4$ mirror of $A_6([1,1,\dots,1],[4,3],[2,2,2,1])$. The {\it balanced} nodes are shaded.}
		\label{fig:A6mirror}
%		\caption{ICIS (57):  
%			$
%			\m=43
%			%,a=,c=
%			$}
\end{figure}

$\left\{ \begin{array}{l} x_1x_2+x_3^{2}+x_4^{2}+x_5^{3}=0\\x_1x_3+x_2^{2}x_4+x_4x_5^{2}=0\end{array}\right.$
\\$(w_1,w_2,w_3,w_4,w_5;1,d)=(\frac{2}{3},\frac{1}{3},\frac{1}{2},\frac{1}{2},\frac{1}{3};1,\frac{7}{6})$
\\$\mu=36,\quad
\m_1={25\over 3},\quad
\A=6,\quad
r=15,\quad
f=6,\quad
a={581\over 24},\quad
c={74\over 3}.$
\begin{figure}[H]
		\centering
		\includegraphics[scale=.9,keepaspectratio]{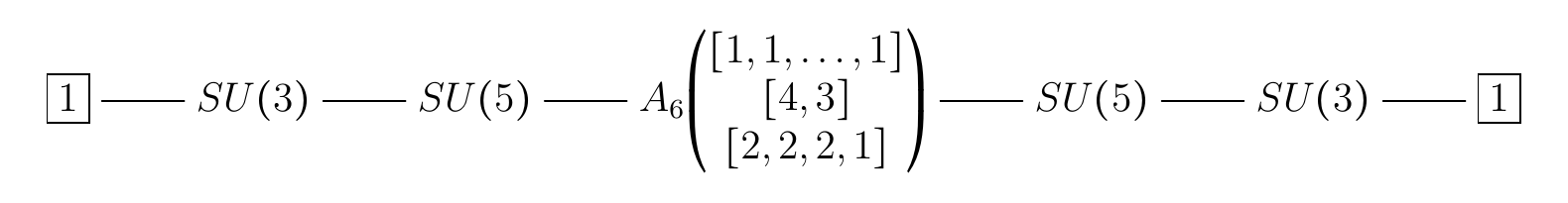}
%		\caption{ICIS (58):  
%			$
%			\m=36
%			%,a=,c=
%			$}
\end{figure}
Here the conformal matter is a Gaiotto type theory $A_6([1,1,\dots,1],[4,3],[2,2,2,1])$ with Coulomb branch spectrum $(7,6,4)$ and (enhanced) flavor symmetry $SU(10)_7\times U(1)$ which can be seen from the $3d$ mirror in Figure~\ref{fig:A6mirror}.

\subsection*{ICIS 
(59)}
$\left\{ \begin{array}{l} x_1x_2+x_3^{2}+x_4^{2}+x_5^{n}=0\\x_1x_5+2x_3^{2}+x_4^{2}+3x_2^{n}=0\end{array}\right.$
\\$n\ge3$
\\$(w_1,w_2,w_3,w_4,w_5;1,d)=(\frac{-1 + n}{n},\frac{1}{n},\frac{1}{2},\frac{1}{2},\frac{1}{n};1,1)$
\\$\mu=-3 + 4 n + n^2,\quad
\m_1=n-1,\quad
\A=n,\quad
r=\begin{cases}
\frac{1}{2} (n (n+3)-6) & n \in 2\mZ
\\
\frac{1}{2} (n-1) (n+4)  & n \in 2\mZ+1
\end{cases},
\newline
f=\begin{cases}
n+3  & n \in 2\mZ
\\
 n+1 & n \in 2\mZ+1
\end{cases},
\quad
a=\begin{cases}
{(n (4 n (n+6)-7)-30)\over 48} & n \in 2\mZ
\\
{(n (4 n (n+6)-7)-17)\over 48}  & n \in 2\mZ+1
\end{cases},
\quad 
c=\begin{cases}
{ (n-1) (n+1) (n+6)\over 12} & n \in 2\mZ
\\
{ (n-1) (n+1) (n+6)+2\over 12} & n \in 2\mZ+1
\end{cases}
$
\begin{figure}[H]
			\centering
			\includegraphics[scale=.9,keepaspectratio]{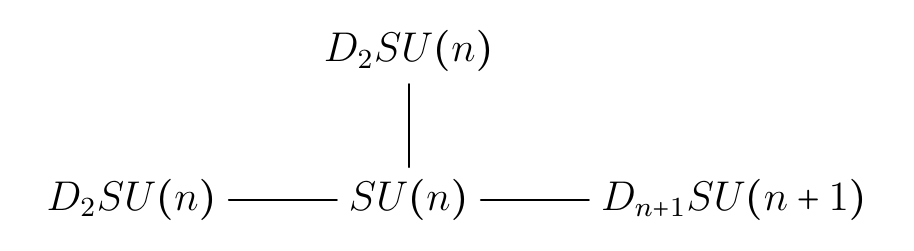}
%			\caption{ICIS (59):  
%				$
%				\m=-3+4n+n^2
%				%,a=,c=
%				$}
		\end{figure}

\subsection*{ICIS 
(60)}$\left\{ \begin{array}{l} x_1x_2+x_3^{2}+x_4^{2}+x_5^{n}=0\\x_1x_5+3x_2^{1 + 2 n}+2x_2x_3^{2}+x_2x_4^{2}=0\end{array}\right.$
\\$n\ge3$
\\$(w_1,w_2,w_3,w_4,w_5;1,d)=(\frac{-1 + 2 n}{2 n},\frac{1}{2 n},\frac{1}{2},\frac{1}{2},\frac{1}{n};1,\frac{1 + 2 n}{2 n})$
\\$\mu=5 + 9 n + 2 n^2,\quad
\m_1={2(n+1)^2\over n},\quad
\A=2n,\quad
r=n^2+4n,
\quad
f=n+5,
\quad
\newline
a= {n (8 n (n+6)+49)+4\over 24},
\quad 
c={n (2 n (n+6)+13)+2\over 6}.
$
	\begin{figure}[H]
		\centering
		\includegraphics[scale=.9,keepaspectratio]{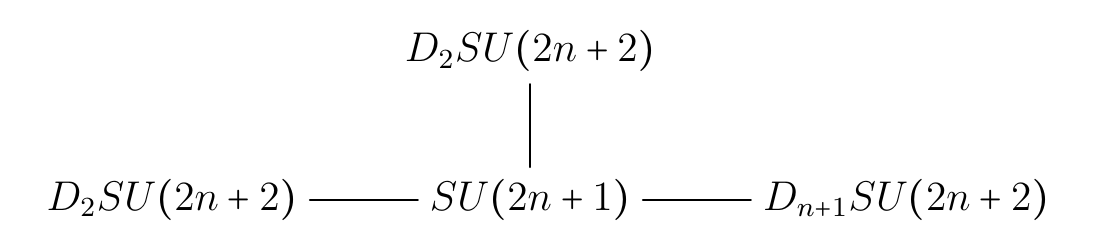}
%		\caption{ICIS (60):  
%			$
%			\m=5+9n+2n^2
%			%,a=,c=
%			$}
	\end{figure}

\subsection*{ICIS 
(74)}
$\left\{ \begin{array}{l} x_1x_2+x_3^{2}+x_4^{2}=0\\x_1x_5+x_2^{2 n}+x_5^{n}+2x_2x_3^{2}+x_2x_4^{2}=0\end{array}\right.$
\\$n\ge3$
\\$(w_1,w_2,w_3,w_4,w_5;1,d)=(\frac{2 (-1 + n)}{-1 + 2 n},\frac{1}{-1 + 2 n},\frac{1}{2},\frac{1}{2},\frac{2}{-1 + 2 n};1,\frac{2 n}{-1 + 2 n})$
\\$\mu=1 + 7 n + 2 n^2,\quad
\m_1={?2n+1)^2\over 2n-1},\quad
\A=2n-1,\quad
r=n(n+3)-1,
\quad
f=n+3,
\quad
\newline
a= { n (4 n (2 n+9)+7)-5\over 24},
\quad 
c={n (2 n (2 n+9)+5)-2\over 12}.
$
\begin{figure}[H]
		\centering
		\includegraphics[scale=.9,keepaspectratio]{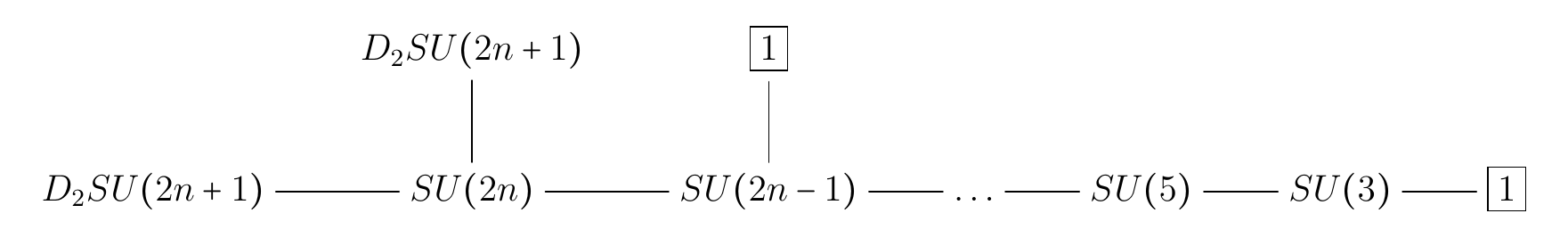}
%		\caption{ICIS (74):  
%			$
%			\m=1+7n+2n^2
%			%,a=,c=
%			$}
\end{figure}

\subsection*{ICIS 
(113)}
$\left\{ \begin{array}{l} x_1x_2+x_4x_5=0\\x_1x_4+x_3^{2}+x_2^{n_2}+x_5^{n_5}=0\end{array}\right.$
\\$n_2 \ge 3~and~2\le n_5 \le n_2$
\\$(w_1,w_2,w_3,w_4,w_5;1,d)=(\frac{n_2 - n_5 + n_2 n_5}{n_2 + n_5 + n_2 n_5},\frac{2 n_5}{n_2 + n_5 + n_2 n_5},\frac{n_2 n_5}{n_2 + n_5 + n_2 n_5},\frac{-n_2 + n_5 + n_2 n_5}{n_2 + n_5 + n_2 n_5},\frac{2 n_2}{n_2 + n_5 + n_2 n_5};1,\frac{2 n_2 n_5}{n_2 + n_5 + n_2 n_5})$
\subsubsection*{For \boldmath $n_2=n_5=n$}
$\mu=(n+1)^2,\quad
\m_1=(n-1)^2,\quad
\A={n+2\over 2},\quad
r=\begin{cases}
\frac{1}{2} (n^2+n-2) & n \in 2\mZ
\\
\frac{1}{2} n(n+1)  & n \in 2\mZ+1
\end{cases},
\newline
f=\begin{cases}
 n+3 & n \in 2\mZ
\\
  n+1 & n \in 2\mZ+1
\end{cases},\quad
a=\begin{cases}
\frac{(n+2) (2 n-1) (2 n+3)}{48} & n \in 2\mZ
\\
{n (2 n+1) (2 n+5)+7\over 48} & n \in 2\mZ+1
\end{cases}
,\quad
c=\begin{cases}
{n (n+1) (n+2)\over 12} & n \in 2\mZ
\\
{n (n+1) (n+2)+2\over 12}  & n \in 2\mZ+1
\end{cases}.
$
\begin{figure}[H]
			\centering
			\includegraphics[scale=.9,keepaspectratio]{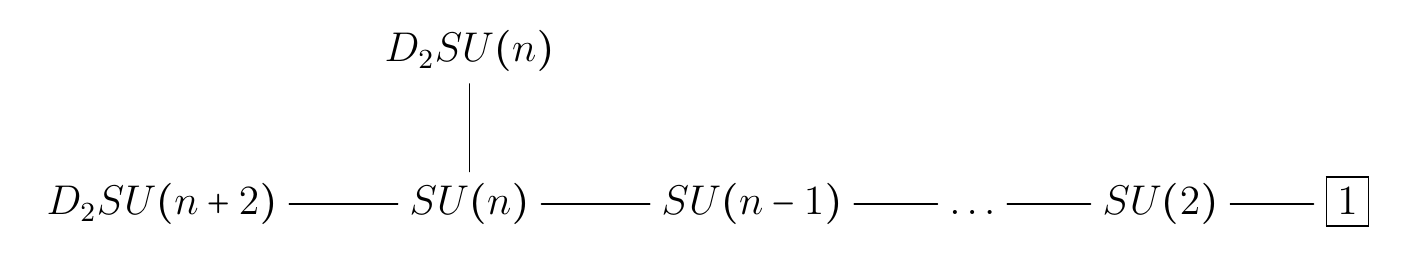}
%			\caption{ICIS (113):  
%				$
%				\m=(1+n)^2
%				%,a=,c=
%				$}
		\end{figure}

\subsection*{ICIS 
(114)}
$\left\{ \begin{array}{l} x_1x_2+x_4x_5=0\\x_1x_4+x_3^{2}+x_2^{n_2}+x_1x_5^{n_5}=0\end{array}\right.$
\\$n_2 \ge 3~and~2\le 2 n_5 \le n_2$
\\$(w_1,w_2,w_3,w_4,w_5;1,d)=(\frac{n_2 - n_5 + n_2 n_5}{(1 + n_2) (1 + n_5)},\frac{1 + 2 n_5}{(1 + n_2) (1 + n_5)},\frac{n_2 + 2 n_2 n_5}{2 (1 + n_2) (1 + n_5)},\frac{n_5}{1 + n_5},\frac{1}{1 + n_5};1,\frac{n_2 + 2 n_2 n_5}{(1 + n_2) (1 + n_5)})$
\subsubsection*{For \boldmath$2n_5=n_2=2n$}
$\mu=(1+2n)^2,\quad
\m_1=(2n-1)^2,\quad
\A=n+1,\quad
r=2n^2+n-1,\quad
\newline
f=2n+3,\quad
a={(n+1)(4n-1)(4n+3)\over 24} ,\quad
c={n(2n+1)(n+1)\over 3}.
$
\begin{figure}[H]
			\centering
			\includegraphics[scale=.9,keepaspectratio]{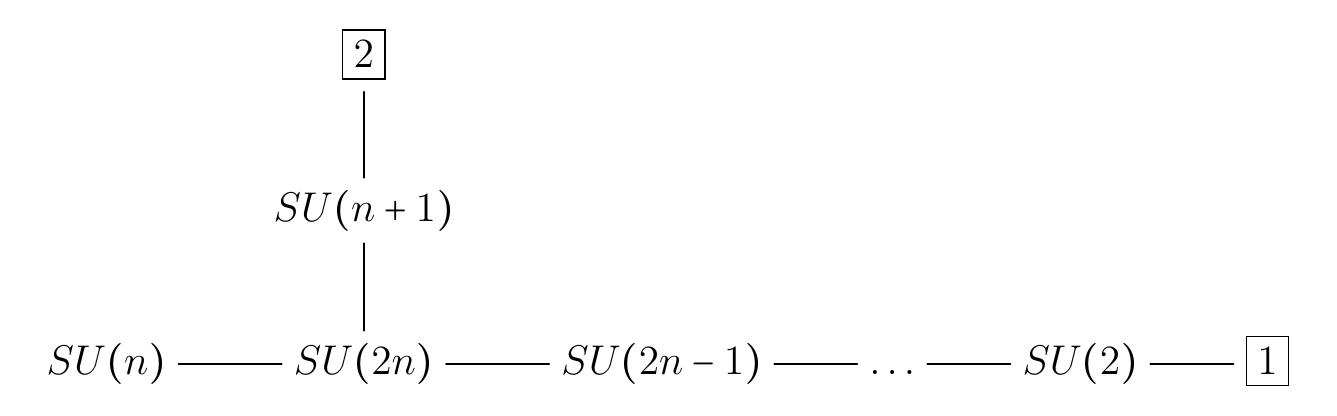}
%			\caption{ICIS (114):  
%				$
%				\m=(1+2n)^2
%				%,a=,c=
%				$}
		\end{figure}
Note that this is a special case of ICIS (113) for $n_2=n_5=2n$.

\subsection*{ICIS 
(115)}
$\left\{ \begin{array}{l} x_1x_2+x_4x_5=0\\x_1x_4+x_3^{2}+x_2^{n_2}x_4+x_5^{n_5}=0\end{array}\right.$
\\$n_2 \ge 2~and~2\le n_5 \le 2 n_2$
\\$(w_1,w_2,w_3,w_4,w_5;1,d)=(\frac{n_2}{1 + n_2},\frac{1}{1 + n_2},\frac{n_5 + 2 n_2 n_5}{2 (1 + n_2) (1 + n_5)},\frac{-n_2 + n_5 + n_2 n_5}{(1 + n_2) (1 + n_5)},\frac{1 + 2 n_2}{(1 + n_2) (1 + n_5)};1,\frac{n_5 + 2 n_2 n_5}{(1 + n_2) (1 + n_5)})$
\\$\mu=1 + 2 (1 + n_2) n_5$

For $n_5=2n_2=2n$ this coincides with ICIS (114).

\subsection*{ICIS 
(116)}
$\left\{ \begin{array}{l} x_1x_2+x_4x_5=0\\x_1x_4+2x_3^{2}+3x_2^{n_2}x_4+4x_1x_5^{n_5}+5x_2^{n_2}x_5^{n_5}=0\end{array}\right.$
\\$n_2 \ge 2~and~1\le n_5 \le n_2$
\\$(w_1,w_2,w_3,w_4,w_5;1,d)=(\frac{n_2}{1 + n_2},\frac{1}{1 + n_2},\frac{n_2 + n_5 + 2 n_2 n_5}{2 (1 + n_2) (1 + n_5)},\frac{n_5}{1 + n_5},\frac{1}{1 + n_5};1,\frac{n_2 + n_5 + 2 n_2 n_5}{(1 + n_2) (1 + n_5)})$
\\$\mu=(1 + 2 n_2) (1 + 2 n_5)$

For $n_5=n_2=n$ this coincides with ICIS (114). 

%
%\subsection*{ICIS 
%(117)}
%$\left\{ \begin{array}{l} x_1x_2+x_4^{2}+x_5^{2}=0\\x_1x_4+x_3^{2}+x_2^{n}=0\end{array}\right.$
%\\$n\ge3$
%\\$(w_1,w_2,w_3,w_4,w_5;1,d)=(\frac{-1 + 2 n}{2 (1 + n)},\frac{3}{2 (1 + n)},\frac{3 n}{4 (1 + n)},\frac{1}{2},\frac{1}{2};1,\frac{3 n}{2 (1 + n)})$
%%\subsubsection*{For \boldmath$n=5$}
%%$\mu=21,\quad
%%\m_1=6,\quad
%%\A={8\over 3},\quad
%%r=10,\quad
%%f=1,\quad
%%a={91\over 12},\quad
%%c={23\over 3}.\quad
%%$
%\subsubsection*{For \boldmath$n=8$}
%$\mu=33,\quad
%\m_1={35\over 3},\quad
%\A={3},\quad
%r=14,\quad
%f=5,\quad
%a={53\over 4},\quad
%c={27\over 2}.\quad
%$
%\begin{figure}[H]
%			\centering
%			\includegraphics[scale=.9,keepaspectratio]{117}
%%			\caption{ICIS (117):  
%%				$
%%				\m=21
%%				%,a=,c=
%%				$}
%		\end{figure}
%

\subsection*{ICIS 
(118)}
$\left\{ \begin{array}{l} x_1x_2+x_4^{2}+x_5^{3}=0\\x_1x_4+x_3^{2}+x_2^{n}=0\end{array}\right.$
\\$n\ge3$
\\$(w_1,w_2,w_3,w_4,w_5;1,d)=(\frac{-1 + 2 n}{2 (1 + n)},\frac{3}{2 (1 + n)},\frac{3 n}{4 (1 + n)},\frac{1}{2},\frac{1}{3};1,\frac{3 n}{2 (1 + n)})$
\subsubsection*{For \boldmath$n=8$}
$\mu=59,\quad
\m_1=21,\quad
\A=6,\quad
r=26,\quad
f=7,\quad
a={527\over 12},\quad
c={133\over 3}.\quad
$
\begin{figure}[H]
			\centering
			\includegraphics[scale=.9,keepaspectratio]{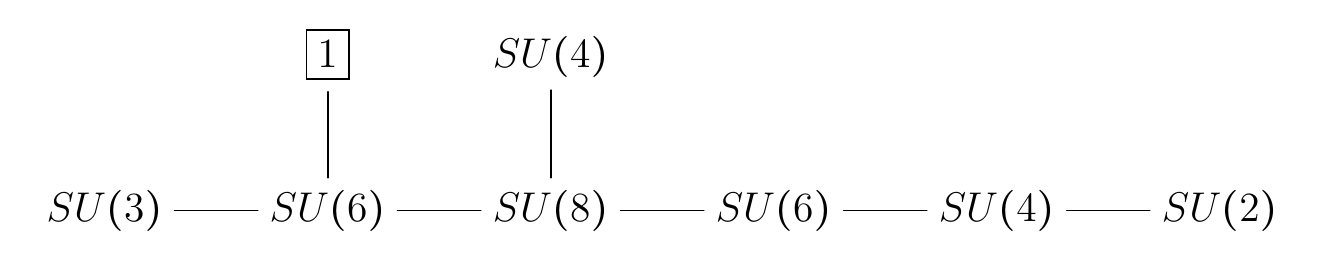}
%			\caption{ICIS (118):  
%				$
%				\m=59
%				%,a=,c=
%				$}
		\end{figure}

\subsection*{ICIS 
(119)}
$\left\{ \begin{array}{l} x_1x_2+x_4^{2}+x_5^{4}=0\\x_1x_4+x_3^{2}+x_2^{n}=0\end{array}\right.$
\\$n\ge3$
\\$(w_1,w_2,w_3,w_4,w_5;1,d)=(\frac{-1 + 2 n}{2 (1 + n)},\frac{3}{2 (1 + n)},\frac{3 n}{4 (1 + n)},\frac{1}{2},\frac{1}{4};1,\frac{3 n}{2 (1 + n)})$
\subsubsection*{For \boldmath$n=6k-1$}
$\mu=60k-5,\quad
\m_1=30k-16+{2\over k},\quad
\A={6k-1\over 4k},\quad
r=3(10k-3),\quad
\newline
f=1,\quad
a=60 k^2-\frac{37 k}{4}+\frac{25}{24},\quad
c=60 k^2-9 k+\frac{5}{6}.\quad
$
\begin{figure}[H]
			\centering
			\includegraphics[scale=.75,keepaspectratio]{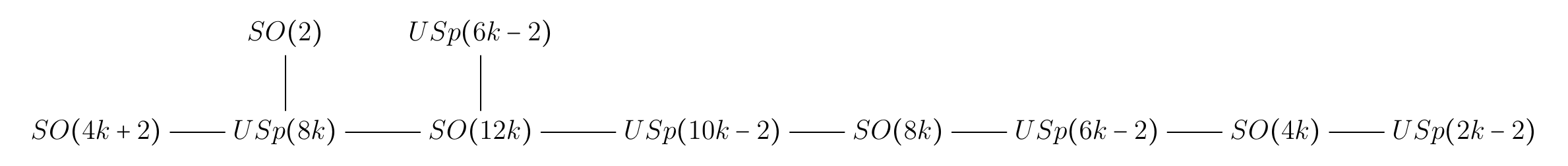}
%			\caption{ICIS (119):  
%				$
%				\m=60k-5
%				%,a=,c=
%				$}
		\end{figure}

\subsection*{ICIS 
(143)}
$\left\{ \begin{array}{l} x_1x_2+x_4^{2}+x_5^{n_5}=0\\x_1x_5+x_3^{2}+x_2^{n_2}=0\end{array}\right.$
\\$n_2 \ge 3~and~3\le n_5 \le n_2$
\\$(w_1,w_2,w_3,w_4,w_5;1,d)=(\frac{-1 + n_2 n_5}{(1 + n_2) n_5},\frac{1 + n_5}{(1 + n_2) n_5},\frac{n_2 + n_2 n_5}{2 (1 + n_2) n_5},\frac{1}{2},\frac{1}{n_5};1,\frac{n_2 + n_2 n_5}{(1 + n_2) n_5})$
\\$\mu=-3 + 2 n_5 + n_2 (2 + n_5)$

For $n_5=n_2=k$ this is the same as ICIS (59).

\subsection*{ICIS 
(144)}
$\left\{ \begin{array}{l} x_1x_2+x_4^{2}+x_5^{n_5}=0\\x_1x_5+x_3^{2}+x_2^{n_2}x_4=0\end{array}\right.$
\\$n_2 \ge 2~and~3\le n_5 \le 2 n_2$
\\$(w_1,w_2,w_3,w_4,w_5;1,d)=(\frac{-2 + n_5 + 2 n_2 n_5}{2 (1 + n_2) n_5},\frac{2 + n_5}{2 (1 + n_2) n_5},\frac{2 n_2 + n_5 + 2 n_2 n_5}{4 (1 + n_2) n_5},\frac{1}{2},\frac{1}{n_5};1,\frac{2 n_2 + n_5 + 2 n_2 n_5}{2 (1 + n_2) n_5})$
\\$\mu=3 (-1 + n_5) + 2 n_2 (1 + n_5)$

For $n_5=2n_2=2k$ this is a subclass of ICIS (59) with $n=2k$.

\subsection*{ICIS (152)}
$\left\{ \begin{array}{l} x_1x_2+x_4^{3}+x_5^{3}=0\\x_1x_4+x_3^{2}+x_2^{3}=0\end{array}\right.$
\\$n=3$
\\$(w_1,w_2,w_3,w_4,w_5;1,d)=({2\over3},\frac{1}{3},\frac{1 }{2 },\frac{1}{3},\frac{1}{3};1,1)$
\\$\mu=34, \quad \mu_1=4, \quad \alpha=6,\quad r=16, \quad f=2, \quad a=\frac{259}{12} ,\quad c=\frac{65}{3}$ 
 \begin{figure}[H]
 \centering
\includegraphics[scale=.9,keepaspectratio]{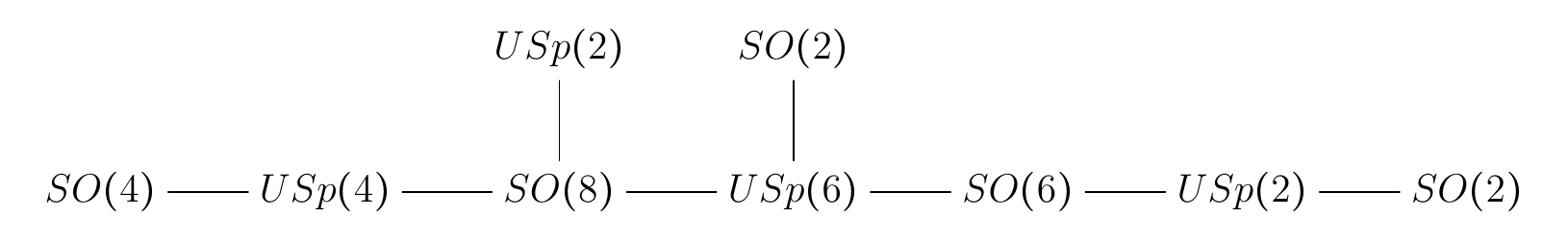} 
\end{figure}

\subsection*{ICIS (154)}		
$\left\{ \begin{array}{l} x_1x_2+x_4^{3}+x_5^{3}=0\\x_1x_4+x_3^{2}+x_2^{n}x_5=0\end{array}\right.$
\\$n\ge2$
\\$(w_1,w_2,w_3,w_4,w_5;1,d)=(\frac{n}{1 + n},\frac{1}{1 + n},\frac{1 + 4 n}{6 (1 + n)},\frac{1}{3},\frac{1}{3};1,\frac{1 + 4 n}{3 (1 + n)})$

\subsubsection*{For \boldmath$n=3k-1$}
$\mu=2(18k-1), \quad 
\mu_1=12k+{2\over k}-10, \quad 
\alpha=6k,\quad 
r=18k-2 \quad
 f=2, \quad
 \newline
  a= 24 k^2-\frac{13 }{4}k+\frac{5}{6},\quad 
  c=24k^2-3k+\frac{2}{3}
  $ 
 \begin{figure}[H]
 \centering
\includegraphics[scale=.7,keepaspectratio]{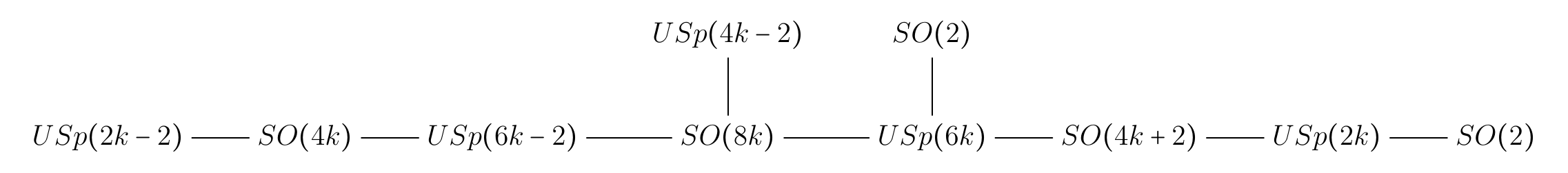} 
\end{figure}

\subsubsection*{For general \boldmath$n$}
$\mu=2(5+6n), \quad 
\mu_1={2n(2n-1)\over n+1}, \quad
 \alpha=2(n+1),\quad
  r=6n+4, \quad 
  f=2, \quad
  \newline
   a={32n^2-77n+55\over 12},\quad 
   c={8n^2-19n+13\over 3}.
   $ 
 \begin{figure}[H]
 \centering
\includegraphics[scale=.7,keepaspectratio]{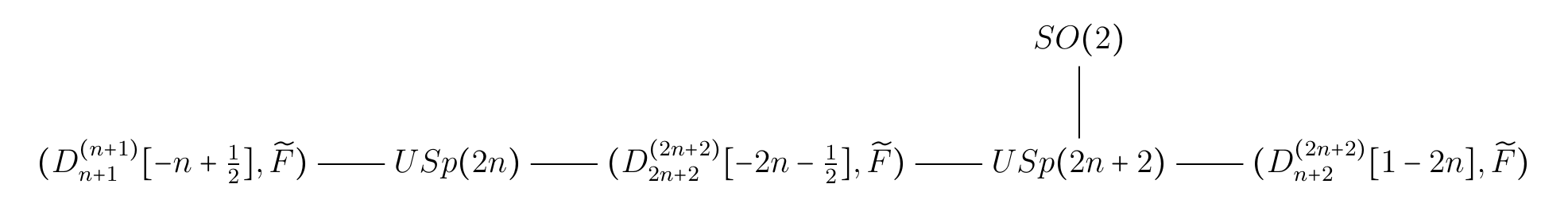} 
\end{figure}
It's straightforward to check that the gauge groups are conformal. Explicitly, let's look at the $USp(2n)$ node. Recall from subsection \eqref{sec:ADmatters} that the $(D_{n+1}^{(n+1)}[k],\widetilde{F})$ has flavor symmetry $USp(2n)$ with flavor central charge $n+1-{n+1\over 2(n+1+k)}$, while $(D_{2n+2}^{(2n+2)}[k],\widetilde{F})$ has non-abelian flavor symmetry $USp(4n+2)$ with flavor central charge $2n+2-{n+1\over 2n+2+k}$ and the Dynkin index of embedding $I_{C_{n}\hookrightarrow C_{2n+1}}=1$. Together the beta function is
\ie
\B_{C_n}=2(n+1)-\left(n+1-{n+1\over 2 (n+1-n+1/2)}\right)-\left(2n+2-{n+1\over 2n+2-2n-1/2}\right)=0
\fe
Similarly the other gauge node $USp(2n+2)$ is also conformal. One can also check that the left over flavor symmetry from the quiver has rank $2$ which agrees with what we see from the singularity.

\subsection*{ICIS (182)}	
$\left\{ \begin{array}{l} x_1x_2+x_3x_4=0\\x_1x_3+x_2^{n}+x_4^{n}+x_5^{2}=0\end{array}\right.$
\\$(w_1,w_2,w_3,w_4,w_5;1,d)=(\frac{n}{2 + n},\frac{2}{2 + n},\frac{n}{2 + n},\frac{2}{2 + n},\frac{n}{2 + n};1,\frac{2 n}{2 + n})$
\subsubsection*{For \boldmath$n=2k+1$}
$\mu=4(k+1)^2, \quad 
\mu_1=4k^2, \quad 
\alpha=k+{3\over 2},\quad 
r=(k+1)(2k+1), \quad 
f=2(k+1), \quad 
\newline
a= { (k+1) (4 k-1) (4 k+3)\over 24},\quad 
c={k (k+1) (2 k+1)\over 3}.$ 
 \begin{figure}[H]
 \centering
\includegraphics[scale=.9,keepaspectratio]{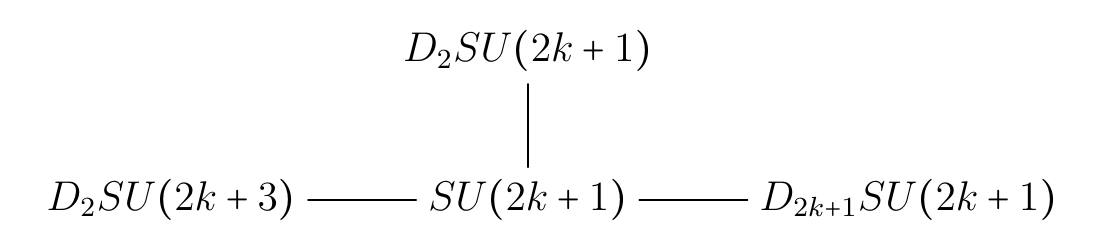} 
\end{figure}
\subsubsection*{For \boldmath$n=2k$}
$\mu=(2k+1)^2, \quad 
\mu_1=(2k-1)^2, \quad 
\alpha=k+1,\quad 
r=2k^2+k-1, \quad 
f=2(k+1)+1, \quad  
\newline
a={k (16 k (k+3)+41)+14\over 24},\quad 
c={k (4 k (k+3)+11)+4\over 6}.$ 
 \begin{figure}[H]
 \centering
\includegraphics[scale=.9,keepaspectratio]{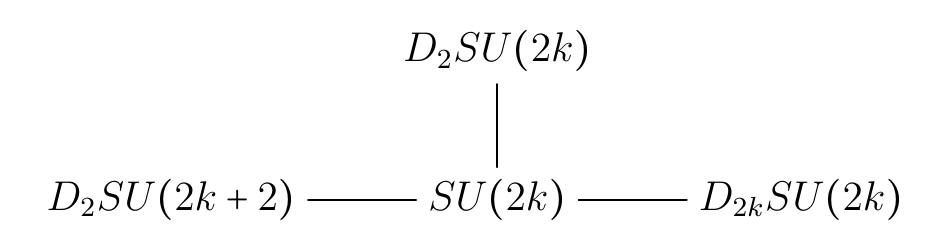} 
\end{figure}

\subsection*{ICIS (234)}		
$\left\{ \begin{array}{l} x_1x_2+x_3x_4+x_5^{n}=0\\x_1x_4+x_3x_5+x_2^{n}=0\end{array}\right.$
\\$n_2 \ge 3~and~2\le n_5 \le n_2$
\\$(w_1,w_2,w_3,w_4,w_5;1,d)=(\frac{n-1}{n},\frac{1 }{n},\frac{n-1}{n},\frac{1}{n},\frac{1 }{n};1,1)$
\\$\mu=1+2n^2, \quad \mu_1=n-1, \quad \alpha=n,\quad r=n^2-n, \quad f=2n+1, \quad 
\newline 
a={ n \left(4 n^2+6 n-7\right)\over 24},\quad
c={ n (n+2) (2 n-1)\over 12}$
 \begin{figure}[H]
 \centering
\includegraphics[scale=.9,keepaspectratio]{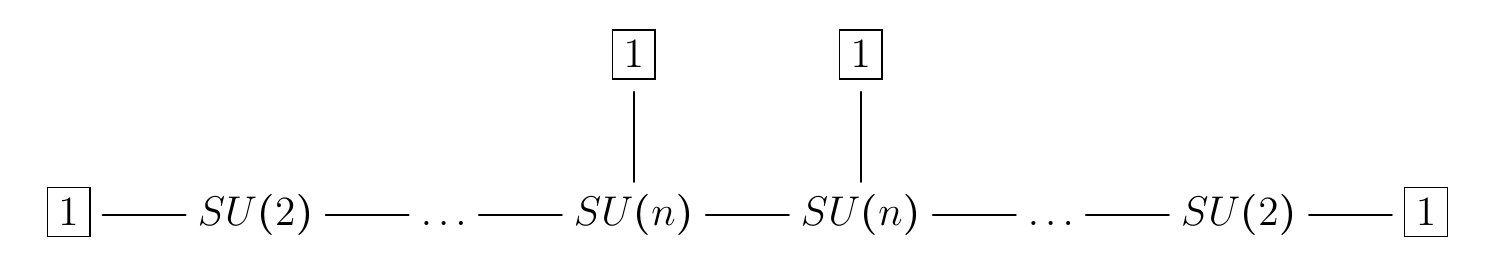} 
\end{figure}

\subsection*{ICIS (253)}	
$\left\{ \begin{array}{l} x_1x_2+x_3x_4=0\\x_1x_4+x_3^{2}+x_2^{4}x_4+x_5^{2}=0\end{array}\right.$
\\$(w_1,w_2,w_3,w_4,w_5;1,d)=(\frac{4}{5},\frac{1}{5},\frac{3}{5},\frac{2 }{5},\frac{3}{5};1,\frac{6}{5})$
\\$\mu=21, \quad \mu_1=5, \quad \alpha={5\over2},\quad r=8, \quad f=5, \quad a=\frac{13}{2}, \quad c=\frac{27}{4}$	
 \begin{figure}[H]
 \centering
\includegraphics[scale=.9,keepaspectratio]{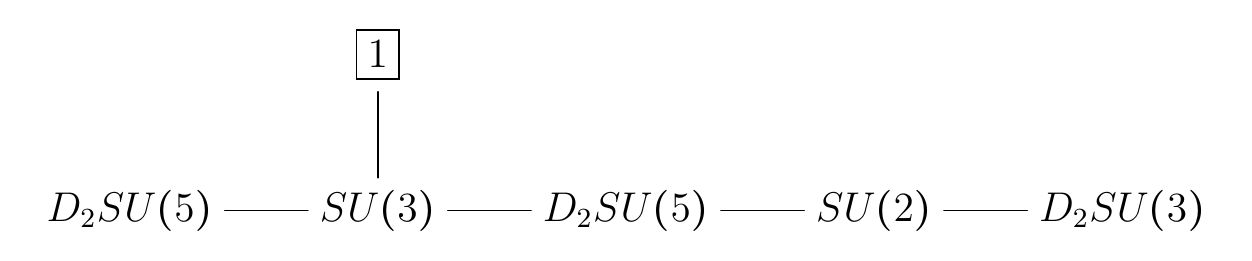} 
\end{figure}

\subsection*{ICIS (261)}	
$\left\{ \begin{array}{l} x_1x_2+x_3x_4+x_5^{2}=0\\x_1x_5+x_3^{2}+x_4^{3}+x_2^{4}=0\end{array}\right.$
\\$(w_1,w_2,w_3,w_4,w_5;1,d)=(\frac{7}{10},\frac{3}{10},\frac{3}{5},\frac{2}{5},\frac{1}{2};1,\frac{6}{5})$
\\$\mu=24, \quad \mu_1=6, \quad \alpha={10\over3},\quad r=11, \quad f=2, \quad a=\frac{241}{24}, \quad c=\frac{61}{6}$	
 \begin{figure}[H]
 \centering
\includegraphics[scale=.9,keepaspectratio]{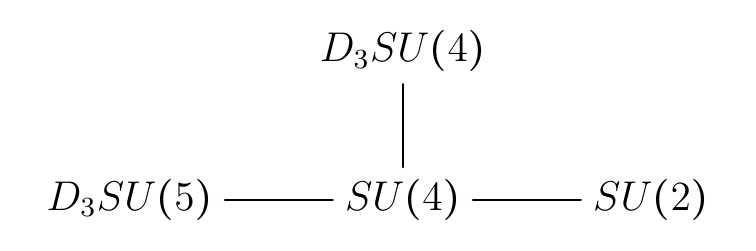}
\end{figure}

\subsection*{ICIS (262)}		
$\left\{ \begin{array}{l} x_1x_2+x_3x_4+x_5^{2}=0\\x_1x_5+x_3^{2}+x_4^{4}+x_2^{8}=0\end{array}\right.$
\\$(w_1,w_2,w_3,w_4,w_5;1,d)=(\frac{5}{6},\frac{1}{6},\frac{2}{3},\frac{1}{3},\frac{1}{2};1,\frac{4}{3})$
\\$\mu=59, \quad \mu_1=21, \quad \alpha=6,\quad r=26, \quad f=7, \quad a=\frac{527}{12}, \quad c=\frac{133}{3}$	
 \begin{figure}[H]
 \centering
\includegraphics[scale=.9,keepaspectratio]{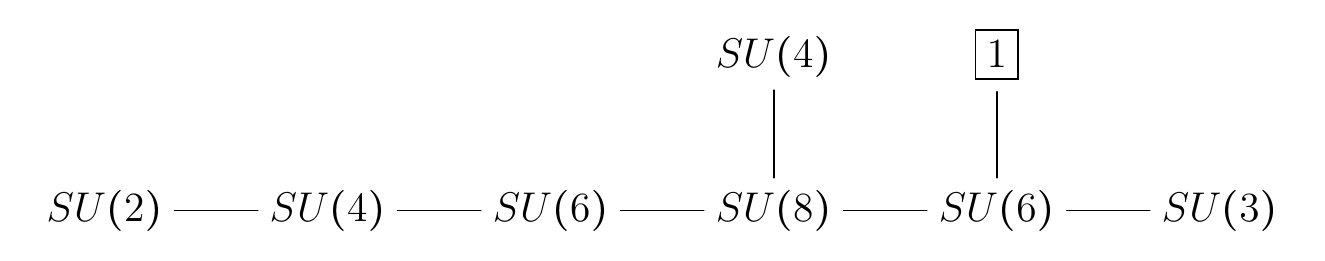}
\end{figure}

\subsection*{ICIS (263)}		
$\left\{ \begin{array}{l} x_1x_2+x_3x_4+x_5^{2}=0\\x_1x_5+x_3^{2}+x_4^{5}+x_2^{20}=0\end{array}\right.$
\\$(w_1,w_2,w_3,w_4,w_5;1,d)=(\frac{13}{14},\frac{1}{14},\frac{5}{7},\frac{2}{7},\frac{1}{2};1,\frac{10}{7})$
\\$\mu=174, \quad \mu_1=76, \quad \alpha=14,\quad r=83, \quad f=8, \quad a=\frac{2439}{8}, \quad c=\frac{611}{2}$	
 \begin{figure}[H]
 \centering
\includegraphics[scale=.9,keepaspectratio]{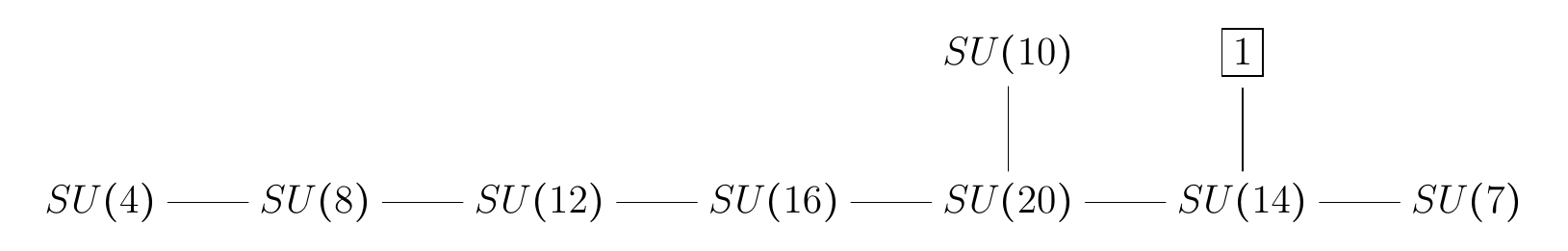}
\end{figure}

\subsection*{ICIS (264)}		
$\left\{ \begin{array}{l} x_1x_2+x_3x_4+x_5^{3}=0\\x_1x_5+x_3^{2}+x_4^{3}+x_2^{9}=0\end{array}\right.$
\\$(w_1,w_2,w_3,w_4,w_5;1,d)=(\frac{13}{15},\frac{2}{15},\frac{3}{5},\frac{2}{5},\frac{1}{3};1,\frac{6}{5})$
\\$\mu=68, \quad \mu_1=32, \quad \alpha={15\over2},\quad r=32, \quad f=4, \quad a=\frac{1381}{24}, \quad c=\frac{347}{6}$	
 \begin{figure}[H]
 \centering
\includegraphics[scale=.9,keepaspectratio]{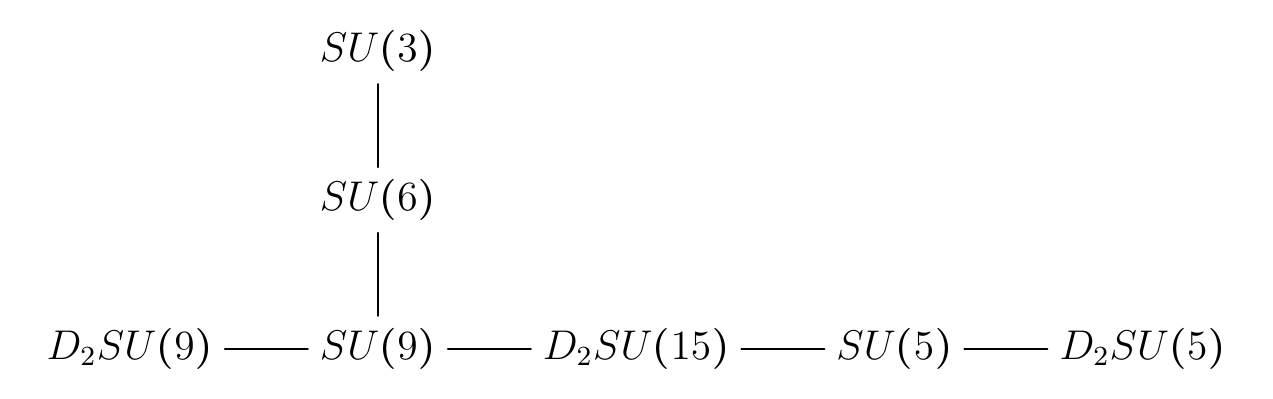}
\end{figure}

\subsection*{ICIS (265)}	
$\left\{ \begin{array}{l} x_1x_2+x_3x_4+x_5^{4}=0\\x_1x_5+x_3^{2}+x_4^{3}+x_2^{24}=0\end{array}\right.$
\\$(w_1,w_2,w_3,w_4,w_5;1,d)=(\frac{19}{20},\frac{1}{20},\frac{3}{5},\frac{2}{5},\frac{1}{4};1,\frac{6}{5})$
\\$\mu=212, \quad \mu_1=46, \quad \alpha=20,\quad r=102, \quad f=8, \quad a=\frac{1785}{4}, \quad c=447$	
 \begin{figure}[H]
 \centering
\includegraphics[scale=.9,keepaspectratio]{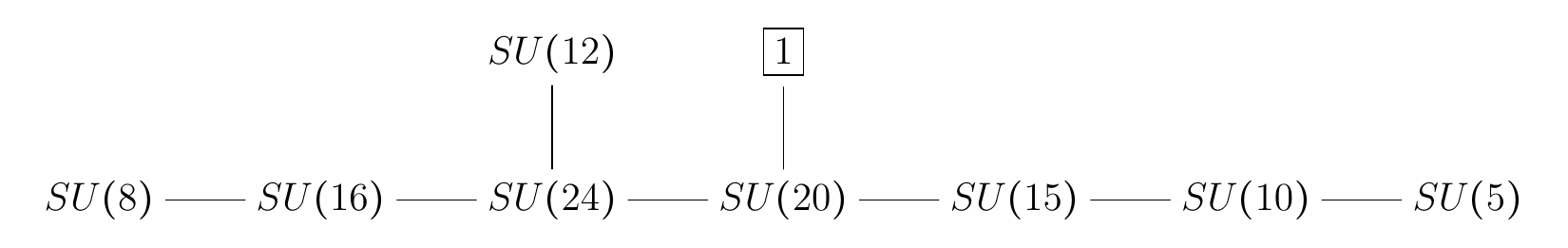}
\end{figure}

\subsection*{ICIS (266)}	
$\left\{ \begin{array}{l} x_1x_2+x_3x_4=0\\x_1x_5+x_3^{2}+x_4^{4}+x_5^{3}+x_2^{12}=0\end{array}\right.$
\\$(w_1,w_2,w_3,w_4,w_5;1,d)=(\frac{8}{9},\frac{1}{9},\frac{2}{3},\frac{1}{3},\frac{4}{9};1,\frac{4}{3})$
\\$\mu=94, \quad \mu_1=33, \quad \alpha=9,\quad r=43, \quad f=8, \quad a=\frac{611}{6}, \quad c=\frac{1229}{12}$	
 \begin{figure}[H]
 \centering
\includegraphics[scale=.9,keepaspectratio]{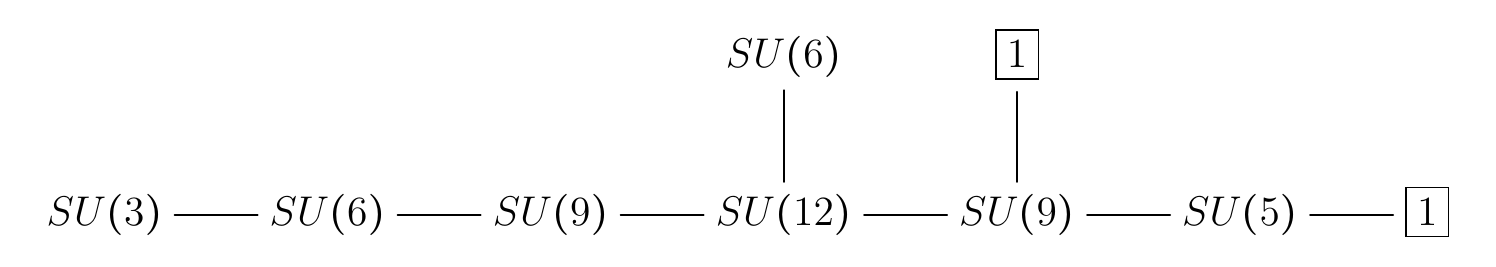}
\end{figure}

\subsection*{ICIS (267)}	
$\left\{ \begin{array}{l} x_1x_2+x_3x_4=0\\x_1x_5+x_3^{2}+x_4^{5}+x_5^{3}+x_2^{30}=0\end{array}\right.$
\\$(w_1,w_2,w_3,w_4,w_5;1,d)=(\frac{20}{21},\frac{1}{21},\frac{5}{7},\frac{2}{7},\frac{10}{21};1,\frac{10}{7})$
\\$\mu=267, \quad \mu_1=116, \quad \alpha=21,\quad r=129, \quad f=9, \quad a=\frac{2763}{4}, \quad c=\frac{2767}{4}$	
 \begin{figure}[H]
 \centering
\includegraphics[scale=.9,keepaspectratio]{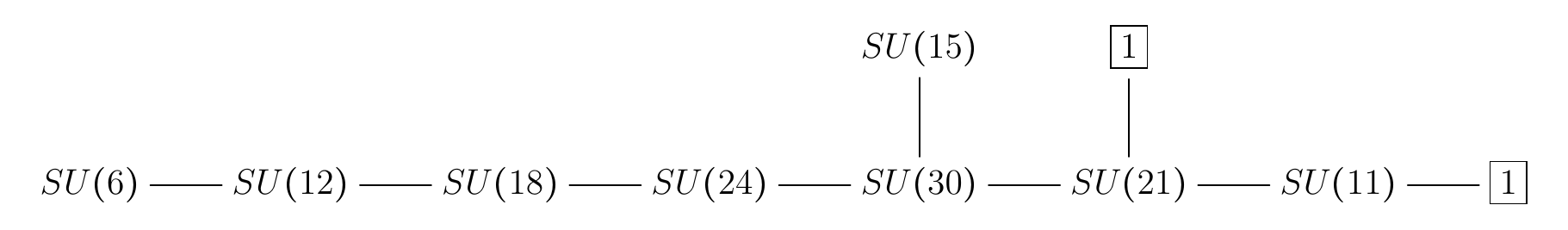}
\end{figure}

\subsection*{ICIS (268)}	
$\left\{ \begin{array}{l} x_1x_2+x_3x_4=0\\x_1x_5+x_3^{2}+x_4^{3}+x_5^{4}+x_2^{12}=0\end{array}\right.$
\\$(w_1,w_2,w_3,w_4,w_5;1,d)=(\frac{9}{10},\frac{1}{10},\frac{3}{5},\frac{2}{5},\frac{3}{10};1,\frac{6}{5})$
\\$\mu=96, \quad \mu_1=22, \quad \alpha=10,\quad r=44, \quad f=8, \quad a=\frac{1261}{12}, \quad c=\frac{317}{3}$	
 \begin{figure}[H]
 \centering
\includegraphics[scale=.9,keepaspectratio]{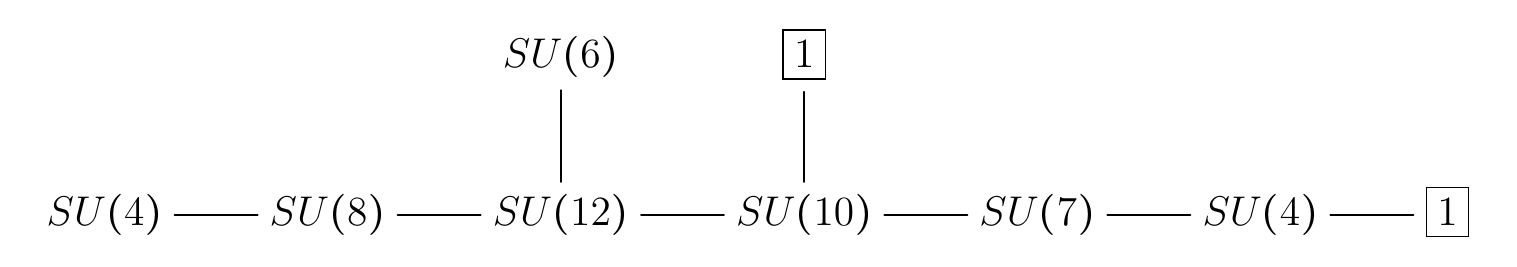}
\end{figure}

\subsection*{ICIS (269)}	
$\left\{ \begin{array}{l} x_1x_2+x_3x_4=0\\x_1x_5+x_3^{2}+x_4^{3}+x_5^{5}+x_2^{30}=0\end{array}\right.$
\\$(w_1,w_2,w_3,w_4,w_5;1,d)=(\frac{24}{25},\frac{1}{25},\frac{3}{5},\frac{2}{5},\frac{6}{25};1,\frac{6}{5})$
\\$\mu=271, \quad \mu_1=58, \quad \alpha=25,\quad r=131, \quad f=9, \quad a=\frac{2825}{4}, \quad c=\frac{2829}{4}$	
 \begin{figure}[H]
 \centering
\includegraphics[scale=.9,keepaspectratio]{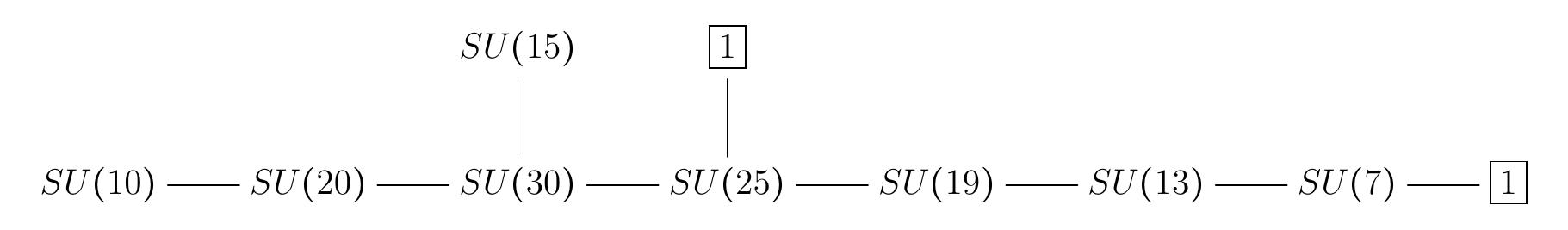} 
\end{figure}

\subsection*{ICIS (271)}		
$\left\{ \begin{array}{l} x_1x_2+x_3x_4+x_5^{1 + 2 n}=0\\x_1x_5+x_3^{2}+x_2x_4^{2}+x_2^{2 + 3 n}=0\end{array}\right.$
\\$n\ge1$
\\$(w_1,w_2,w_3,w_4,w_5;1,d)=(\frac{1 + 6 n}{3 (1 + 2 n)},\frac{2}{3 (1 + 2 n)},\frac{2 + 3 n}{3 (1 + 2 n)},\frac{1 + 3 n}{3 (1 + 2 n)},\frac{1}{1 + 2 n};1,\frac{2 (2 + 3 n)}{3 (1 + 2 n)})$	
\\$\mu=9+19n+6n^2, \quad 
\mu_1=3(n+1), \quad 
\alpha={6n+3\over 2},\quad
 r=3 (n (n+3)+1), \quad 
 f=3, \quad 
 \newline
 a=\frac{3}{8} (n+1) (2 n (2 n+7)+5), \quad
  c=\frac{1}{4} (n+1) (3 n (2 n+7)+8).
  $
 \begin{figure}[H]
 \centering
\includegraphics[scale=.9,keepaspectratio]{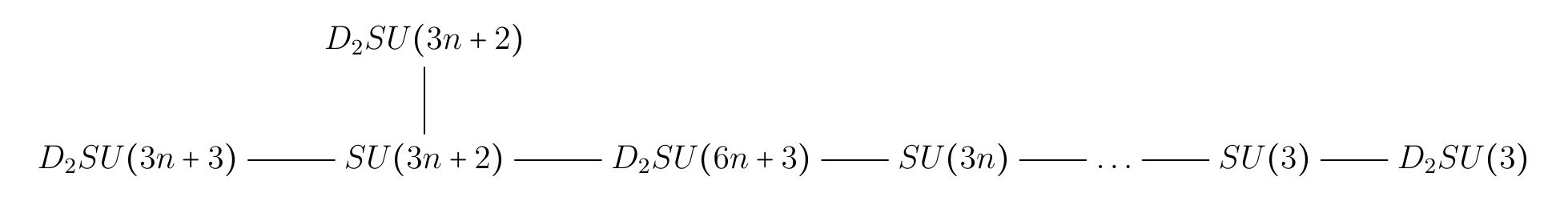} 
\end{figure}	
	
\subsection*{ICIS (272)}	
$\left\{ \begin{array}{l} x_1x_2+x_3x_4+x_5^{2}=0\\x_1x_5+x_3^{2}+x_2x_4^{3}+x_2^{6}=0\end{array}\right.$
\\$(w_1,w_2,w_3,w_4,w_5;1,d)=(\frac{11}{14},\frac{3}{14},\frac{9}{14},\frac{5}{14},\frac{1}{2};1,\frac{9}{7})$	
\\$\mu=41, \quad \mu_1=13, \quad \alpha={14\over3} ,\quad r=19, \quad f=3, \quad a=\frac{575}{24}, \quad c=\frac{145}{6}$	
 \begin{figure}[H]
 \centering
\includegraphics[scale=0.9,keepaspectratio]{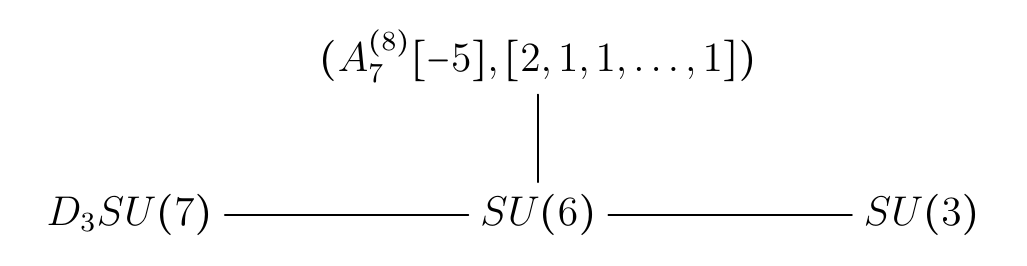} 
\end{figure}		
The AD matter theory $(A_{7}^{(8)}[-5],[2,1^6])$ is defined by six dimensional $A_{7}$ $(2,0)$ theory 
on a sphere with irregular puncture $ A_{7}^{(8)}[-5]$  and a regular puncture labeled by Young Tableaux $[2, 1^6]$. This 
theory has flavor symmetry $SU(6)\times U(1)$, and the flavor central charge of $SU(6)$ is ${13\over3}$. The Coulomb branch 
spectrum is $({13\over3},{10\over3},{8\over3},{7\over3},{5\over3},{4\over3})$, and the central charges are $a=\frac{83}{12}, c=\frac{15}{2}$. 
 The theory $D_3(SU(7))$ 
has flavor symmetry $SU(7)$, and the flavor central charge is ${14\over 3}$.  The Coulomb branch spectrum of this theory is $({14\over3},{11\over3},{8\over3},{7\over3},{5\over3},{4\over3})$.
Using the above data, one can check that the 
gauging for gauge group $SU(6)$ is conformal.

\subsection*{ICIS (273)}		
$\left\{ \begin{array}{l} x_1x_2+x_3x_4+x_5^{2}=0\\x_1x_5+x_3^{2}+x_2x_4^{4}+x_2^{11}=0\end{array}\right.$
\\$(w_1,w_2,w_3,w_4,w_5;1,d)=(\frac{7}{8},\frac{1}{8},\frac{11}{16},\frac{5}{16},\frac{1}{2};1,\frac{11}{8})$	
\\$\mu=87, \quad \mu_1=34, \quad \alpha=8 ,\quad r=87, \quad f=5, \quad a=\frac{697}{8}, \quad c=\frac{175}{2}$	
 \begin{figure}[H]
 \centering
\includegraphics[scale=.9,keepaspectratio]{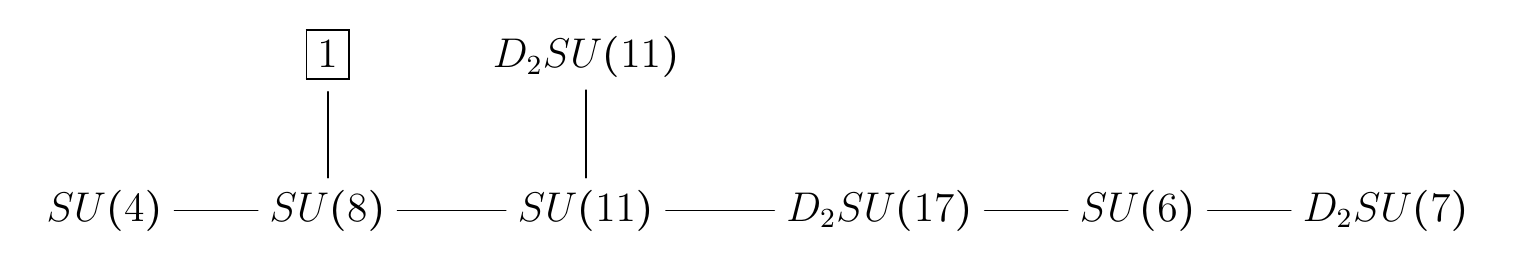}
\end{figure}

\subsection*{ICIS (274)}	
$\left\{ \begin{array}{l} x_1x_2+x_3x_4+x_5^{2}=0\\x_1x_5+x_3^{2}+x_2x_4^{5}+x_2^{26}=0\end{array}\right.$
\\$(w_1,w_2,w_3,w_4,w_5;1,d)=(\frac{17}{18},\frac{1}{18},\frac{13}{18},\frac{5}{18},\frac{1}{2};1,\frac{13}{9})$	
\\$\mu=233, \quad \mu_1=105, \quad \alpha=18 ,\quad r=112, \quad f=9, \quad a=\frac{3149}{6}, \quad c=\frac{1577}{3}$	
 \begin{figure}[H]
 \centering
\includegraphics[scale=.9,keepaspectratio]{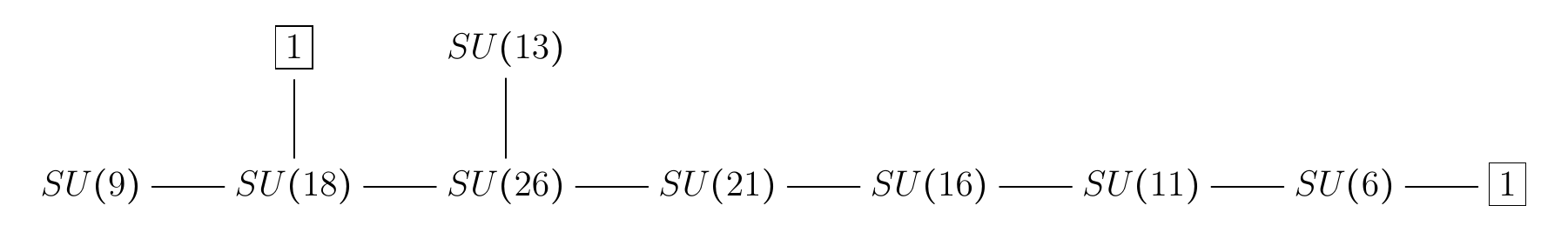}
\end{figure}

\subsection*{ICIS (275)}	
$\left\{ \begin{array}{l} x_1x_2+x_3x_4+x_5^{3}=0\\x_1x_5+x_3^{2}+x_2x_4^{3}+x_2^{13}=0\end{array}\right.$
\\$(w_1,w_2,w_3,w_4,w_5;1,d)=(\frac{19}{21},\frac{2}{21},\frac{13}{21},\frac{8}{21},\frac{1}{3};1,\frac{26}{21})$
\\$\mu=103, \quad \mu_1=27, \quad \alpha={21\over2} ,\quad r=49, \quad f=5, \quad a=\frac{2915}{24}, \quad c=\frac{1463}{12}$	
 \begin{figure}[H]
 \centering
\includegraphics[scale=.9,keepaspectratio]{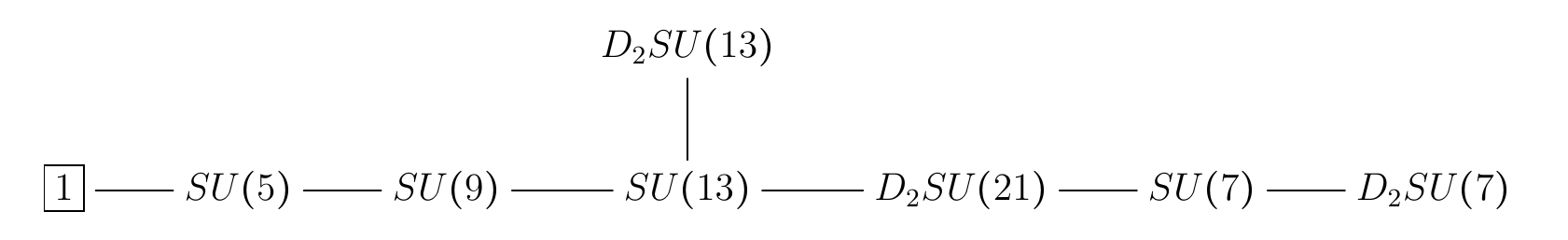}
\end{figure}

\subsection*{ICIS (276)}
$\left\{ \begin{array}{l} x_1x_2+x_3x_4+x_5^{4}=0\\x_1x_5+x_3^{2}+x_2x_4^{3}+x_2^{34}=0\end{array}\right.$
\\$(w_1,w_2,w_3,w_4,w_5;1,d)=(\frac{27}{28},\frac{1}{28},\frac{17}{28},\frac{11}{28},\frac{1}{4};1,\frac{17}{14})$
\\$\mu=305, \quad \mu_1=69, \quad \alpha=28 ,\quad r=148, \quad f=9, \quad a=\frac{5377}{6}, \quad c=\frac{2692}{3}$	
 \begin{figure}[H]
 \centering
\includegraphics[scale=.9,keepaspectratio]{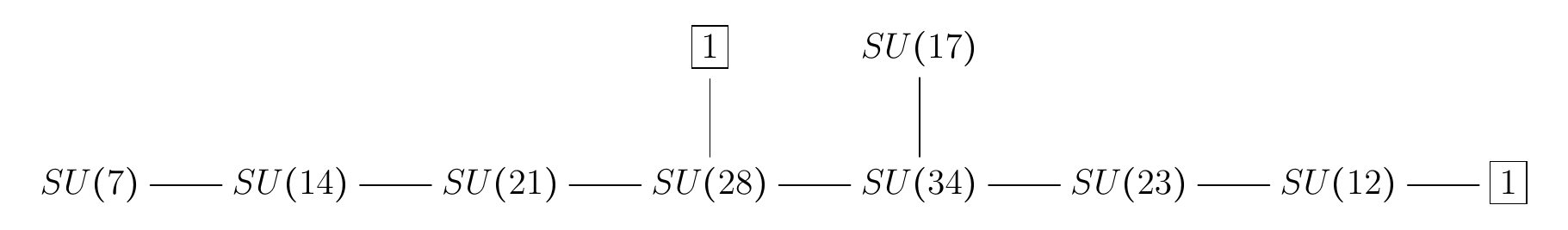}
\end{figure}

\subsection*{ICIS (278)}	
$\left\{ \begin{array}{l} x_1x_2+x_3x_4=0\\x_1x_5+x_3^{2}+x_2x_4^{2}+x_5^{2 n}+x_2^{3 n}=0\end{array}\right.$
\\$n\ge2$
\\$(w_1,w_2,w_3,w_4,w_5;1,d)=(\frac{3 (-1 + 2 n)}{-1 + 6 n},\frac{2}{-1 + 6 n},\frac{3 n}{-1 + 6 n},\frac{-1 + 3 n}{-1 + 6 n},\frac{3}{-1 + 6 n};1,\frac{6 n}{-1 + 6 n})$	
\\$\mu=-1 + 11 n + 6 n^2, \quad \mu_1=1 + 3 n, \quad \alpha=3 n-\frac{1}{2},\quad r=(3n-1)(n+2), \quad f=n+3, \quad a=\frac{1}{24} \left(36 n^3+90 n^2+3 n-10\right), \quad c=\frac{1}{12} \left(18 n^3+45 n^2+3 n-4\right)$	
 \begin{figure}[H]
 \centering
\includegraphics[scale=.9,keepaspectratio]{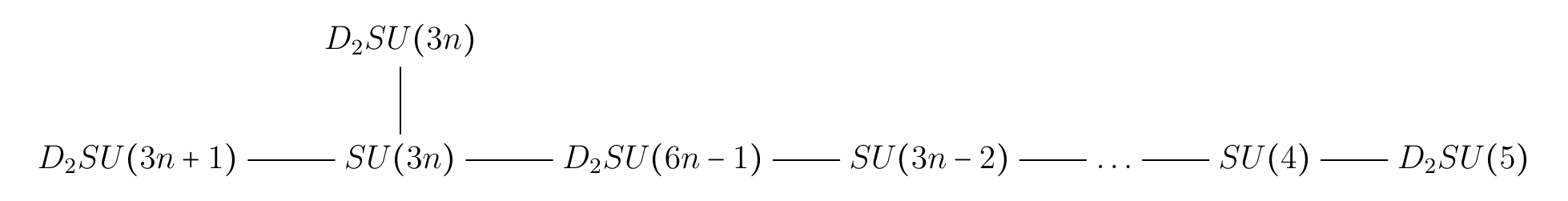}
\end{figure}

\subsection*{ICIS (279)}	
$\left\{ \begin{array}{l} x_1x_2+x_3x_4=0\\x_1x_5+x_3^{2}+x_2x_4^{3}+x_5^{3}+x_2^{8}=0\end{array}\right.$
\\$(w_1,w_2,w_3,w_4,w_5;1,d)=(\frac{16}{19},\frac{3}{19},\frac{12}{19},\frac{7}{19},\frac{8}{19};1,\frac{24}{19})$	
\\$\mu=58, \quad \mu_1=17, \quad \alpha={19\over3} ,\quad r=27, \quad f=4, \quad a=\frac{175}{4}, \quad c=\frac{529}{12}$	
 \begin{figure}[H]
 \centering
\includegraphics[scale=.9,keepaspectratio]{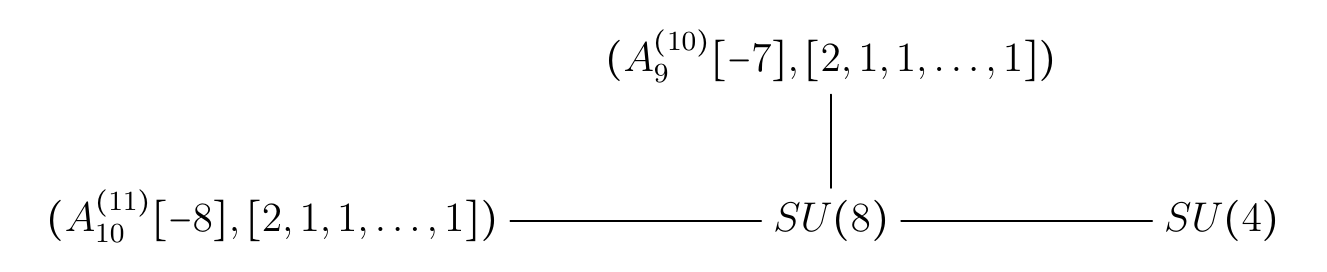}
\end{figure}
The theory $(A_{10}^{(11)}[-8],[2,1^9])$ is defined by six dimensional $A_{10}$ $(2,0)$ theory 
on a sphere with irregular puncture $ A_{10}^{(11)}[-8]$ and a  regular puncture labeled by Young Tableaux $[2, 1^9]$. This 
theory has flavor symmetry $SU(9)\times U(1)$, and the flavor central charge of $SU(9)$ is ${19\over3}$. The Coulomb branch spectrum is $({19\over3},{16\over3},{13\over3},{11\over3},{10\over3},{8\over3},{7\over3},{5\over3},{4\over3})$, and 
the central charge is $a={29\over2}, c={63\over4}$.
The theory $(A_9^{(10)}[-7],[2,1^8])$ is defined 
by six dimensional $A_{10}$ $(2,0)$ theory 
on a sphere with irregular puncture $ A_9^{(10)}[-7]$ and a regular puncture  $[2, 1^8]$, and the flavor symmetry is $SU(8)\times U(1)$.
The flavor central charge of $SU(8)$ is ${17\over3}$. The Coulomb branch spectrum is $({17\over3},{14\over3},{11\over3},{10\over3},{8\over3},{7\over3},{5\over3},{4\over3})$, and the central charge is $a=\frac{35}{3},c=\frac{38}{3}$.
Using the above data, one can check that the 
gauging for gauge group $SU(8)$ is conformal.

\subsection*{ICIS (280)}	
$\left\{ \begin{array}{l} x_1x_2+x_3x_4=0\\x_1x_5+x_3^{2}+x_2x_4^{3}+x_5^{4}+x_2^{16}=0\end{array}\right.$
\\$(w_1,w_2,w_3,w_4,w_5;1,d)=(\frac{12}{13},\frac{1}{13},\frac{8}{13},\frac{5}{13},\frac{4}{13};1,\frac{16}{13})$
\\$\mu=131, \quad \mu_1=33, \quad \alpha=13 ,\quad r=61, \quad f=9, \quad a=\frac{4489}{24}, \quad c=\frac{1127}{6}$	
 \begin{figure}[H]
 \centering
\includegraphics[scale=.9,keepaspectratio]{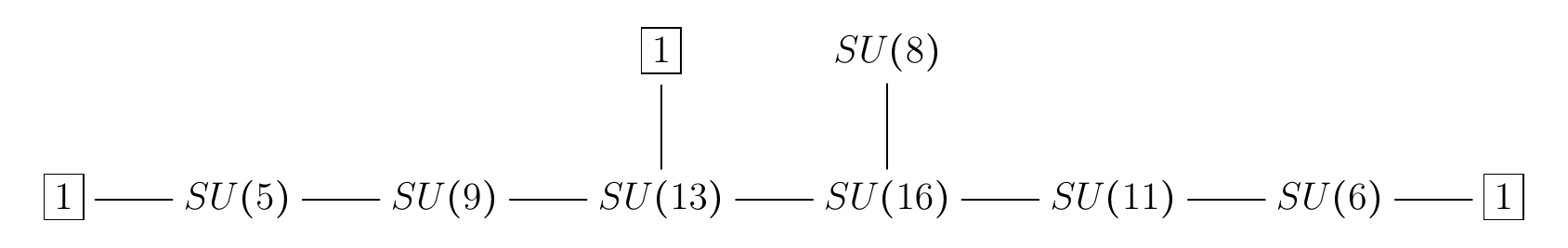}
\end{figure}

\subsection*{ICIS (281)}	
$\left\{ \begin{array}{l} x_1x_2+x_3x_4=0\\x_1x_5+x_3^{2}+x_2x_4^{3}+x_5^{5}+x_2^{40}=0\end{array}\right.$
\\$(w_1,w_2,w_3,w_4,w_5;1,d)=(\frac{32}{33},\frac{1}{33},\frac{20}{33},\frac{13}{33},\frac{8}{33};1,\frac{40}{33})$
\\$\mu=364, \quad \mu_1=81, \quad \alpha=33 ,\quad r=177, \quad f=10, \quad a=\frac{5007}{4}, \quad c=\frac{5013}{4}$	
 \begin{figure}[H]
 \centering
\includegraphics[scale=.9,keepaspectratio]{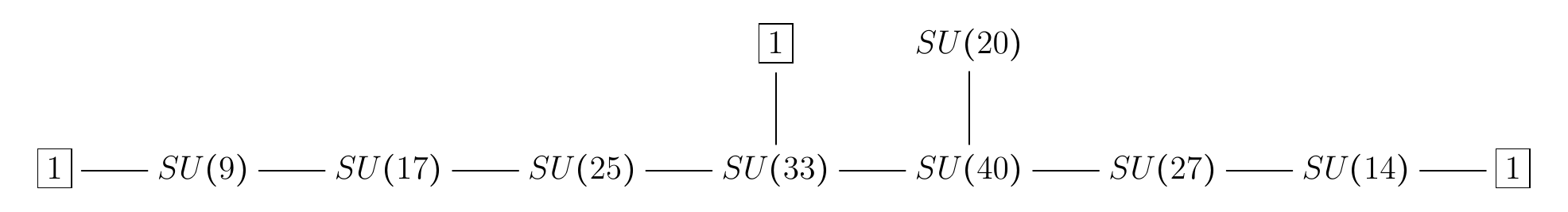}
\end{figure}	
	
\subsection*{ICIS (282)}		
$\left\{ \begin{array}{l} x_1x_2+x_3x_4=0\\x_1x_5+x_3^{2}+x_2x_4^{4}+x_5^{3}+x_2^{15}=0\end{array}\right.$
\\$(w_1,w_2,w_3,w_4,w_5;1,d)=(\frac{10}{11},\frac{1}{11},\frac{15}{22},\frac{7}{22},\frac{5}{11};1,\frac{15}{11})$		
\\$\mu=122, \quad \mu_1=46, \quad \alpha=11 ,\quad r=58, \quad f=6, \quad a=\frac{1957}{12}, \quad c=\frac{491}{3}$	
 \begin{figure}[H]
 \centering
\includegraphics[scale=.9,keepaspectratio]{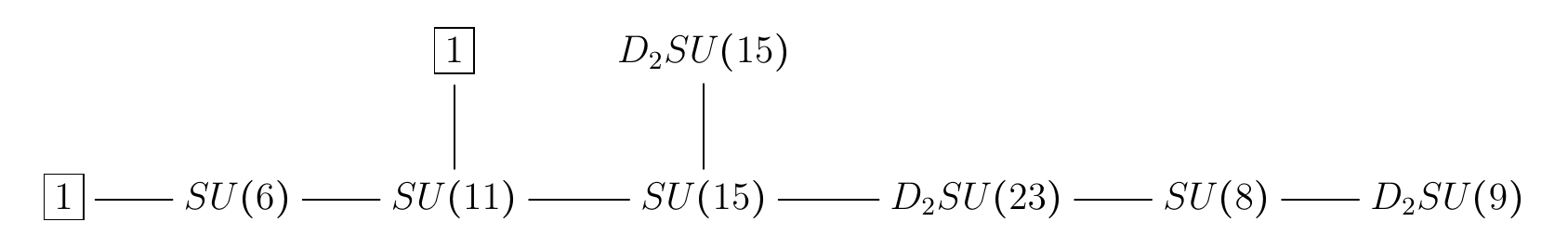}
\end{figure}

\subsection*{ICIS (283)}		
$\left\{ \begin{array}{l} x_1x_2+x_3x_4=0\\x_1x_5+x_3^{2}+x_2x_4^{5}+x_5^{3}+x_2^{36}=0\end{array}\right.$
\\$(w_1,w_2,w_3,w_4,w_5;1,d)=(\frac{24}{25},\frac{1}{25},\frac{18}{25},\frac{7}{25},\frac{12}{25};1,\frac{36}{25})$
\\$\mu=326, \quad \mu_1=145, \quad \alpha=25 ,\quad r=158, \quad f=10, \quad a=\frac{24151}{24}, \quad c=\frac{12091}{12}$	
 \begin{figure}[H]
 \centering
\includegraphics[scale=.85,keepaspectratio]{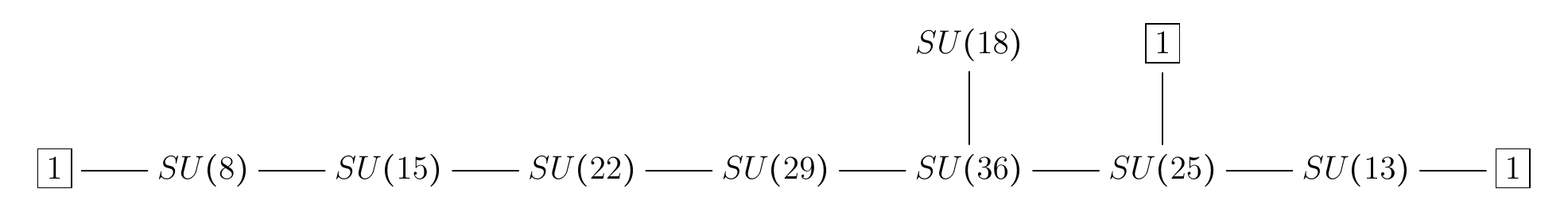}
\end{figure}

\subsection*{ICIS (284)}		
$\left\{ \begin{array}{l} x_1x_2+x_3x_4+x_5^{2}=0\\x_1x_5+x_2x_3^{2}+x_4^{3}+x_2^{7}=0\end{array}\right.$
\\$(w_1,w_2,w_3,w_4,w_5;1,d)=(\frac{13}{16},\frac{3}{16},\frac{9}{16},\frac{7}{16},\frac{1}{2};1,\frac{21}{16})$	
\\$\mu=47, \quad \mu_1=16, \quad \alpha={16\over3} ,\quad r=22, \quad f=3, \quad a=\frac{377}{12}, \quad c=\frac{95}{3}$	
 \begin{figure}[H]
 \centering
\includegraphics[scale=.9,keepaspectratio]{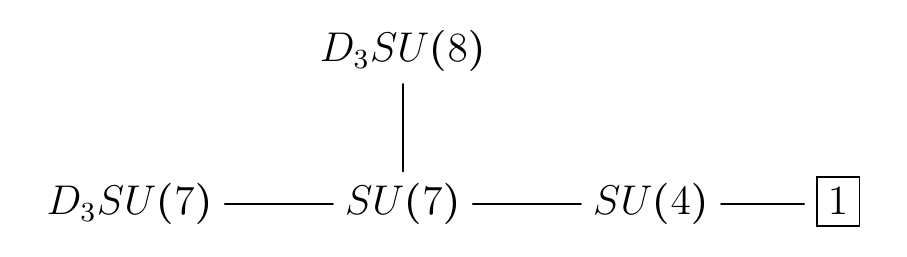}
\end{figure}

\subsection*{ICIS (285)}		
$\left\{ \begin{array}{l} x_1x_2+x_3x_4+x_5^{2}=0\\x_1x_5+x_2x_3^{2}+x_4^{4}+x_2^{14}=0\end{array}\right.$
\\$(w_1,w_2,w_3,w_4,w_5;1,d)=(\frac{9}{10},\frac{1}{10},\frac{13}{20},\frac{7}{20},\frac{1}{2};1,\frac{7}{5})$	
\\$\mu=109, \quad \mu_1=45, \quad \alpha=10 ,\quad r=52, \quad f=5, \quad a=\frac{273}{2}, \quad c=137$	
 \begin{figure}[H]
 \centering
\includegraphics[scale=.9,keepaspectratio]{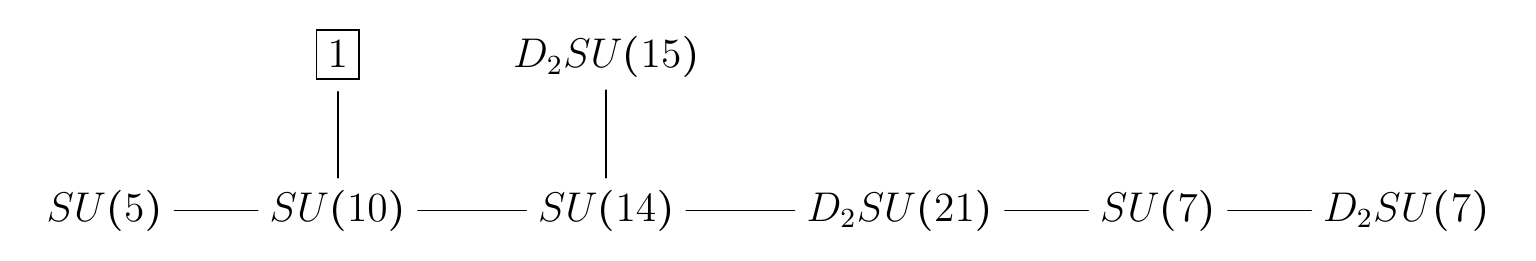}
\end{figure}

\subsection*{ICIS (286)}	
$\left\{ \begin{array}{l} x_1x_2+x_3x_4+x_5^{2}=0\\x_1x_5+x_2x_3^{2}+x_4^{5}+x_2^{35}=0\end{array}\right.$
\\$(w_1,w_2,w_3,w_4,w_5;1,d)=(\frac{23}{24},\frac{1}{24},\frac{17}{24},\frac{7}{24},\frac{1}{2};1,\frac{35}{24})$	
\\$\mu=311, \quad \mu_1=144, \quad \alpha=24 ,\quad r=151, \quad f=9, \quad a=\frac{22415}{24}, \quad c=\frac{5611}{6}$	
 \begin{figure}[H]
 \centering
\includegraphics[scale=.9,keepaspectratio]{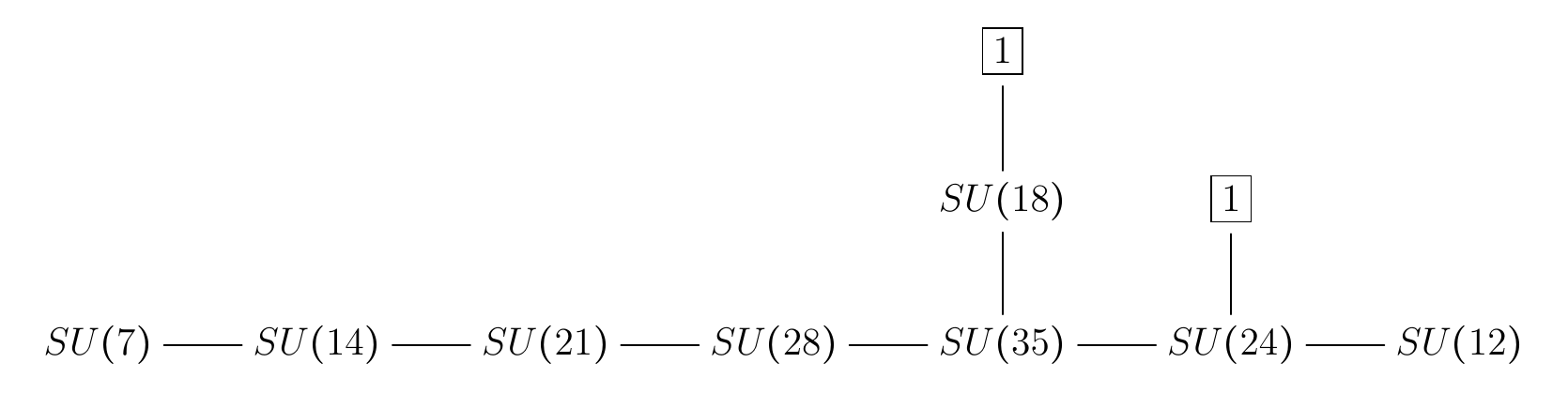}
\end{figure}	
	
\subsection*{ICIS (287)}	
$\left\{ \begin{array}{l} x_1x_2+x_3x_4+x_5^{3}=0\\x_1x_5+x_2x_3^{2}+x_4^{3}+x_2^{15}=0\end{array}\right.$
\\$(w_1,w_2,w_3,w_4,w_5;1,d)=(\frac{11}{12},\frac{1}{12},\frac{7}{12},\frac{5}{12},\frac{1}{3};1,\frac{5}{4})$	
\\$\mu=118, \quad \mu_1=32, \quad \alpha=12 ,\quad r=55, \quad f=8, \quad a=\frac{3803}{24}, \quad c=\frac{955}{6}$	
 \begin{figure}[H]
 \centering
\includegraphics[scale=.9,keepaspectratio]{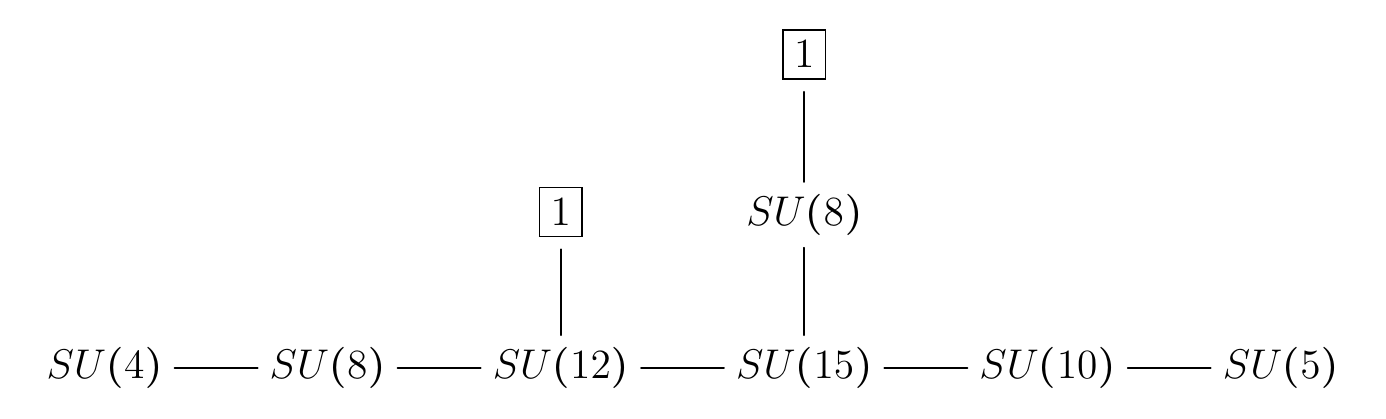}
\end{figure}

\subsection*{ICIS (288)}		
$\left\{ \begin{array}{l} x_1x_2+x_3x_4+x_5^{4}=0\\x_1x_5+x_2x_3^{2}+x_4^{3}+x_2^{39}=0\end{array}\right.$
\\$(w_1,w_2,w_3,w_4,w_5;1,d)=(\frac{31}{32},\frac{1}{32},\frac{19}{32},\frac{13}{32},\frac{1}{4};1,\frac{39}{32})$	
\\$\mu=349, \quad \mu_1=80, \quad \alpha=32 ,\quad r=170, \quad f=9, \quad a=\frac{14051}{12}, \quad c=\frac{3517}{3}$	
 \begin{figure}[H]
 \centering
\includegraphics[scale=.9,keepaspectratio]{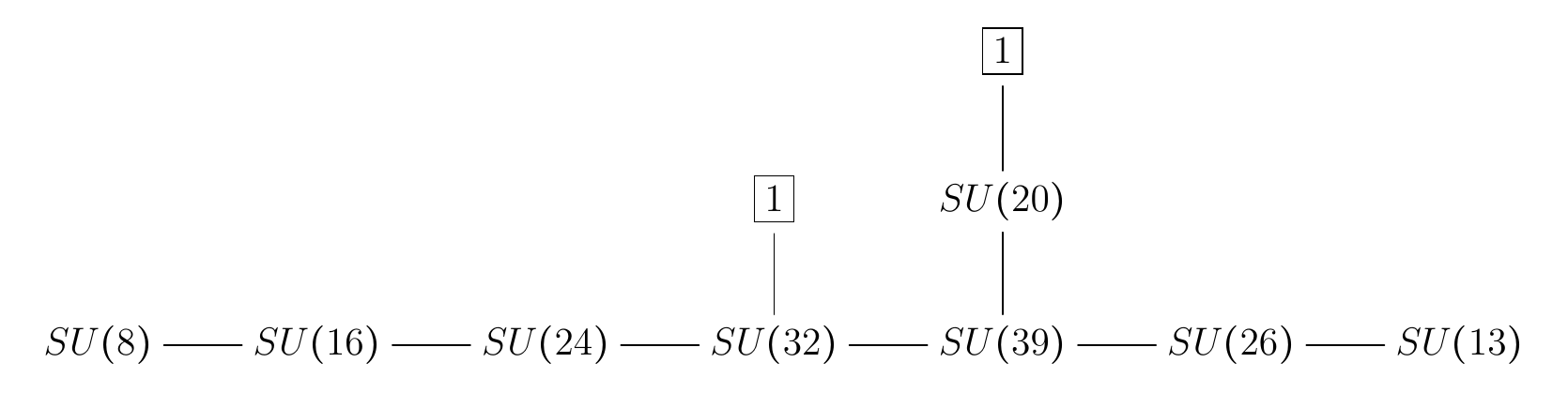}
\end{figure}	
	
\subsection*{ICIS (289)}		
$\left\{ \begin{array}{l} x_1x_2+x_3x_4=0\\x_1x_5+x_2x_3^{2}+x_4^{3}+x_5^{3}+x_2^{9}=0\end{array}\right.$
\\$(w_1,w_2,w_3,w_4,w_5;1,d)=(\frac{6}{7},\frac{1}{7},\frac{4}{7},\frac{3}{7},\frac{3}{7};1,\frac{9}{7})$	
\\$\mu=64, \quad \mu_1=20, \quad \alpha=7 ,\quad r=28, \quad f=8, \quad a=\frac{637}{12}, \quad c=\frac{161}{3}$	
 \begin{figure}[H]
 \centering
\includegraphics[scale=.9,keepaspectratio]{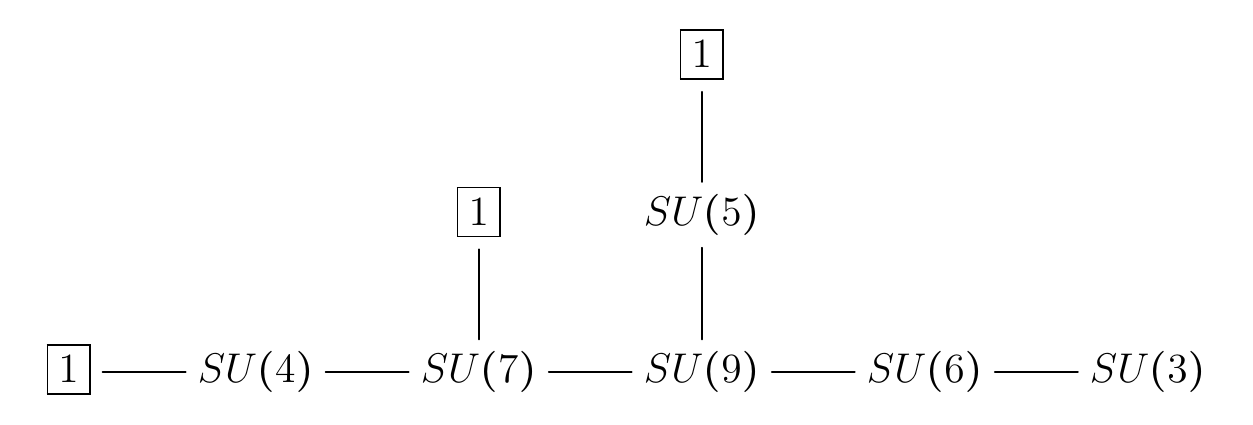}
\end{figure}

\subsection*{ICIS (290)}	
$\left\{ \begin{array}{l} x_1x_2+x_3x_4=0\\x_1x_5+x_2x_3^{2}+x_4^{4}+x_5^{3}+x_2^{18}=0\end{array}\right.$
\\$(w_1,w_2,w_3,w_4,w_5;1,d)=(\frac{12}{13},\frac{1}{13},\frac{17}{26},\frac{9}{26},\frac{6}{13};1,\frac{18}{13})$
\\$\mu=144, \quad \mu_1=57, \quad \alpha=13 ,\quad r=69, \quad f=6, \quad a=\frac{457}{2}, \quad c=\frac{917}{4}$	
 \begin{figure}[H]
 \centering
\includegraphics[scale=.9,keepaspectratio]{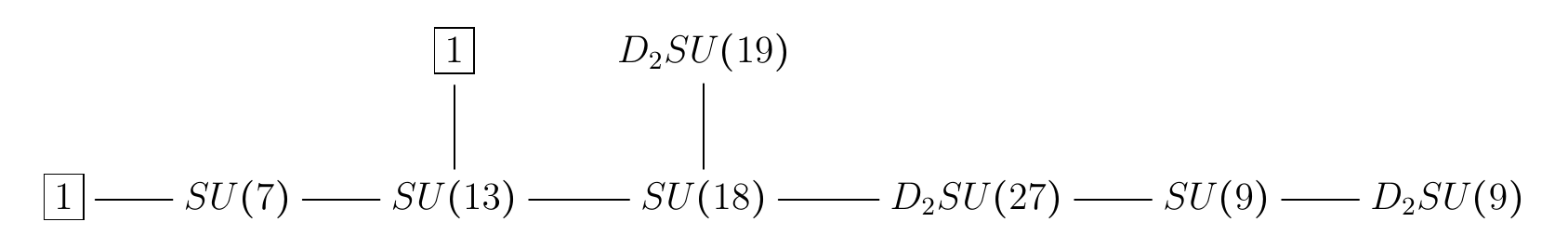}
\end{figure}

\subsection*{ICIS (291)}		
$\left\{ \begin{array}{l} x_1x_2+x_3x_4=0\\x_1x_5+x_2x_3^{2}+x_4^{5}+x_5^{3}+x_2^{45}=0\end{array}\right.$
\\$(w_1,w_2,w_3,w_4,w_5;1,d)=(\frac{30}{31},\frac{1}{31},\frac{22}{31},\frac{9}{31},\frac{15}{31};1,\frac{45}{31})$		
\\$\mu=404, \quad \mu_1=184, \quad \alpha=31 ,\quad r=197, \quad f=10, \quad a=\frac{37201}{24}, \quad c=\frac{9311}{6}$	
 \begin{figure}[H]
 \centering
\includegraphics[scale=.9,keepaspectratio]{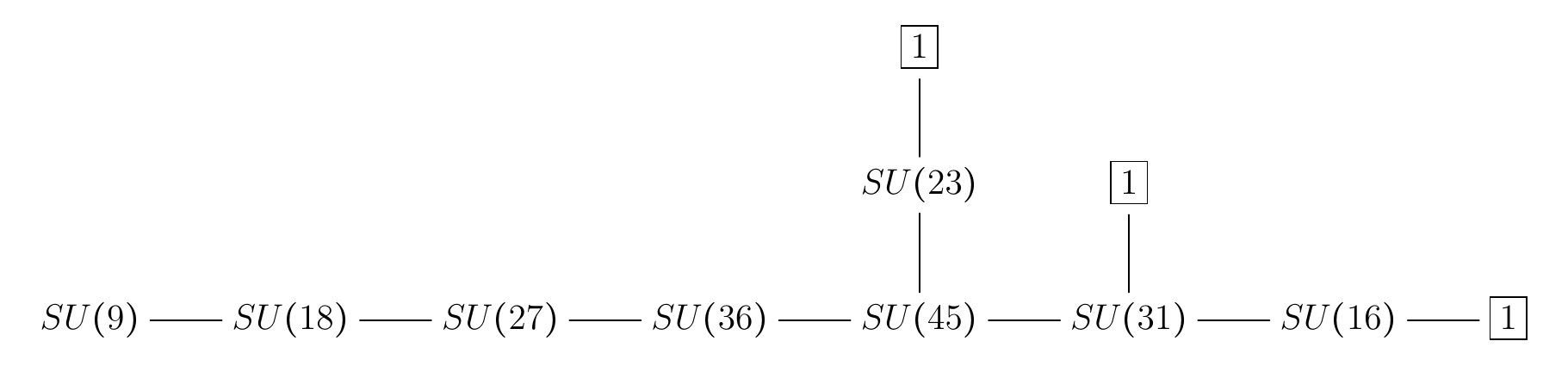}
\end{figure}

\subsection*{ICIS (292)}	
$\left\{ \begin{array}{l} x_1x_2+x_3x_4=0\\x_1x_5+x_2x_3^{2}+x_4^{3}+x_5^{4}+x_2^{18}=0\end{array}\right.$
\\$(w_1,w_2,w_3,w_4,w_5;1,d)=(\frac{27}{29},\frac{2}{29},\frac{17}{29},\frac{12}{29},\frac{9}{29};1,\frac{36}{29})$
\\$\mu=146, \quad \mu_1=38, \quad \alpha={29\over2} ,\quad r=70, \quad f=6, \quad a=\frac{933}{4}, \quad c=234$	
 \begin{figure}[H]
 \centering
\includegraphics[scale=.9,keepaspectratio]{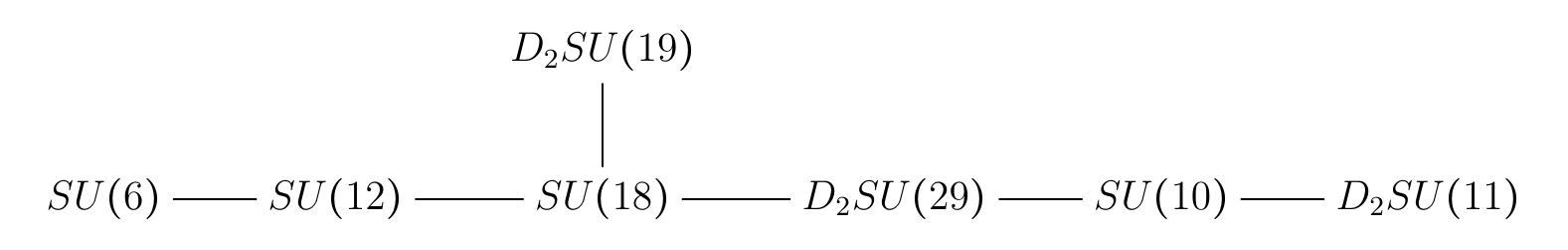}
\end{figure}

\subsection*{ICIS (293)}
$\left\{ \begin{array}{l} x_1x_2+x_3x_4=0\\x_1x_5+x_2x_3^{2}+x_4^{3}+x_5^{5}+x_2^{45}=0\end{array}\right.$
\\$(w_1,w_2,w_3,w_4,w_5;1,d)=(\frac{36}{37},\frac{1}{37},\frac{22}{37},\frac{15}{37},\frac{9}{37};1,\frac{45}{37})$
\\$\mu=408, \quad \mu_1=92, \quad \alpha=37 ,\quad r=199, \quad f=10, \quad a=\frac{37753}{24}, \quad c=\frac{9449}{6}$	
 \begin{figure}[H]
 \centering
\includegraphics[scale=.9,keepaspectratio]{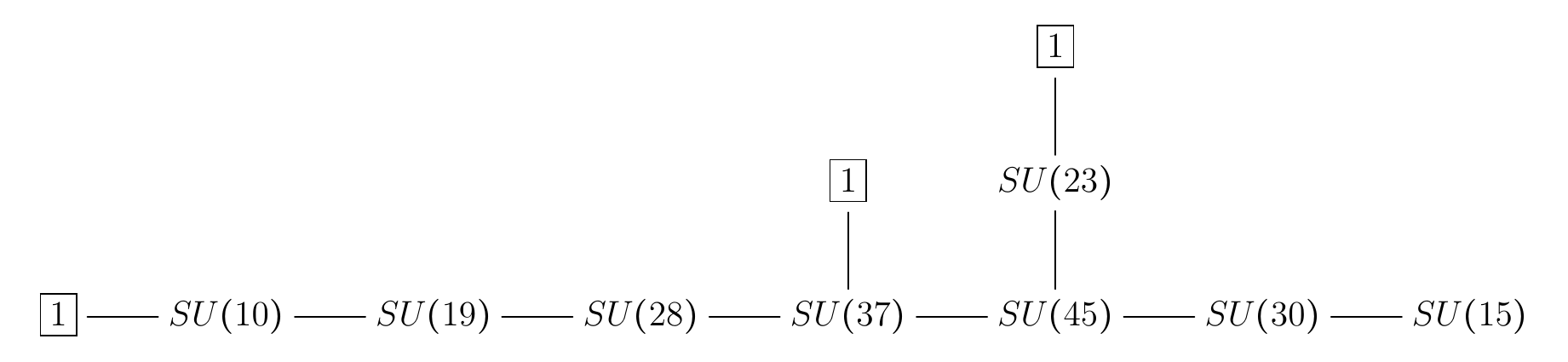}
\end{figure}

\subsection*{ICIS (294)}	
$\left\{ \begin{array}{l} x_1x_2+x_3x_4+x_5^{2}=0\\x_1x_5+x_2x_3^{2}+x_2x_4^{3}+x_2^{9}=0\end{array}\right.$
\\$(w_1,w_2,w_3,w_4,w_5;1,d)=(\frac{17}{20},\frac{3}{20},\frac{3}{5},\frac{2}{5},\frac{1}{2};1,\frac{27}{20})$
\\$\mu=64, \quad \mu_1=\frac{95}{4}, \quad \alpha={20\over3} ,\quad r=30, \quad f=4, \quad a=\frac{427}{8}, \quad c=\frac{215}{4}$	
 \begin{figure}[H]
 \centering
\includegraphics[scale=.9,keepaspectratio]{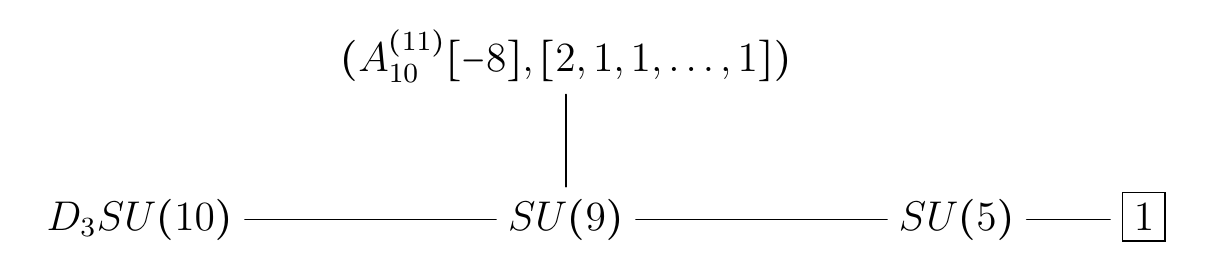}
\end{figure}	
Let's explain some details about  AD matter $(A_{10}^{(11)}[-8],[2,1^9])$, which is defined by six dimensional $A_{10}$ $(2,0)$ theory 
on a sphere with irregular puncture $ A_{10}^{(11)}[-8]$ and the regular puncture is labeled by Young Tableaux $[2, 1^9]$. This 
theory has flavor symmetry $SU(9)\times U(1)$, and the flavor central charge of $SU(9)$ is ${19\over3}$. The Coulomb branch spectrum 
of this theory is $({19\over3},{16\over3},{13\over3},{11\over3},{10\over3},{8\over3},{7\over3},{5\over3},{4\over3})$, and the central charges are
$a={29\over2},c=\frac{63}{4}$. The flavor central 
charge of  $SU(10)$ flavor group of $D_3 SU(10)$ theory is ${20\over3}$. The Coulomb branch spectrum is 
$({20\over3},{17\over3},{14\over3},{11\over3},{10\over3},{8\over3},{7\over3},{5\over3},{4\over3})$, and the central charges are $a=\frac{121}{8},c=\frac{33}{2}$. Using the above data, one can check that the 
gauging for gauge group $SU(9)$ is conformal.

\subsection*{ICIS (295)}	
$\left\{ \begin{array}{l} x_1x_2+x_3x_4+x_5^{3}=0\\x_1x_5+x_2x_3^{2}+x_2x_4^{3}+x_2^{19}=0\end{array}\right.$
\\$(w_1,w_2,w_3,w_4,w_5;1,d)=(\frac{14}{15},\frac{1}{15},\frac{3}{5},\frac{2}{5},\frac{1}{3};1,\frac{19}{15})$
\\$\mu=153, \quad \mu_1=\frac{130}{3}, \quad \alpha=15 ,\quad r=72, \quad f=9, \quad a=\frac{6155}{24}, \quad c=\frac{3089}{12}$	
 \begin{figure}[H]
 \centering
\includegraphics[scale=.9,keepaspectratio]{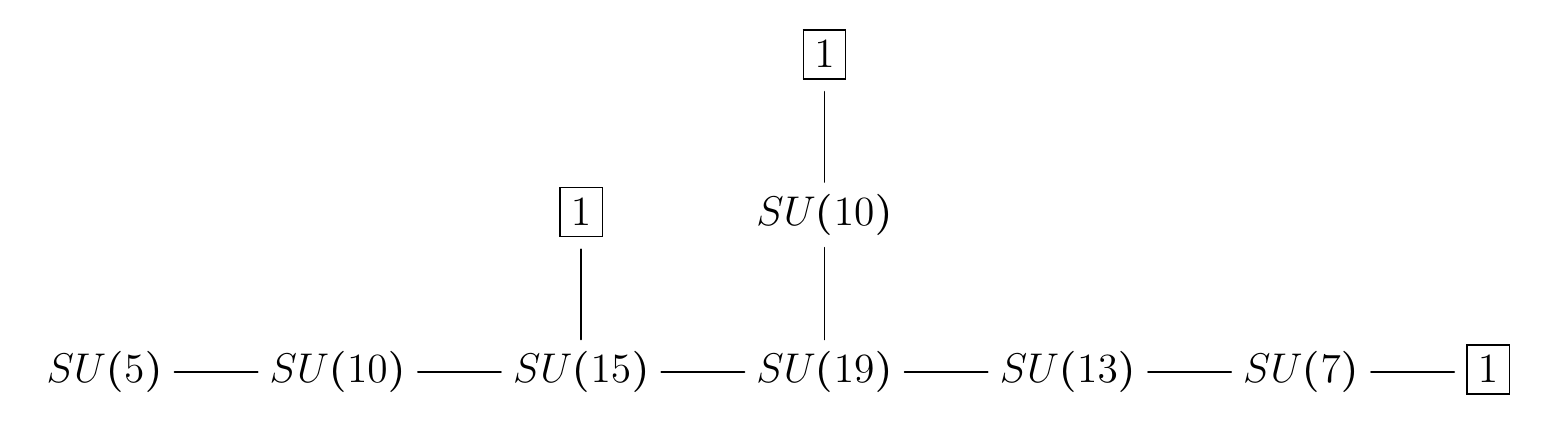}
\end{figure}

\subsection*{ICIS (296)}	
(296)$\left\{ \begin{array}{l} x_1x_2+x_3x_4+x_5^{4}=0\\x_1x_5+x_2x_3^{2}+x_2x_4^{3}+x_2^{49}=0\end{array}\right.$
\\$(w_1,w_2,w_3,w_4,w_5;1,d)=(\frac{39}{40},\frac{1}{40},\frac{3}{5},\frac{2}{5},\frac{1}{4};1,\frac{49}{40})$
\\$\mu=442, \quad \mu_1=\frac{825}{8}, \quad \alpha=40 ,\quad r=216, \quad f=10, \quad a=\frac{44425}{24}, \quad c=\frac{22237}{12}$	
 \begin{figure}[H]
 \centering
\includegraphics[scale=.9,keepaspectratio]{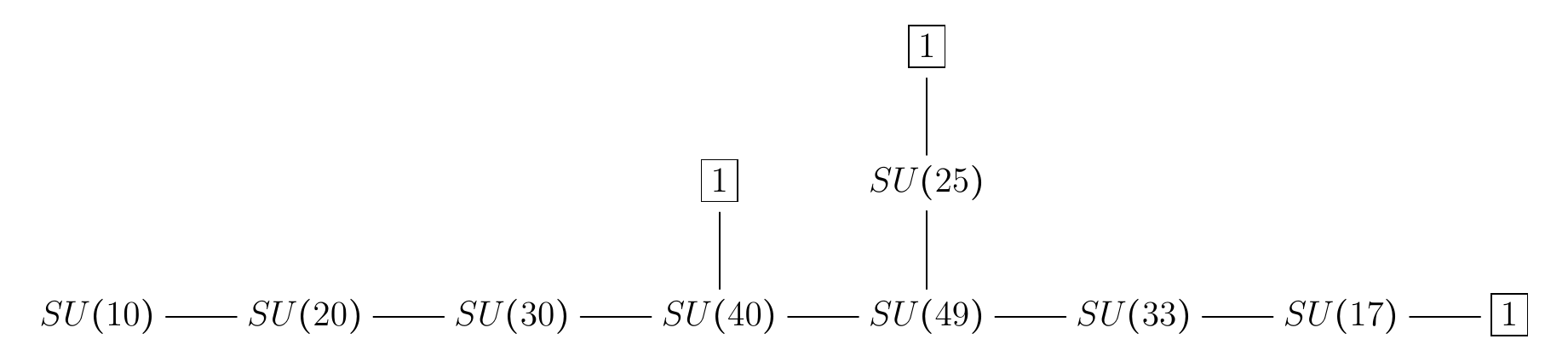}
\end{figure}

\subsection*{ICIS (297)}
$\left\{ \begin{array}{l} x_1x_2+x_3x_4+x_5^{2}=0\\x_1x_5+x_2x_3^{2}+x_2x_4^{4}+x_2^{17}=0\end{array}\right.$
\\$(w_1,w_2,w_3,w_4,w_5;1,d)=(\frac{11}{12},\frac{1}{12},\frac{2}{3},\frac{1}{3},\frac{1}{2};1,\frac{17}{12})$
\\$\mu=137, \quad \mu_1=\frac{117}{2}, \quad \alpha=12 ,\quad r=64, \quad f=9, \quad a=\frac{616}{3}, \quad c=\frac{1237}{6}$	
 \begin{figure}[H]
 \centering
\includegraphics[scale=.9,keepaspectratio]{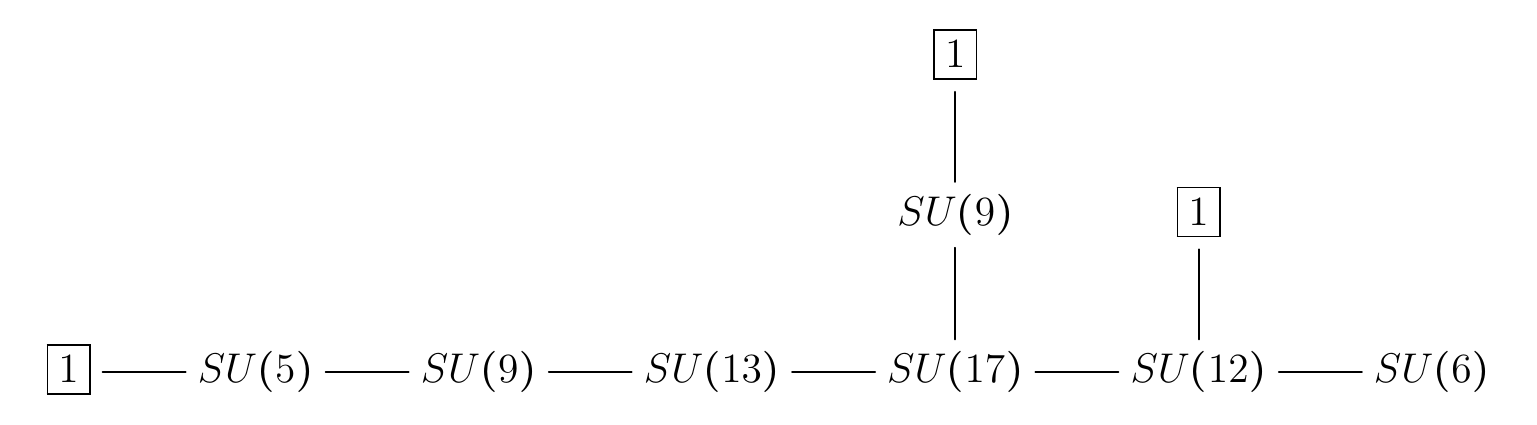}
\end{figure}

\subsection*{ICIS (298)}
$\left\{ \begin{array}{l} x_1x_2+x_3x_4+x_5^{2}=0\\x_1x_5+x_2x_3^{2}+x_2x_4^{5}+x_2^{41}=0\end{array}\right.$
\\$(w_1,w_2,w_3,w_4,w_5;1,d)=(\frac{27}{28},\frac{1}{28},\frac{5}{7},\frac{2}{7},\frac{1}{2};1,\frac{41}{28})$
\\$\mu=81, \quad \mu_1={693\over 4}, \quad \alpha=28 ,\quad r=180, \quad f=10, \quad a=\frac{31105}{24}, \quad c=\frac{15571}{12}$	
 \begin{figure}[H]
 \centering
\includegraphics[scale=.9,keepaspectratio]{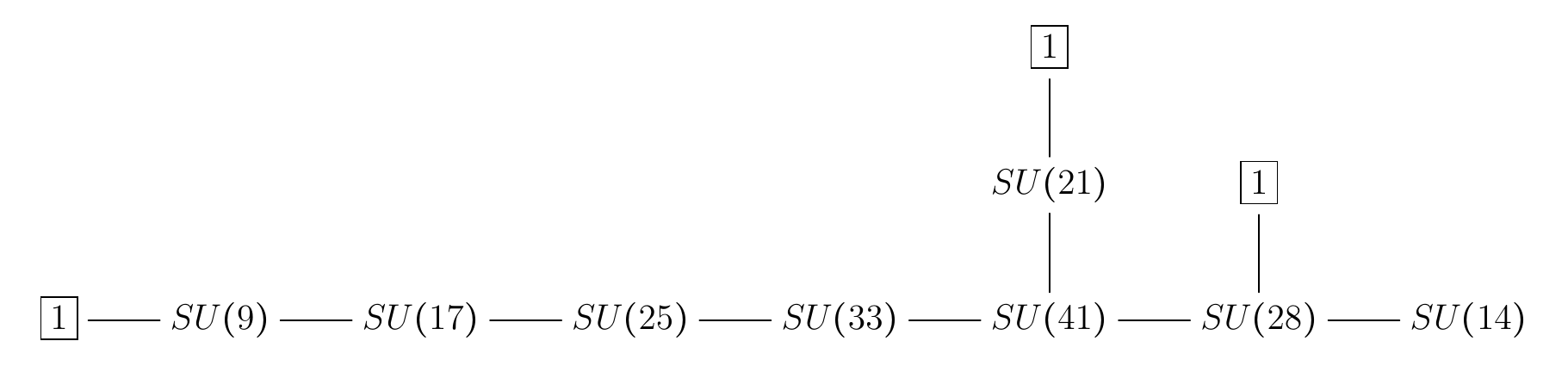}
\end{figure}

\subsection*{ICIS (299)}
$\left\{ \begin{array}{l} x_1x_2+x_3x_4=0\\x_1x_5+x_2x_3^{2}+x_2x_4^{3}+x_5^{3}+x_2^{11}=0\end{array}\right.$
\\$(w_1,w_2,w_3,w_4,w_5;1,d)=(\frac{22}{25},\frac{3}{25},\frac{3}{5},\frac{2}{5},\frac{11}{25};1,\frac{33}{25})$
\\$\mu=81, \quad \mu_1=\frac{138}{5}, \quad \alpha=\frac{25}{3} ,\quad r=38, \quad f=5, \quad a=\frac{1949}{24}, \quad c=\frac{327}{4}$	
 \begin{figure}[H]
 \centering
\includegraphics[scale=.9,keepaspectratio]{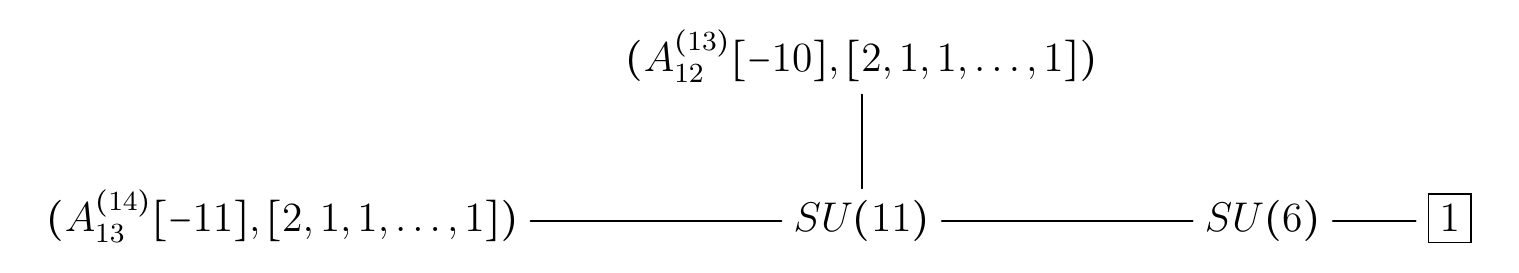}
\end{figure}	
In above case, the AD matters are engineered using six dimensional $A_{12}$ and $A_{13}$ 
theory with the irregular puncture type $A_{12}^{(13)}[-10]$ and $A_{13}^{(14)}[-11]$, and the regular puncture has the the Young Tableaux type $[2,1^{11}]$ and $[2, 1^{12}]$ respectively. 
The theory $(A_{12}^{(13)}[-10],[2,1^{11}])$ has flavor symmetry $SU(11)\times U(1)$, and the flavor central charge for $SU(12)$ flavor group is ${23\over 3}$.
The Coulomb branch spectrum is $({23\over3},{20\over3},{17\over3},{14\over3},{13\over3},{11\over 3},{10\over3},{8\over3},{7\over3},{5\over3},{4\over3})$, and the central charge is $a=\frac{253}{12},~ c=\frac{275}{12}$.
The theory $(A_{13}^{(14)}[-11],[2,1^{12}])$ has flavor symmetry $SU(12)\times U(1)$, and the flavor central charge for $SU(12)$ flavor group is ${25\over 3}$. The Coulomb branch spectrum is $({25\over3}, {22\over3},{19\over3},{16\over3},{14\over3},{13\over3},{11\over3},{10\over3},{8\over3},{7\over3},{5\over3},{4\over3})$, and the central charge is $a=\frac{149}{6}, ~c=27$.
Using the above data, one can check that the gauging is conformal for $SU(11)$ gauge group.

\subsection*{ICIS (300)}
$\left\{ \begin{array}{l} x_1x_2+x_3x_4=0\\x_1x_5+x_2x_3^{2}+x_2x_4^{3}+x_5^{4}+x_2^{22}=0\end{array}\right.$
\\$(w_1,w_2,w_3,w_4,w_5;1,d)=(\frac{33}{35},\frac{2}{35},\frac{3}{5},\frac{2}{5},\frac{11}{35};1,\frac{44}{35})$
\\$\mu=181, \quad \mu_1=\frac{345}{7},\quad \alpha={35\over2}, \quad r=87, \quad f=7, \quad a=\frac{8383}{24}, \quad c=\frac{1051}{3}$	
 \begin{figure}[H]
 \centering
\includegraphics[scale=.9,keepaspectratio]{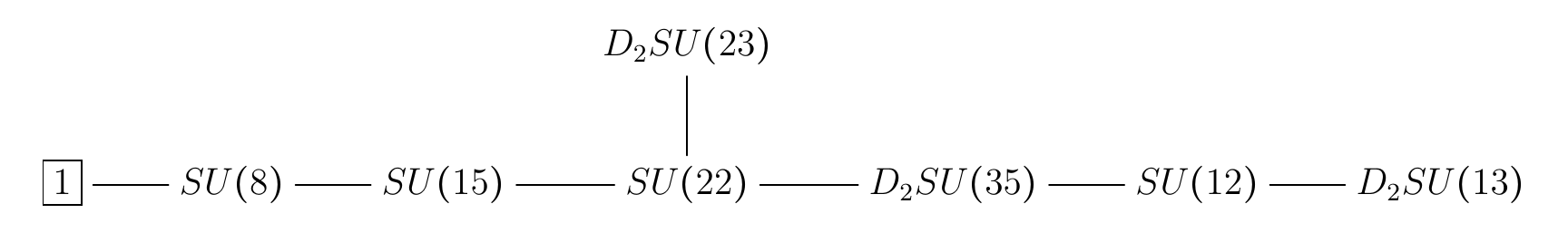}
\end{figure}

\subsection*{ICIS (301)}
$\left\{ \begin{array}{l} x_1x_2+x_3x_4=0\\x_1x_5+x_2x_3^{2}+x_2x_4^{3}+x_5^{5}+x_2^{55}=0\end{array}\right.$
\\$(w_1,w_2,w_3,w_4,w_5;1,d)=(\frac{44}{45},\frac{1}{45},\frac{3}{5},\frac{2}{5},\frac{11}{45};1,\frac{11}{9})$
\\$\mu=501,\quad \mu_1=\frac{1036}{9}, \quad \alpha=45, \quad r=245, \quad f=11, \quad a=\frac{9395}{4}, \quad c=\frac{9405}{4}$	
 \begin{figure}[H]
 \centering
\includegraphics[scale=.85,keepaspectratio]{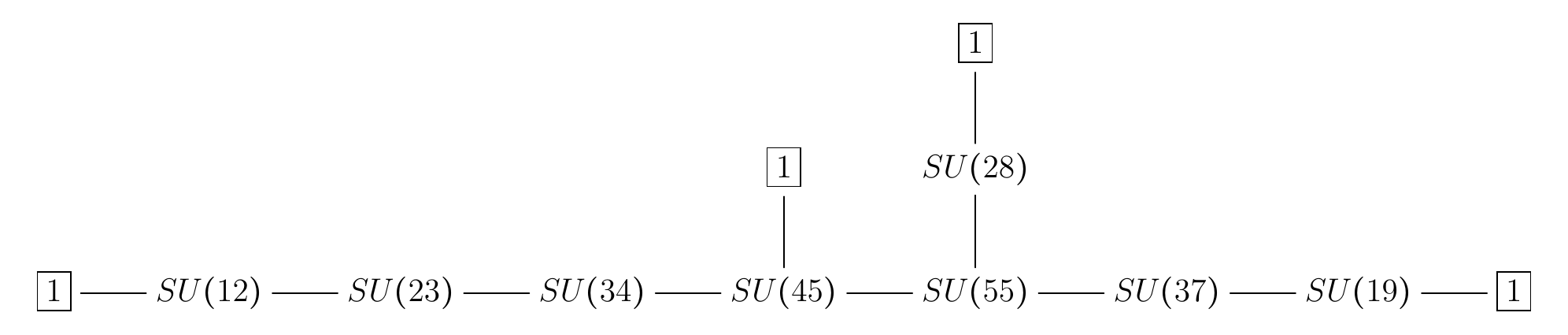}
\end{figure}

\subsection*{ICIS (302)}

$\left\{ \begin{array}{l} x_1x_2+x_3x_4=0\\x_1x_5+x_2x_3^{2}+x_2x_4^{4}+x_5^{3}+x_2^{21}=0\end{array}\right.$
\\$(w_1,w_2,w_3,w_4,w_5;1,d)=(\frac{14}{15},\frac{1}{15},\frac{2}{3},\frac{1}{3},\frac{7}{15};1,\frac{7}{5})$
\\$\mu=172, \quad \mu_1={352\over5}, \quad \alpha=15,  \quad r=81, \quad f=10, \quad a=\frac{2523}{8}, \quad c=\frac{633}{2}$	
 \begin{figure}[H]
 \centering
\includegraphics[scale=.9,keepaspectratio]{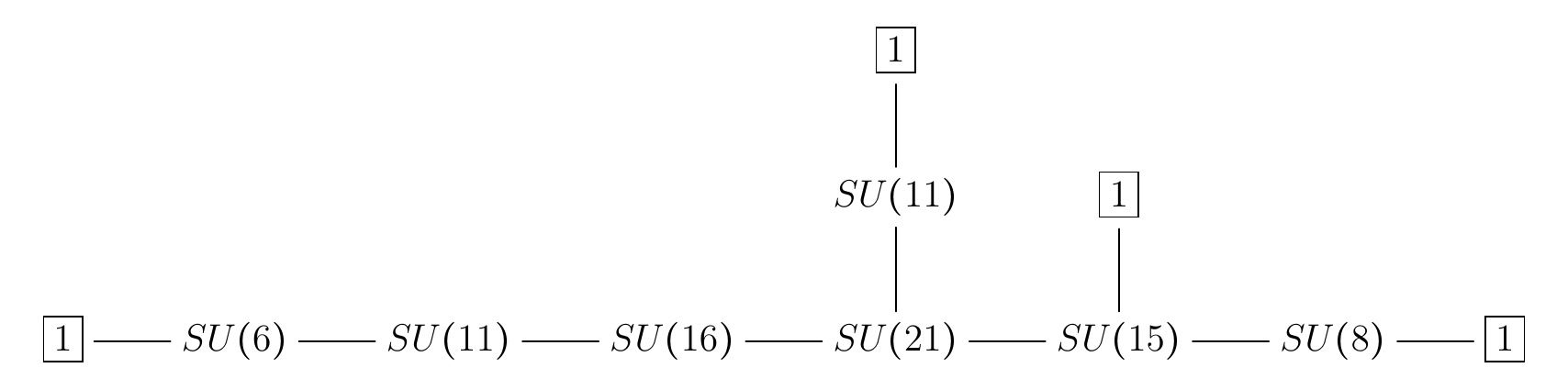}
\end{figure}

\subsection*{ICIS (303)}
%\ie
%&x_1x_2+x_3x_4=0\\
%&x_1x_5+x_2x_3^{2}+x_2x_4^{5}+x_5^{3}+x_2^{51}=0
%\fe
$\left\{ \begin{array}{l} x_1x_2+x_3x_4=0\\x_1x_5+x_2x_3^{2}+x_2x_4^{5}+x_5^{3}+x_2^{51}=0\end{array}\right.$
\\$(w_1,w_2,w_3,w_4,w_5;1,d)=(\frac{34}{35},\frac{1}{35},\frac{5}{7},\frac{2}{7},\frac{17}{35};1,\frac{51}{35})$
\\$\mu=463, \quad \mu_1=\frac{1066}{5}, \quad \alpha=35, \quad r=226,\quad f=11,\quad a=\frac{48185}{24},\quad c=\frac{24119}{12}$
 \begin{figure}[H]
 \centering
\includegraphics[scale=.85,keepaspectratio]{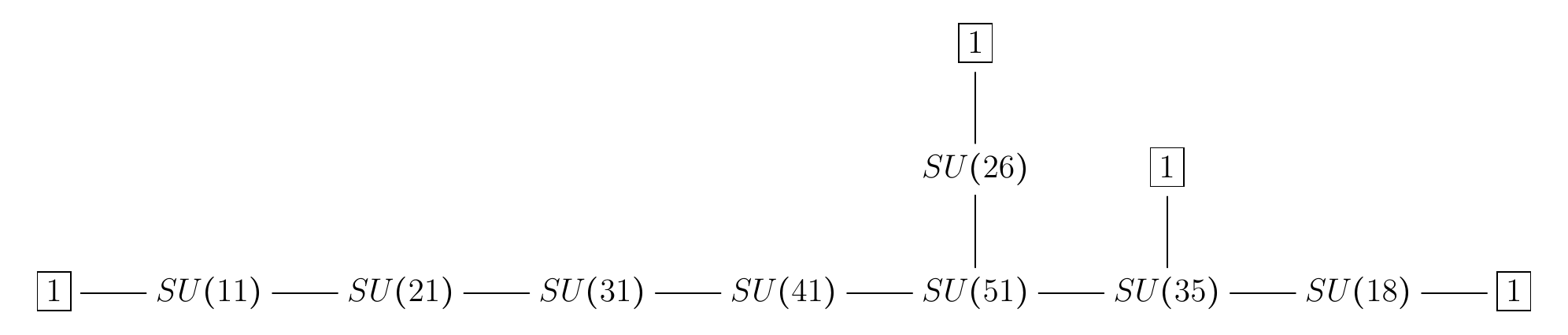}
\end{figure}

 \section{Discussion}\label{dis}	
 In this paper, we studied physical properties of four dimensional $\mathcal{N}=2$ SCFT defined by three-fold  isolated complete intersection
 singularities with a $\mC^*$ action which is classified in \cite{Chen:2016bzh}. Let's summarize the major findings:
 \begin{itemize}
 \item The Seiberg-Witten solution is identified with the mini-versal deformation of the singularity, see formula \eqref{mf}, and  the Seiberg-Witten differential 
 is given by formula \eqref{h3}. 
 \item The Coulomb branch spectrum (the scaling dimensions of Coulomb branch parameters) can be found from the basis of Jacobi module and 
the charges associated with the $\mC^*$ action, see formula \eqref{scale}.
 \item The dimension of charge lattice is given by the weights and degrees of the defining equation of the singularity, see formula \eqref{milnor}.
 \item The central charges $a$ and $c$ can be found using formula (\eqref{eq:acformula}, \eqref{c2}, \eqref{alpha}, \eqref{RB} and \eqref{c3}). 
 \end{itemize}
 We also provide a type IIB string theory realization of our SCFTs using the defining data of ICIS.   
 
 Many SCFTs studied in this paper have exactly marginal deformations, and it is expected that one can find the weakly coupled gauge theory descriptions.
 Using the Coulomb branch spectrum, we identify such descriptions for many SCFTs. Those quiver gauge theories often take the form 
 of the (affine) $D$ or $E$ shape quivers (see \cite{Katz:1997eq,Chacaltana:2012ch} for other descriptions of these theories). 
 We compute various physical quantities such as the central charges which are in complete agreement with the results from singularity theory,
 which provides strong evidence for the correctness of the singularity approach.
 
 There are some interesting questions that we would like to study in the future: a) We have identified the weakly coupled gauge theory descriptions by guessing, it 
 is desirable to have a more systematic way of finding all the weakly coupled duality frames, perhaps the hypersurface examples studied in \cite{Xie:2016uqq} can be useful. 
b): In the hypersurface case \cite{Xie:2015rpa}, we identify the BPS quiver as the intersection form of the vanishing cycles of the Milnor fibration; the naive generalization 
to ICIS case does not work so well as the number of the vanishing cycles is bigger than the dimension of the middle homology of the Milnor fibration \cite{ebeling2014monodromy}, it is interesting 
to further investigate this issue. c): The renormalization flow of SCFTs defined by hypersurface singularities has a remarkable semicontinuity property \cite{Xie:2015xva}, similar 
semicontinuity property of ICIS was also studied in \cite{ebeling1998spectral}, however, the details of ICIS case is much less understood, and it is interesting to further study RG 
flows along this line.

\section*{Acknowledgements} 
 YW is supported in part by the U.S. Department of Energy under grant Contract Number  DE-SC00012567.  The work of DX is supported by Center for Mathematical Sciences and Applications at Harvard University, and in part by the Fundamental Laws Initiative of
the Center for the Fundamental Laws of Nature, Harvard University. The work of S.T Yau is supported by  NSF grant  DMS-1159412, NSF grant PHY-
0937443, and NSF grant DMS-0804454. The work of Stephen Yau is supported by NSFC (grant nos. 11401335, 11531007) and  Tsinghua University Initiative Scientific Research Program.

	\bibliography{CIrefs} 
	\bibliographystyle{JHEP}
\end{document}